\documentclass[11pt,a4paper]{article}
\usepackage{jheppub}
\usepackage[utf8]{inputenc}\usepackage[T1]{fontenc} 
\usepackage{amsmath}
\usepackage{amssymb}
\usepackage{graphicx}
\usepackage{slashed}
\usepackage{cases}
\usepackage[toc,page]{appendix}
\usepackage{empheq}
\usepackage[normalem]{ulem} 
\usepackage{xcolor} 
\usepackage{afterpage}
\usepackage{float}
\usepackage{bbold}
\usepackage{hypbmsec}
\usepackage{slashed}                     
\usepackage{braket}  
\usepackage{comment} 

\usepackage[inline,shortlabels]{enumitem}

\allowdisplaybreaks

\newcommand{\wrehbar}{\bar{w}_{\rm re}}
\newcommand{\wreh}{w_{\rm re}}
\newcommand{\Nreh}{N_{\rm re}}
\newcommand{\Treh}{T_{\rm re}}
\newcommand{\Tmax}{T_{\rm max}}
\newcommand{\GG}{\Gamma_\phi}
\newcommand{\GX}{\Gamma_X}
\newcommand{\Nk}{N_{\rm k}}
\newcommand{\NSZ}{\Nk}
\renewcommand{\c}{{\rm c}}
\newcommand{\DD}{D}
\newcommand{\n}{ j}
\newcommand{\epsilonvarM}{\frac{m_\phi}{\mpl}}
\newcommand{\epsilonvar}{\frac{m_\phi}{\phi_{\rm end}}}
\newcommand{\vv}{\mathsf{v}}

\newcommand{\sigmar}{\sigma_r}
\newcommand{\sigmans}{\sigma_{n_s}}
\newcommand{\nsbar}{\bar{n}_s}
\newcommand{\rbar}{\bar{r}}

\newcommand{\lambdaphi }{\lambda_\phi}
\newcommand{\EFTscale}{\Lambda}
\newcommand{\V}{\mathcal{V}}
\newcommand{\Vend}{\mathcal{V}_{\rm end}}
\newcommand{\VI}{\Vend}
\newcommand{\mpl}{m_\mathrm{Pl}}

\renewcommand{\d}{\mathrm{d}}
\newcommand{\g}{\mathsf{g}}
\newcommand{\y}{\mathsf{a}}
\newcommand{\M}{M}
\newcommand{\Vphi}{\V}
\newcommand{\fF}{f_{\rm F}}
\newcommand{\Mpsi}{M_\Psi}
\newcommand{\gPsi}{g}
\newcommand{\yX}{y_X}
\newcommand{\YX}{\mathcal{G}_X}

\newcommand{\X}{X}

\begin{document}

\title{Thermal effects on Dark Matter production during cosmic reheating}

\author[a,b]{Marco Drewes,}
\emailAdd{marco.drewes@uclouvain.be}
\author[c,a]{Yannis Georis,}
\emailAdd{yannis.georis@ipmu.jp}
\author[a]{Mubarak A. S. Mohammed,} 
\emailAdd{mubarak.abdallah@uclouvain.be}
\author[d,e]{Sebastian Zell} 
\emailAdd{sebastian.zell@lmu.de}

\affiliation[a]{Centre for Cosmology, Particle Physics, and Phenomenology, Université catholique de Louvain, Louvain-la-Neuve B-1348, Belgium}
\affiliation[b]{Physik Department, James-Franck-Stra\ss e~1,
Technische Universit\"at M\"unchen,\\
D--85748 Garching, Germany}
\affiliation[c]{Kavli IPMU (WPI), UTIAS, University of Tokyo, Kashiwa, 277-8583, Japan}
\affiliation[d]{Arnold Sommerfeld Center, Ludwig-Maximilians-Universität, Theresienstraße 37, 80333 München, Germany}
\affiliation[e]{Max-Planck-Institut für Physik, Boltzmannstraße 8, 85748 Garching bei München, Germany}

\abstract{
The relic abundance of Dark Matter (DM) produced via thermal freeze-in is sensitive to the thermal history during and after cosmic reheating. In minimal models, this opens up the possibility to make predictions for collider observables by combining the requirement to match the DM relic abundance with observations of the Cosmic Microwave Background (CMB). We assess the impact of thermal corrections to the rate of cosmic reheating and the rate of thermal DM production on CMB observables and the relic abundance. We find that such corrections are generally small in the regime where they can be computed by means of finite-temperature field theory. We construct counter-examples where this general rule is violated. 
}

\maketitle

\section{Introduction}

Dark Matter (DM)  appears to make up more than 80\% of the matter density in the observable universe, but its composition remains unknown and continues to be one of the central unresolved puzzles in particle physics and cosmology \cite{Cirelli:2024ssz}. 
An important avenue to test theories is to establish a predictive connection between the observed abundance and distribution of DM on one hand and experimental signatures in the laboratory on the other. In the simplest scenarios, where DM is produced from non-relativistic \emph{thermal freeze-out}, this connection is very clear \cite{Kolb:1990vq}. However, since the lack of discovery of any new particles with masses and couplings around the electroweak scale has ruled out the simplest scenarios that employ this mechanism within supersymmetric theories and beyond,
theory provides little objective guidance  in the vast landscape of models greatly outnumbering the signatures \cite{Bertone:2018krk}. 

Roughly speaking, one may explain the lack of experimental new particle discoveries in at least two ways: Either the DM particles and mediators are too heavy to be produced at the Large Hadron Collider (LHC), or their couplings are so feeble that their interactions with ordinary matter are too rare to have been observed. In the present work, we focus on the second scenario, which can be realised in many different ways \cite{Albertus:2026fbe}. 
The DM can either be composed of an isolated singlet state or be part of an extended \emph{hidden sector} that communicates with the Standard Model (SM) only through so-called \emph{portals}. 
While the DM particles themselves are detector-stable, the feeble portal couplings can give rise to striking signatures from long-lived mediators, such as displaced vertices, disappearing tracks or missing transverse energy \cite{Alekhin:2015byh,Curtin:2018mvb,Beacham:2019nyx,Alimena:2019zri,Agrawal:2021dbo,Antel:2023hkf}.

In the early universe, the feeble interactions can prevent the hidden particles from reaching thermal equilibrium. The probably simplest way of realizing this is a \emph{thermal freeze-in} \cite{Dodelson:1993je,Hall:2009bx}, i.e., a slow production from a bath in thermal equilibrium. 
Alternatively the DM can be produced in the decay of a heavy particle or condensate or gravitational interactions, see e.g.~\cite{Chu:2011be,Drewes:2016upu,Bernal:2017kxu,Boyarsky:2018tvu,Kolb:2023ydq} for some reviews. 
In such scenarios, the DM particles' abundance and momentum distribution do not only carry information about their interaction, but also about the underlying cosmic history, including the epoch of \emph{cosmic reheating} ~\cite{Albrecht:1982mp,Dolgov:1989us,Traschen:1990sw,Shtanov:1994ce,Kofman:1994rk,Boyanovsky:1996sq,Kofman:1997yn}, i.e., the transitional period between cosmic inflation \cite{Starobinsky:1980te,Guth:1980zm,Linde:1981mu} and radiation domination during which dissipative processes transferred energy from the inflaton into the hot primordial plasma.
It has been know for a long time that reheating may leave an imprint in both the DM abundance \cite{Chung:1998rq,Giudice:2000ex} and the Cosmic Microwave Background (CMB) \cite{Lidsey:1995np}. 
This opens up the possibility to combine the two in minimal models to make predictions for collider experiments, see e.g.~\cite{Bhattiprolu:2022sdd,Becker:2023tvd,Barman:2024tjt,Bhattacharya:2025wef,Mondal:2025awq,Belanger:2026ctm}.

The sensitivity of the CMB to the reheating temperature $\Treh$ primarily comes from the impact of the \emph{expansion history} during reheating -- the time evolution of the scale factor $a$ --  on the redshifting of cosmological perturbations
\cite{Martin:2010kz,Adshead:2010mc,Mielczarek:2010ag,Easther:2011yq,Dai:2014jja,Drewes:2015coa}.\footnote{The thermal gravitational wave background \cite{Ghiglieri:2015nfa} in principle provides a more direct probe of $\Treh$ \cite{Ghiglieri:2020mhm,Ringwald:2020ist},  but detecting it is extremely challenging practice \cite{Drewes:2023oxg}  (unless processes other than thermal emission contribute, cf.~\cite{Caprini:2018mtu,Roshan:2024qnv,Xu:2025wjq} and references therein).}   
This can be parametrised in terms of the average equation of state $\wrehbar$ and number of $e$-folds $\Nreh$ during reheating. In a given model of inflation, $\Nreh$ is the only quantity that is not fixed by the choice of parameters in the inflaton potential $\V(\phi)$, as the duration of reheating is governed by the rate of energy transfer $\GG$ from the inflaton to other degrees of freedom, which in turn depends on the set of microphysical coupling constants $\{\g_i\}$ associated with the inflaton interactions that fuel reheating. 
Specifying $\Nreh$ fixes $\Treh$ at the onset of the radiation dominated era for a given $\V(\phi)$, see \eqref{Tre} below.

However, constraining $\Nreh$ from CMB observations does not necessarily fix the \emph{thermal history} -- the time evolution of the temperature $T$ -- throughout the entire reheating epoch, which determines the amount of thermally produced DM. 
The thermal history during reheating can be reconstructed if the inflaton couplings $\{\g_i\}$ 
can be measured in CMB observations \cite{Drewes:2015coa,Drewes:2019rxn}, and if these same couplings dominate $\GG$ throughout the entire reheating epoch. 
The former is possible if either $\GG$ depends only on a small set of parameters \cite{Drewes:2019rxn} (ideally only a single $\g \in \{\g_i\}$) or if essentially all but a few parameters are already known (e.g.~because the inflaton $\phi$ directly couples to Standard Model particles, as in 
Starobinsky inflation \cite{Starobinsky:1980te},
Higgs inflation \cite{Bezrukov:2007ep} or QCD-driven warm inflation \cite{Berghaus:2025dqi}).
If the thermal history can be reconstructed, the DM abundance can be predicted for a given set of DM couplings $\{\y_i\}$ if the DM is primarily produced from thermal processes. 

In the present work, we 
consider scenarios where the aforementioned conditions are met, i.e., 
\begin{enumerate*}[1)]
\item the same inflaton coupling $\g$ dominates $\GG$ throughout the entire reheating period, 
\item $\g$ is small enough to be measured in CMB observations and 
\item DM is produced via thermal processes.
\end{enumerate*}
Our goal is to investigate whether the relationship between the relic abundance, CMB observables and signatures in collider experiments are modified by thermal corrections. By \emph{thermal corrections} we mean quantum statistics or screening effects in the plasma. 
While such effects are typically negligible when particles freeze out non-relativistically, they can have a sizeable impact on $\GX$ in freeze-in scenarios when the production occurs at temperatures exceeding the DM mass.
Within the regime where perturbation theory can be applied,\footnote{It is well-known that similar corrections are of great importance in the non-perturbative regime \cite{Lebedev:2021tas,Ahmed:2022tfm,Garcia:2021iag,Kainulainen:2024etd,Sopov:2025hhr,Garcia:2025rut}. However, in such a regime it is very difficult to relate collider signatures to cosmological observables in any case, as strong feedback effects tend to introduce a dependence on a large set of unknown parameters, cf.~Sec.~\ref{sec:DMprod&reheating}. } 
thermal corrections can impact observables in two ways: They can either modify the efficiency of reheating characterised by $\GG$ \cite{Kolb:2003ke,Yokoyama:2004pf,Yokoyama:2005dv,Drewes:2010pf,Mukaida:2012qn,Mukaida:2012bz,Mukaida:2013xxa,Drewes:2013iaa,Drewes:2014pfa,Adshead:2016xxj,Tanin:2017bzm,Drewes:2017fmn,Garcia:2020wiy,Ai:2021gtg,Ming:2021eut,Adshead:2019uwj,Ai:2023ahr,Wang:2022mvv,Ai:2023qnr,Minami:2025waa,Bernal:2026dsu}, or they can impact the DM production rate $\GX$ \cite{Beneke:2014gla,Harigaya:2014waa,Drewes:2015eoa,Ghiglieri:2015jua,Binder:2018znk,Hambye:2019dwd,Biondini:2020ric,Binder:2021otw,Binder:2020efn,Biondini:2023hek,Bringmann:2021sth,Becker:2023vwd,Ai:2023qnr,Bouzoud:2024bom,Becker:2025lkc,Becker:2025yvb,Biondini:2025jvp,Biondini:2025ihi,Binder:2026fwe}.

For the sake of definitenesses, we consider plateau-type inflationary models -- such as $\alpha$-attractor T models ($\alpha$-T) \cite{Kallosh:2013maa,Carrasco:2015pla,Carrasco:2015rva}, 
Radion Gauge Inflation (RGI) \cite{Fairbairn:2003yx}
and Mutated Hilltop Inflation (MHI) \cite{Pal:2009sd,Pal:2010eb}
-- for which the phenomenology of reheating has already been studied in detail \cite{Ueno:2016dim,Nozari:2017rta,DiMarco:2017zek,Drewes:2017fmn,Maity:2018dgy,Rashidi:2018ois,German:2020cbw,Mishra:2021wkm,Ellis:2021kad,Drewes:2022nhu,Drewes:2023bbs,Haque:2025uri,Chakraborty:2023ocr,Mondal:2025kur,German:2024rmn},
but little attention has been given to thermal effects. 
In \cite{Drewes:2017fmn,Haque:2023yra} it was found in a specific set of scenarios that thermal corrections to $\GG$ do not affect CMB observables within the regime where perturbation theory can be applied. One goal of the present work is to assess how general this conclusion is. A second goal is to investigate the role of thermal corrections to $\GX$ and their impact on the DM relic abundance. 
Finally, we estimate at what level upcoming CMB observations can make predictions for collider experiments.

The structure of this paper is the following. 
In Sec.~\ref{sec:DMprod&reheating}, we detail the set of equations used to describe the reheating dynamics as well as the DM production. 
In addition, we briefly describe the qualitatively different regimes that reheating can undergo and their implication for our ability to reconstruct the complete thermal history of the universe.
Then, in Sec.~\ref{sec:ThermaleffectsInflaton}, we discuss the conditions under which finite-temperature corrections to the inflaton decay rate $\Gamma_\phi$ can significantly affect the overall DM yield. 
In the next section, Sec.~\ref{sec:ThermaleffectsDMprod}, we investigate whether thermal corrections to the DM production rate itself can impact the predictions for the DM relic abundance. 
Finally, in Sec.~\ref{sec:counterexamples}, we provide a set of three counter-examples to the general rule derived in Sec.~\ref{sec:ThermaleffectsInflaton}, showing the thermal corrections to $\Gamma_\phi$ can in specific case drastically modify the expected present DM abundance.
We conclude in Sec.~\ref{sec:conclusion}. We also provide three appendices discussing how to relate $\Treh$ to CMB observables (appendix \ref{TrefromCMB}), deriving 
useful relations for 
plateau models of inflation (appendix \ref{app:PlateauModels}) as well as the form of the fundamental coupling of the inflaton to fermions in $\alpha$-attractor T models (appendix \ref{YukawainalphaT}), respectively.

\section{DM production during and after reheating}
\label{sec:DMprod&reheating}

We start by briefly reviewing the description of reheating and of DM production within this epoch.
The time evolution of the inflaton field during inflation and the subsequent reheating can be 
described by the following differential equation\footnote{\label{EoMfoot}A priori it is not obvious that a Markovian equation of the form \eqref{eom} accurately describes the evolution throughout inflation and reheating, but it can be justified for the present purpose, see appendix C in \cite{Drewes:2019rxn}. }
\begin{align}\label{eom}
   \ddot{\phi} + (3H+\GG)\dot{\phi} + \partial_\phi \V(\phi) = 0 \ ,
\end{align}
where $\V$ is the 
effective 
inflaton potential, $H \equiv \dot{a}/a$, with $a$ being the scale factor and $\dot{a}$ its time derivative, denotes the usual Hubble rate and $\GG$ the inflaton friction rate. 
Assuming a standard  (cold) single field inflation scenario, 
we can distinguish three epochs, neglecting the brief transitional periods between them:
An extended phase of cosmic inflation ($\ddot{a} > 0$, $\V \gg \rho_R$) lasting at least several tens of $e$-folds $\Nk$,
a period of cosmic reheating ($\ddot{a} < 0$, $\V \gg \rho_R$) lasting $\Nreh$ $e$-folds, 
and, finally, the epoch of radiation domination ($\ddot{a} < 0$, $\V \ll \rho_R$), with $\rho_R$ the radiation energy density.
Roughly speaking, reheating begins when the slow-roll parameters 
$\epsilon = \frac{\mpl^2}{2}\left(\partial_\phi \V/\V\right)^2$ and $\eta  =  \mpl^2\partial_\phi^2 \V/\V$ become of order unity,\footnote{This statement is only valid at leading order in the slow-roll parameters, but suffices for the present purpose, see appendix B in \cite{Drewes:2023bbs}. } 
and it ends when $\GG = H$.
If $\Nreh$ is sizeable enough to leave an observable imprint in the CMB, the inflaton oscillates many times during reheating and we can use the average
of $\rho_\phi \equiv \dot{\phi}^2/2+\V(\phi)$
instead of tracking individual oscillations, yielding the well-known equation
\begin{align}
    \dot{\rho}_\phi + (3H+\GG)(\rho_\phi+P_\phi) &= 0\  ,
    \end{align}
with $P_\phi = w \rho_\phi$ the effective pressure of the inflaton field and $w$ the equation of state parameter. 
In the present work, we focus on the regime where 
the inflaton field predominantly decays perturbatively\footnote{
We use the word \emph{perturbative} whenever a quantity can be computed by expanding in some small parameter, no matter whether its dependence on this parameter is a series of integer powers or not. This in particular includes resummed perturbation theory and time-dependent perturbation theory.}
into radiation, i.e., particles with relativistic momenta. 
The radiation density $\rho_R$ evolves according to
 \begin{align}
 \label{eq:radiation_evolutionequation}
     \dot{\rho}_R  +4H\rho_R-\GG \left(\rho_\phi+P_\phi\right) = 0\ .
 \end{align}
If the interactions in this plasma are strong enough to establish kinetic equilibration \cite{Berges:2008wm,Mazumdar:2013gya,Harigaya:2013vwa,Harigaya:2014waa,Mukaida:2015ria,Mukaida:2024jiz} instantaneously (i.e., on a time scale much shorter than all other relevant timescales), one can describe the occupation numbers in terms of an effective temperature  $\frac{\pi^2 g_\star}{30}T^4 \equiv \rho_R$, with $g_\star$ parametrising the effective number of degrees of freedom contributing to the energy density of the thermal bath. 

We further assume that the DM particles $X$ are produced thermally from this radiation bath, and 
that the evolution of their number density $n_X$ follows the equation
\begin{align}
    \dot{n}_X + 3H n_X &= \mathcal{C}_{X}\ .
\end{align}
Here, $\mathcal{C}_{X}$ denotes the collision term. For the examples used in this work, it can be parametrised as \cite{Drewes:2012qw}\footnote{Note that $\GX$ is in general a functional of $f_X$.}
\begin{eqnarray}\label{CollInTermsOfGammaX}
\mathcal{C}_{X} = \int \GX (f_X^{\rm eq} - f_X) \  .
\end{eqnarray}
If the DM is composed of a single or a few particle species while the radiation bath is composed of a large number of degrees of freedom, the change in $\rho_R$ due to the DM production can be safely neglected.

For practical purposes, it is useful to rewrite these three equations in terms of the dimensionless variables $x \equiv m_\phi a$, $\Phi \equiv \rho_\phi a^3/m_\phi = \rho_\phi x^3/m_\phi^4$, $R \equiv \rho_R a^4 = \rho_R x^4/m_\phi^4$ and $X \equiv n_X a^3= n_X x^3/m_\phi^3$, where $m_\phi$ is the inflaton mass at the minimum of its potential.
Here the inflaton mass $m_\phi$ is defined by the expansion $ \V = \sum_\n\frac{\vv_{\n}}{\n !} \frac{\phi^\n}{\EFTscale^{\n-4}}  = \frac{1}{2}m_\phi^2\phi^2+\frac{g_\phi}{3!}\phi^3+\frac{\lambdaphi }{4!}\phi^4 + \ldots$, with $\EFTscale$ an appropriately chosen mass scale.
The dimensionless scale factor $x$ is chosen to be $1$ at the beginning of reheating.
The aforementioned equations then become
\begin{align} \label{eq:reducedform_withDM}
    \frac{\mathrm{d}}{\mathrm{d} x}\Phi = -\frac{\GG}{H x}\Phi \ , \quad 
    \frac{\mathrm{d}}{\mathrm{d} x}R &= \frac{\GG}{H}\Phi \ , \quad 
    \frac{\mathrm{d}}{\mathrm{d} x}X = \frac{\mathcal{C}_{X}x^2}{H m_\phi^3}\ .
    \end{align}
The Hubble rate and temperature can self-consistently be obtained from the following equations
\begin{subequations}\label{HandT}
    \begin{align}
        H &=\frac{m_\phi^2}{\sqrt{3}\mpl}\sqrt{\frac{R}{x^4}+\frac{\Phi}{x^3}+\frac{X}{x^2}\frac{m_{X}}{m_\phi}}\ ,\label{eq:redshifting}\\
        T &= \frac{m_\phi}{x}\left(\frac{30 R}{\pi^2 g_\star}\right)^{1/4}\ ,
    \end{align}
\end{subequations}
where $\mpl \approx 2.43 \cdot 10^{18}$ GeV denotes the reduced Planck mass.
In Eq.~\eqref{eq:redshifting} we have assumed that the inflaton condensate redshifts like matter, implying $w=0$, 
which can be justified if $\V$ is approximately parabolic near its minimum.\footnote{We assure the validity of this approximation by imposing the condition \eqref{GeneralScalingSelf} below. Nevertheless, modelling the primordial universe as a sum of components with $w=0$ and $w=1/3$ of course represents an approximation, cf.~\cite{Lozanov:2016hid,Saha:2020bis,Antusch:2020iyq,Antusch:2025ewc}.}
Moreover, we set the effective number of radiation degrees of freedom to its SM value $g_\star = 106.75$.
Under the aforementioned assumptions, the set of equations~\eqref{eq:reducedform_withDM} and \eqref{HandT} is closed; solving them permits to reconstruct both the expansion history and the thermal history of the universe.
Of particular interest is a determination of the maximal temperature during reheating $\Tmax$  and the maximal temperature in the radiation epoch $\Treh$. The latter is commonly referred to as \emph{reheating temperature} and can be defined as the value of $T$ at the moment when $\rho_R=\rho_\phi$ for the last time, i.e.,
\begin{align}
\label{eq:defreheatingtemp}
\rho_R = \rho_\phi \ {\rm when} \ T = \Treh \ .
\end{align}
It can be related to $\Nreh$ via the standard redshifting relation 
\begin{equation}\label{Tre}
	\Treh=\exp\left[-\frac{3(1+\wrehbar)}{4}N_{\rm re}\right]\left(\frac{40 \ \Vend }{g_\star\pi^2}\right)^{1/4}\ ,
\end{equation}
where $\Vend$ denotes the value of the inflaton potential at the end of inflation, $\wrehbar$ the averaged equation of state parameter during reheating and $\rho_{\rm end} \simeq 4\Vend/3$ at the end of inflation accounting for the non-vanishing kinetic energy of the inflaton field.

Predicting the relation between a given DM interaction (which may be probed in collider experiments) and the relic abundance requires knowledge of the thermal history during reheating.
Within the framework considered here, information on $\Treh$ can always be obtained 
by relating the quantities on the RHS of Eq.~\eqref{Tre}
to CMB observations --
parametrised by the amplitude $A_s$ and the spectral index $n_s$ of the scalar perturbations as well as the tensor-to-scalar ratio $r$ -- 
using the well-known relations reviewed in appendix \ref{TrefromCMB}.
With three observables at hand, observations can fix at most three parameters. Most of the inflationary models that are consistent with present CMB observations 
have at least two parameters in the potential $\V$ \cite{Martin:2013tda}; in the examples given in Appendix \ref{app:CMBconstraints_inflationaryparam}  -- $\alpha$-T, RGI and MHI -- these are the scale $\M$ of the inflaton potential
and a parameter $\alpha$ that controls the ratio $m_\phi/\M$. Hence, at most one coupling $\g$ can be measured with CMB observations. 

Regarding the ability to reconstruct the complete thermal history during reheating from CMB observations in a given model of inflation, one can distinguish three regimes \cite{Liu:2025sut}. 
\begin{enumerate}[(i)]
	\item \label{it:reg1}
    For a sufficiently small inflaton coupling $\g$, non-perturbative particle production due to instabilities in the field equations (such as parametric resonance, tachyonic instabilities or inflaton fragmentation) is either non-existent or subdominant at all times.
    If $\V$ is approximately parabolic near its minimum, the system can be treated as a perturbed harmonic oscillator \cite{Ai:2021gtg}.
    Moreover, Hubble expansion ensures that the occupation numbers in the radiation bath remain sufficiently low so that their backreaction on $\GG$ can be neglected. Then $\GG$ 
    is calculable by means of perturbation theory 
    during the entire reheating epoch
    from the knowledge of $\g$ and the inflaton mass $m_\phi$ alone.
    For constant $\GG$, the set of equations formed by \eqref{eq:reducedform_withDM} and \eqref{HandT} can be solved analytically \cite{Giudice:2000ex}, yielding
    	\begin{eqnarray}
		\Treh \approx \sqrt{\GG \mpl }\left(\frac{90}{\pi^2g_\star}\right)^{1/4}\Big|_{\GG=H}\ .
        \label{eq:approxTr}
	\end{eqnarray}
and 
\begin{align}
\label{eq:Tmaxrelation}
          \Tmax & \approx 0.9 \left(\GG^2 \mpl^2 \VI/g_\star^2\right)^{1/8}\ ,
\end{align}
where we approximated the value of $\V(\phi)$ at the onset of reheating by its value $\Vend$ at the end of inflation. 
	In this regime, observational constraints on $\Treh$ can directly be related to $\GG$, 
\begin{equation}
	\GG
	\approx \frac{\Treh^2}{\mpl}\frac{\sqrt{g_\star}}{3}\ ,\label{GammaConstraint}
\end{equation}
    which in turn can be expressed in terms of the inflaton mass $m_\phi$ and coupling $\g$.  
	\item \label{it:reg2} 
    For intermediate values of $\g$, instabilities and backreaction from the produced particles affect $\GG$ during the early stages of reheating, but do not alter the moment when $\GG=H$. Hence, they modify the thermal history during reheating, but leave the expansion history unaffected.
    $\Treh$ is still given by \eqref{eq:approxTr}, but $\Tmax$ can significantly differ from \eqref{eq:Tmaxrelation}.
    The value of $\GG$ at the end of the reheating can still be obtained from \eqref{GammaConstraint} and $\g$ may be extracted from this, but the thermal history during reheating can only be reconstructed from this knowledge if instabilities and backreaction can be modelled. 
    This feedback typically depends not only on $\V$ and $\g$, but also the properties of the produced particles and their interactions with each other,\footnote{For instance, the occupation numbers of particles directly produced in inflaton decays depend on their interactions with each other and  with other particles that do not directly couple to $\phi$. The latter determine how quickly their occupation numbers are depleted by decays into secondary decay products, the former control the rate of their kinetic equilibration as well as their thermal masses that impact the inflaton decay phase space.} and thereby introduces a dependence of $\GG$ on a set of microphysical parameters that is too large for all of them to be constrained from only three observables $\{A_s,n_s,r\}$.
    However, this is still possible in scenarios where the leading effects come from quantum statistics alone, or if they only depend on at most three parameters from the set $\{\g, \vv_i\}$
    (in addition to known SM interactions).
    \item \label{it:reg3}  
    For large $\g$, non-linearities and backreactions modify both the thermal history and the expansion history. 
While in principle it is possible that these effects only depend on at most three parameters from the set $\{\g, \vv_i\}$ that can be constrained from observations, 
    in practice it is hard to confine their impact to such a restricted sector if they are strong enough to modify $\Treh$.  
    Hence $\GG$ is practically not calculable  during the whole reheating process without further model assumptions  even if the same inflaton coupling $\g$ dominates reheating throughout the entire process, and the thermal history during reheating cannot be reconstructed.\footnote{Reconstructing the thermal history in regime \ref{it:reg3} is possible in minimal models with few or no unknown parameters beyond $\V$ and $\g$, such as in
    Starobinsky inflation \cite{Starobinsky:1980te},
    Higgs inflation \cite{Bezrukov:2007ep} (see \cite{Rubio:2019ypq,Dux:2022kuk}) or QCD-driven warm inflation \cite{Berghaus:2025dqi}.\label{foot:exceptions}} 
\end{enumerate}
From a methodological viewpoint, the computation of $\GG$ in the different regimes requires different approaches.
In regime \ref{it:reg1}, $\GG$ is in good approximation given by the vacuum decay rate for inflaton particles if the interaction is linear in $\phi$ and can be computed by means of time-dependent perturbation theory otherwise \cite{Ichikawa:2008ne}. Corrections to this can be systematically computed by means of multi-scale perturbation theory \cite{Ai:2021gtg}.
In regime \ref{it:reg2}, 
a decisive question is whether the thermalisation of the decay products occurs faster or slower than the frequency of the inflaton oscillations. 
In the former case,\footnote{This requirement can be relaxed to the rate of thermalisation being faster than $1/\GG$ once the effective masses of quasiparticles in the plasma are dominated by forward scattering (rather than their coupling to the inflaton condensate).} 
corrections to $\GG$ can be computed by means of thermal quantum field theory \cite{Mukaida:2012qn,Mukaida:2012bz,Drewes:2013iaa} and the set of equations \eqref{eq:reducedform_withDM} may be solved analytically \cite{Drewes:2014pfa}.
If this is not the case, a set of coupled quantum kinetic equations must be solved, which can e.g.~be derived in the framework of the Closed Time Path (CTP) formalism \cite{Ai:2023qnr}. 
In regime \ref{it:reg3}, a perturbative treatment is in general not possible. Fundamentally, one should solve the coupled set of equations for the condensate and correlation functions in nonequilibrium quantum field theory \cite{Berges:2002cz,Berges:2010zv,Mukaida:2013xxa,Kainulainen:2024etd} or resort to analogue systems \cite{Chatrchyan:2020cxs,Steinhauer:2021fhb}. In practice it is common to approximate the former by classical lattice simulations, see e.g.~\cite{Lebedev:2021tas,Ahmed:2022tfm,Garcia:2021iag,Sopov:2025hhr,Garcia:2025rut}.\footnote{The application of the perturbative approach from \cite{Ichikawa:2008ne,Almeida:2018oid} beyond regime \ref{it:reg1} to study DM production is conceptually questionable and introduces an error the size of which cannot be quantified within this scheme. The range of validity of perturbative methods can be extended by the introduction of effective time dependent masses and decay rates \cite{Garcia:2020wiy} in the averaged equations of motion \eqref{eq:reducedform_withDM}, though the results found in \cite{Garcia:2021iag} indicate that this approach can already fail within regime \ref{it:reg2}, implying that the results obtained for several DM candidates may need revision, cf.~Sec.~\ref{sec:counterexamples}.}
Apart from the technical difficulty of accurately simulating the reheating process, regime \ref{it:reg3} also poses the aforementioned conceptual challenge that the thermal history depends on a set of microphysical parameter that is too large to be constrained from observation even within a given model of inflation.

The value of $\g$ that marks the transition between regimes \ref{it:reg2} and \ref{it:reg3} 
for a general inflaton interaction of the form $\g \phi^\n \EFTscale^{4-\DD}\mathcal{O}[\{\mathcal{X}_i\}]$ can be conservatively estimated as \cite{Drewes:2019rxn}
\begin{subequations}
\label{PertReh}
\begin{eqnarray}
	|\g| &\ll&\left(\frac{m_\phi}{\phi_{\rm end}}\right)^{\n-\frac{1}{2}}
	{\rm min}\left(
	\sqrt{\epsilonvarM}
	,
	\sqrt{\epsilonvar}
	\right)
	\left(\frac{m_\phi}{\EFTscale}\right)^{4-\DD} \ , \label{GeneralScaling}\\
    	|\vv_\n| &\ll& \left(\frac{m_\phi}{\phi_{\rm end}}\right)^{\n-\frac{5
		}{2}}
	{\rm min}\left(
	\sqrt{\epsilonvarM}
	,
	\sqrt{\epsilonvar}
	\right)
	\left(\frac{m_\phi}{\EFTscale}\right)^{4-\n}\ .\label{GeneralScalingSelf}
\end{eqnarray}
\end{subequations}
Here $\mathcal{O}[\{\mathcal{X}_i\}]$ is a general operator of mass dimension $\DD-\n$ involving fields $\mathcal{X}_i$ of arbitrary spin. 
The former condition \eqref{GeneralScaling} directly refers to the value of $\g$ while the latter relation \eqref{GeneralScalingSelf} assures that there are no strong feedback effects from inflaton self-interactions, i.e., \emph{fragmentation}.

In \cite{Becker:2023tvd} it has been shown that, in a given model of inflation and for a specified DM model,
collider signatures can be predicted by combining CMB data with the observed relic abundance in regime \ref{it:reg1}.\footnote{While the authors of \cite{Becker:2023tvd} have chosen values for $\g$ that are small enough to justify a perturbative treatment, the part of their study using potentials $\V(\phi) \propto \phi^\n$ with $\n>2$ violates condition \eqref{GeneralScalingSelf}, implying that inflaton fragmentation may introduce unaccounted errors \cite{Garcia:2023dyf,Bhusal:2025oqg}. 
} 
This is possible because in regime \ref{it:reg1} $\GG$ is approximately constant, so that the entire thermal history during reheating can be reconstructed with the help of \eqref{GammaConstraint}, provided that $\Treh$ has been determined by relating the RHS of \eqref{Tre} to CMB observables, cf.~appendix \ref{TrefromCMB}.
In the present work, we address the question to what degree this approach can be extended beyond regime \ref{it:reg1}. 
A reconstruction of the thermal history during reheating based on the limited set of observables $\{A_s,n_s,r\}$ is usually impossible in regime \ref{it:reg3}.

The three relevant physical scales are the scale of the inflaton potential $\M$,
the inflaton mass $m_\phi$ and the energy density at the end of reheating \eqref{Tre}, which we parametrise by $\Treh$.  
The scale $\M$ is determined by the value of $r$, 
in plateau models one finds\footnote{This general result can be obtained from Eq.~\eqref{H_k} in the appendix; for the three examples considered there one finds \eqref{M3}, \eqref{M2}, \eqref{M1}.}
  \begin{eqnarray}\label{Mgeneral}
   \M \approx
     \mpl
     \left(
     \frac{3\pi^2}{2}
     A_s r
     \right)^{1/4} \lesssim 1.5 \cdot 10^{16} \mathrm{ GeV}\ ,
\end{eqnarray}
where we used that $A_s \approx 2 \cdot 10^{-9}$ and $r\lesssim 0.03$ \cite{Planck:2018vyg,BICEP:2021xfz}.
 We show in appendix \ref{mphigenericPlateauPotential} 
that $m_\phi$ is generally given for two-parameter inflationary scenarios by
\begin{equation} \label{mPhiPlateau}
m_\phi \approx \frac{\sqrt{24 \pi^2 A_s}}{\NSZ} \mpl \approx 10^{13} \mbox{~GeV}\ ,
\end{equation}
see \eqref{mPhiPlateauAppendix},
and confirm this explicitly in three different models in appendix \ref{app:CMBconstraints_inflationaryparam}.
Here $\NSZ$ is the number of $e$-foldings between the generation of CMB perturbations and the end of inflation. 
We note in passing that Eq.~\eqref{mPhiPlateau} also allows to express the inflaton energy density at the beginning of the reheating era as
\begin{equation}\label{PhiInitialMax}
    \Phi = \frac{\V}{m_\phi^4} \approx \frac{r \NSZ^4}{384\pi^2 A_s} \ .
\end{equation}
Since the tensor-to-scalar ratio is bounded as $r\lesssim 0.03$, we see that $\Phi \lesssim 9 \cdot 10^{10}$ in the regime of validity of Eq.~\eqref{mPhiPlateau}. 
Constraints on $\Treh$ can be extracted from observational data by means of the well-known approach summarised in appendix \ref{TrefromCMB}. 
The best sensitivity to $r$ in the foreseeable future can be expected from the
extended Simons Observatory \cite{SimonsObservatory:2025avm}
and the Japanese LiteBIRD satellite \cite{LiteBIRD:2022cnt}.
The Chinese AliCPT-1 array's \cite{Li:2017drr,Li:2018rwc} 
sensitivity to reheating in the currently foreseen configuration \cite{Liu:2025sut} is not competitive with LiteBIRD,
but the array has the potential to be extended and is complementary to the Simons Observatory due to its geographic location in the Northern Hemisphere.
We here take LiteBIRD as our benchmark experiment. To estimate its sensitivity to reheating in Tab.~\ref{tab:sensitivities}, 
we use the analytic approach from \cite{Drewes:2022nhu}
which has been verified by comparing to Monte
Carlo Markov Chain based forecasts in \cite{Drewes:2023bbs}.
We provide the basic ingredients in appendix \ref{TrefromCMB}.

\bgroup
\def\arraystretch{1.5}
\begin{table}
\centering
\begin{tabular}{c|c|c||c|c|c|c||c}
benchmark &model  & $\alpha$ & $\nsbar$ & $\sigmans$ & $\rbar$ & $\sigmar$ & ${\rm log}_{10}(\frac{T_{\text{re}}}{\text{GeV}})$\\
\hline
\hline
  \begin{enumerate*}[A]
\item \label{it:A}
\end{enumerate*}
&
  $\alpha$-T & $2$ &$0.9651$ & $0.002$ & $0.00696$ & $0.001$ & $13.7^{+1.2}_{-2.0}$ \\
\hline
  \begin{enumerate*}[B]
\item \label{it:B}
\end{enumerate*}
&
  RGI   & $20$ & $0.9607$ & $0.002$ & $0.0265$ & $0.00175$ & $-0.8^{+1.5}_{-1.0}$ \\
 \hline 
   \begin{enumerate*}[C]
\item \label{it:C}
\end{enumerate*}
&
MHI & 2 & $0.9623$ & $0.002$ & $0.00947$ & $0.0013$ & $ 6.8^{+3.0}_{-2.7}$ \\
\end{tabular} 
\caption{Forecasted $1\sigma$ sensitivities of LiteBIRD to the reheating temperature $\Treh$ for three benchmark scenarios \ref{it:A}-\ref{it:C}. 
To obtain these constraints, we followed the procedure highlighted in \cite{Drewes:2022nhu}, using projected LiteBIRD sensitivities from \cite{LiteBIRD:2022cnt}.}
  \label{tab:sensitivities}
\end{table}

If $\GG$ is constant, as assumed in \cite{Becker:2023tvd}, 
this can directly be converted into knowledge on the rate $\GG$ that is used in \eqref{eq:reducedform_withDM} to reconstruct the thermal history during reheating. 
The assumption of constant $\GG$ can generally only be justified in regime \ref{it:reg1}. 
However, as long as $\GG$ is calculable in terms of the parameters in $\V$ and $\g$ alone, a constraint on $\Treh$ can still be translated into information on $\g$ in regime \ref{it:reg2}, which in turn allows to compute $\GG$ as a function of $T$ throughout reheating. This motivates two questions. 
\begin{itemize}
\item In Sec.~\ref{sec:ThermaleffectsInflaton}, we address the question whether the approach used in \cite{Becker:2023tvd} can be extended into regime \ref{it:reg2}, i.e., if there is a regime where thermal corrections to $\GG$ are both calculable and relevant in the context of DM production. 
More specifically, we ask whether within regime \ref{it:reg2} there exists a range of parameter values where $\GG$ as a function of $T$ can be computed from knowledge of $m_\phi$ and $\g$ alone by means of finite-temperature Quantum Field Theory (QFT).

\item Thermal corrections to $\GX$ can potentially be relevant in all three regimes. In particular, they can strongly modify $\GX$ during the period when $T$ considerably exceeds the masses of the involved particles. In Sec.~\ref{sec:ThermaleffectsDMprod}, we investigate whether thermal corrections can significantly modify the relic abundance within the range of parameter values where they can be computed by means of finite-temperature QFT.
\end{itemize}
We find that thermal corrections to both $\GG$ and $\GX$ generically only lead to sub-leading changes in the DM relic density in regime \ref{it:reg2}. In Sec.~\ref{sec:counterexamples}, we present three exceptions from this general rule.

\section{Thermal effects on the inflaton decay rate and related DM abundance}
\label{sec:ThermaleffectsInflaton}
A necessary condition to predict the abundance of thermally produced DM is the knowledge of the thermal history during the period when it was produced.
In the following, we focus on scenarios in which the interactions that dominate the dissipative energy transfer from the inflaton condensate to other degrees of freedom can be characterised by a single microphysical coupling constant throughout the reheating process. Since most viable models of inflation require two parameters in $\V$ -- $\M$ and $\alpha$ in the models considered here -- the three observables $\{A_s, n_s, r\}$ can at most fix one additional parameter $\g$.
Extending the approach from \cite{Becker:2023tvd} into regime \ref{it:reg2} 
and observing the imprint of finite-temperature corrections to the inflaton decay rate on the DM relic abundance relies on three conditions. 

\begin{enumerate}
\item \label{cond1} Firstly, not being in regime \ref{it:reg1} -- where the impact of thermal corrections to $\GG$ on the DM relic density is by definition negligible and the results from \cite{Becker:2023tvd} can be applied --  imposes the lower bound \eqref{NotIn1} on $\g$. 
For a thermalised plasma, a simple criterion to estimate the transition between regimes \ref{it:reg1} and \ref{it:reg2} is $\Tmax = m_\phi$.
Assuming that dissipation in regime \ref{it:reg1} is dominated by perturbative inflaton decays,
$\GG$ can be parametrised as\footnote{
Since $\phi_{\text{end}}$ can be of the order of $\mpl$ in plateau models,
cf.~appendix \ref{app:PlateauModels}, the decay products typically receive a huge time dependent mass due to their coupling to the oscillating inflaton condensate. However, while this in principle invalidates the use of the perturbative vacuum decay rate to estimate $\GG$, it turns out that 
Eq.~\eqref{GammaPerturbative} 
provides a reasonable approximation for the averaged decay rate in regime \ref{it:reg1} \cite{Garcia:2020wiy,Garcia:2021iag}.} 
\begin{eqnarray}\label{GammaPerturbative}
\GG = \g^2 m_\phi/\c \ ,
\end{eqnarray}
with $\c$ a numerical constant,
and~\eqref{eq:Tmaxrelation} then yields the condition
\begin{eqnarray}\label{NotIn1}
    \g > \sqrt{\c g_\star \frac{m_\phi}{\mpl}}\frac{m_\phi}{\Vend^{1/4}}
\end{eqnarray}
in order to lie in regime \ref{it:reg2}. In plateau models as those considered here one typically finds 
$\Vend\approx \M^4$ and $\M\approx 10^{16}$ GeV, cf.~\eqref{Mgeneral}.
Assuming in addition that the radiation bath is primarily composed of SM particles (i.e., $g_\star=106.75$), this implies\footnote{We note in passing that, even if one considers a more minimal bath with $g_\star=1$, this lower bound on the inflaton coupling is only relaxed by a factor of 3 approximately.} that the bound \eqref{NotIn1} reads
\begin{align}
\label{eq:lowerboundcoupling}
    \g \gtrsim 1.5 \cdot 10^{-4} \left(\frac{m_\phi}{10^{13} \mbox{~GeV}}\right)^{3/2}
\end{align}
in order for thermal effects to matter.
Combining Eq.~\eqref{mPhiPlateau} with Eq.~\eqref{eq:lowerboundcoupling} implies
\begin{eqnarray}\label{gLowerBound}
\g \gtrsim 1.5 \cdot 10^{-4}.
\end{eqnarray}
\item Avoiding regime \ref{it:reg3} requires to fulfil conditions \eqref{GeneralScaling} and \eqref{GeneralScalingSelf}, which for plateau models roughly read \cite{Drewes:2019rxn} 
\begin{eqnarray}\label{YippieYaYaySchweinebacke}
	|\vv_\n| \ll \left(3\pi^2  r A_s\right)^{(\n-2)/2} \ , \qquad
	|\g| \ll \left(3\pi^2  r A_s\right)^{\n/2}\ .
\end{eqnarray}
Since $A_s \approx 2 \cdot 10^{-9}$ and $r<0.03$, this implies 
\begin{eqnarray}\label{gUpperBound}
\g \lesssim 10^{-4}\ , 
\end{eqnarray}
which is in tension with \eqref{gLowerBound}. Concluding that regime \ref{it:reg2} does not exist at all would be premature, as we have neglected various numerical factors, and the conditions \eqref{GeneralScaling} and \eqref{GeneralScalingSelf} from which \eqref{YippieYaYaySchweinebacke} was obtained are conservative.
However, one may safely conclude that the range of $\g$ in which
thermal corrections to $\GG$ affect the DM relic density and one can  rely on perturbative computations to quantify them without further model assumptions is very limited.\footnote{This is e.g.~confirmed by lattice simulations in figure 4 in \cite{Garcia:2021iag}.}

\item \label{it:UVFreezeIn} Finally, 
in order for thermal corrections to $\GG$ to affect the DM relic density, 
a significant fraction of the presently observed DM must have been produced during the part of the reheating epoch when thermal effects do matter. 
When DM is produced from renormalisable interactions during the radiation dominated epoch, it is well known \cite{Elahi:2014fsa}  that the current relic abundance is primarily produced towards the end of the freeze-in process, as the ratio between $\GX\propto T$ and $H\propto T^2$ increases with time. 
This scenarios is often referred to as \emph{infrared (IR) dominated freeze-in}.
If the freeze-in is terminated by the Boltzmann suppression of a mediator or a threshold effect, then thermal corrections to $\GX$ tend to be suppressed due to the smallness of $T$ compared to the concerned particles' masses.  
In order to realise an \emph{ultraviolet (UV) dominated freeze-in} -- a scenario where the relic abundance is primarily produced at the earliest times -- the DM production rate must grow sufficiently quickly with $T$. In the radiation dominated epoch this requires at least $\GX\propto T^{2+\delta}$, $\delta$ being a positive number. 
During reheating, it is more difficult to realise UV-dominated freeze-in due to the different equation of state $\wreh$ compared to radiation domination.
Even though the interaction strength of $d>4$-operators scales with a positive power of temperature, it is still possible that DM production is not UV-dominated, i.e., that particles produced around $\Tmax$ give a negligible contribution to the final DM abundance.
Schematically, if DM is produced from an operator of dimension $d$, one expects $\Gamma_X \propto T^{2(d-4)+1}$ and 
\begin{align}
\label{eq:CXScaling}
    \mathcal{C}_{X} \propto \GX n_X^{\rm eq} \propto T^{2(d-4)+4}
\end{align}
in the relativistic regime. 
Considering an averaged equation of state parameter during reheating $\wrehbar$, the dimensionless inflaton energy density $\Phi$ scales as 
\begin{align}
\label{eq:PhiScalingGeneral}
    \Phi \sim x^{-3\wrehbar}
\end{align}
such that 
\begin{align}
\label{eq:Hubblereheating}
    H \sim \sqrt{\frac{\Phi}{x^3}} \sim x^{-3(1+\wrehbar)/2}\ .
\end{align}
Hence, inserting the scalings \eqref{eq:PhiScalingGeneral} and \eqref{eq:Hubblereheating} into Eq.~\eqref{eq:reducedform_withDM}, we obtain
\begin{align}
\label{eq:TemperatureScaling_reheatinggeneral}
    R \sim x^{(5-3\wrehbar)/2} \mbox{ and } T \sim x^{-3(1+\wrehbar)/8}\ .
\end{align}
Finally, assuming a temperature-independent inflaton decay rate $\GG$ in regime \ref{it:reg1} and combining Eqs.~\eqref{eq:CXScaling}, \eqref{eq:Hubblereheating} and \eqref{eq:TemperatureScaling_reheatinggeneral}, we obtain that 
\begin{align}
\label{Gleichung39}
    \frac{\mathrm{d}}{\mathrm{d} x}X \propto x^{(20-3d-3\wrehbar(d-4))/4} \ .
\end{align}
Hence, DM production is UV-dominated if\footnote{Of course the simplified scaling relation \eqref{Dfine1} can only be used if the equation of state parameter $w$ varies only mildly during reheating and does not deviate much from its average value $\wrehbar$.  }
\begin{eqnarray}\label{Dfine1}
    d \geq \frac{4(2+\wrehbar)}{1+\wrehbar}\ .
\end{eqnarray}
For a monomial potential $\Vphi(\phi)\propto \phi^\n$, the averaged equation of state parameter is $\wrehbar = (\n-2)/(\n+2)$ \cite{Turner:1983he}, and \eqref{Dfine1} reads $d \geq 6 + 4/\n$.
In the case of a purely quadratic potential 
$\wrehbar = 0$, \eqref{Dfine1} yields $d\geq 8$ \cite{Chen:2017kvz}, while in a quartic potential with $\wrehbar = 1/3$ one finds $d\geq 7$.
\end{enumerate}
The last point \ref{it:UVFreezeIn} is physically a result of the modified time evolution of $T$.
Focusing for simplicity on the case of a quadratic potential $\wrehbar = 0$, one may think at first sight that the different expansion rate during reheating ($H\propto x^{-3/2}$ instead of $H\propto x^{-2}$) favours early production of DM because it leads to less dilution. 
However, this is overcompensated by the fact that the relation between physical time $t$ and temperature $T$ also changes.
Because the temperature redshifts more slowly than in radiation domination, the DM production at later physical times, when the temperature is lower, becomes relatively more important compared to the radiation domination case. 
This is reflected by the scaling $\mathcal{C}_X \propto T^{2(d-4)+4} \propto x^{-3(d-2)/4}$,  instead of $x^{-2(d-2)}$, 
cf.~\eqref{Gleichung39}.
Moreover, after the freeze-in terminates ($\GX \ll H$), the DM density is further diluted relative to that of other constituents of the radiation bath, as the inflaton keeps dumping energy into the latter, which would not be the case during the radiation dominated epoch.

Pushing the DM production further into the UV requires a strengthening of the temperature’s redshifting or, equivalently, a relative enhancement of the inflaton decay rate $\GG$ at early times. 
While this can in principle be achieved by thermal effects, it typically occurs when $\phi$ couples to bosons (due to induced transitions), while $\GG$ is suppressed by Pauli-blocking if $\phi$ couples to fermions, see e.g.~\cite{Drewes:2013iaa} and references therein. 
However, for bosonic coupling the combination of the bounds \eqref{gLowerBound} and \eqref{gUpperBound} implies that regime \ref{it:reg2} is either non-existent or very narrow,
i.e.,
the range of $\g$-values for which thermal corrections to $\GG$ actually matter (affect the DM relic abundance) and can be quantified from near-future CMB observations is extremely limited. 

Hence, we conclude that feedback effects of the produced particles on $\GG$ generally do not affect the DM relic density in the regime where they can be computed by means of resummed thermal field theory.
Even if they enhance $\GG$ in the first moments of reheating, they are either not efficient for a sufficiently long time to overcome the dominance of the later-time DM production, or they are sub-dominant compared to the non-perturbative effects found in regime \ref{it:reg3}. 
In Sec.~\ref{sec:counterexamples} we discuss three ways of circumventing this conclusion. All three employ UV-dominated freeze-in, but this is realised in different ways:
\begin{itemize}
    \item[$\star$] Sec.~\ref{Sec:dim9}: DM production through operators with sufficiently high mass dimension $d\geq 9$.
    \item[$\star$] Sec.~\ref{Sec:SuperHeavy}: Super-heavy DM whose production starts to be Boltzmann suppressed during the reheating era.
    \item[$\star$] Sec.~\ref{Sec:Threshold}: Screening effects that kinematically suppress the decay of a particle below a threshold temperature.
\end{itemize} 
We note in passing that our conclusions can likely also be circumvented by considering more complicated inflationary scenarios beyond the two-parameter models considered here.
Finally, the large dissipative effects intrinsic to warm inflation \cite{Berera:1995ie} scenarios can modify the standard relation \eqref{eq:Tmaxrelation} and lead to a different scaling of the CMB constraints, cf.~\cite{Freese:2024ogj}. 
This can be of particular interest in models with so few parameter that $\g$ can potentially be determined from CMB observations or experiments \cite{Berghaus:2025dqi}.

\section{Impact of thermal effects on the DM production rate}
\label{sec:ThermaleffectsDMprod}

The relation between the DM relic density and the properties of new particles can potentially be affected by thermal corrections to $\GX$ in all three regimes \ref{it:reg1}-\ref{it:reg3}. In principle, the freedom is much larger than in the case of $\GG$, as the DM particles' mass $m_X$ in freeze-in scenarios can be almost arbitrarily dialled to match the observed relic density, and one may be tempted to think that it would be impossible to make any general statements. However, it turns out to be difficult to construct a simple scenario where thermal effects substantially modify the relic abundance. 

In view of the results of the previous Sec.~\ref{sec:ThermaleffectsInflaton}, we can investigate the impact of thermal effects on $\GX$ in regime \ref{it:reg1} for simplicity. 
We will therefore not specify the inflaton potential in this section and will only assume that the potential is quadratic close to its minimum such that the reheating remains perturbative.
In essence, our only assumption is that the inflaton behaves during reheating as a perturbed damped harmonic oscillator.
For specific incarnations of realistic inflationary scenarios, we refer the reader to Sec.~\ref{sec:ThermaleffectsInflaton} and appendix~\ref{app:CMBconstraints_inflationaryparam}. 
As we will see, the impact on the final DM abundance remains comparatively small, of the order of a few tens of percent, consistent with the findings of \cite{Becker:2023vwd} for various scenarios of freeze-in DM production during the radiation-domination epoch.

\subsection{DM production from a scalar particle decay}
\label{sec:1to2DMprod}

As a first example, we consider the DM interaction studied in Ref.~\cite{Becker:2023tvd}, 
where DM is produced
from the decay of a scalar particle $P$ coupled to a (Majorana) DM candidate $X$ and a right-handed SM lepton $f_R$ through the Yukawa coupling
\begin{align}
\label{eq:Lagrangian1to2decay}
    \mathcal{L}_X \supset \yX P\bar{f}_R X \ .
\end{align}
Assuming one can neglect the daughter particle's masses $m_X, m_{f_R} \ll m_P$, the parent particle $P$ decay rate in its rest frame is
\begin{align}
    \Gamma_P = \frac{\yX^2}{16\pi}m_P \ .
\end{align}
This decay rate can be extracted from collider measurements, in particular from searches for long-lived particles. 
Determining the event rates in collider experiments requires specifying the couplings of $P$ to SM particles.\footnote{While these interactions can in principle open new channels for the production of $X$, e.g.~through scatterings, we here neglect them to better highlight the impact of thermal effects. These should however be accounted for to put accurate constraints on specific models. 
} 
Concrete models considered in \cite{Becker:2023tvd} include a muonphilic Majorana DM model \cite{Calibbi:2021fld} and a  leptophilic scalar singlet DM model \cite{Belanger:2018sti}.

\subsubsection{Impact of thermal corrections}

In the early universe, the parent particle may itself be in or out of thermal equilibrium \cite{Konig:2016dzg}. 
In both cases, thermal corrections can be computed systematically in the CTP formalism \cite{Drewes:2015eoa}.
In the present work, we consider the case that the parent particle is in thermal equilibrium. We further assume $f_X \ll 1$ in all modes, so that 
inverse decays can be neglected for $X$. 
Then the momentum-dependent DM production rate including  thermal effects is given by \cite{Laine:2013lka,Ghisoiu:2014ena},
\begin{align}
\label{eq:DMprodrate_renormalisable_fullthermal}
    \GX(q) &= \yX^2\frac{M_P^2 T}{8\pi |\mathbf{q}|q_0}\operatorname{ln}\left[\frac{\sinh\left(E_{\rm max}/2T\right)}{\sinh\left(E_{\rm min}/2T\right)}\frac{\cosh\left((q_0-E_{\rm min})/2T\right)}{\cosh\left((q_0-E_{\rm max})/2T\right)}\right] \ ,
\end{align}
where $q$ is the DM 4-momentum, $M_P$ ($M_X$) the effective in-medium mass of $P$ ($X$) and 
\begin{align}
    E_{\rm max/min} = \frac{q_0\pm |\mathbf{q}|}{2}\left(\frac{M_P}{M_X}\right)^2\ .
\end{align}
While \eqref{eq:DMprodrate_renormalisable_fullthermal} takes full account of quantum statistical effects and the thermal mass of $P$, more subtle effects -- such as the inclusion of fully resummed thermal propagators for the produced particles, thermal vertex corrections or the 
Landau-Pomeranchuk-Migdal effect -- are not included because they at higher order; some discussion on these matters can e.g.~be found in \cite{Garbrecht:2013bia,Ghisoiu:2014ena,Drewes:2015eoa,Becker:2023vwd,Becker:2025lkc}.
As we are interested solely in the evolution of the total number density and since all particles outside of $X$ are assumed to be in complete thermodynamic equilibrium, it is sufficient to consider the momentum averaged version of this result for the collision integral, i.e., 
\begin{align}\label{MomAvColl}
    \mathcal{C}_X = \int \GX(q) f_X^{\rm eq}(q) \frac{\d^3 \mathbf{q}}{(2\pi)^3} \ .
\end{align}
In the Maxwell-Boltzmann limit, i.e. in the limit where one can neglect quantum statistics and approximate Fermi-Dirac distributions $\fF(E) = 1/(e^{E/T}+1)$ by Maxwell-Boltzmann ones $f_{\rm MB}(E) = e^{-E/T}$, it is possible to express the collision integral $\mathcal{C}_X$ capturing the production of $X$ in terms of the vacuum decay rate $\Gamma_P$,\footnote{$X$ can also be produced from scatterings mediated by the right-handed SM lepton $f_R$, but this channel exhibits the same Boltzmann suppression for $m_P/T \gg 1$ as the leading $1\rightarrow 2$ production mode and is therefore expected to be subdominant at the temperatures where DM production peaks \cite{Drewes:2015eoa}.} 
i.e., 
\begin{align}
\label{eq:CDM_MBapprox}
    \mathcal{C}_{X} = m_P \frac{g_P}{2\pi^2} \Gamma_P \int_0^{+\infty} \mathrm{d} p \frac{p^2}{E} e^{-E/T} &= \frac{\Gamma_P}{\pi^2}  m_P^2  T K_1(m_P/T)\ .
\end{align}

We first compare in Fig.~\ref{fig:Thermal effects_DMprodrate} the impact of using Eq.~\eqref{MomAvColl} instead of its Maxwell-Boltzmann approximation \eqref{eq:CDM_MBapprox} for a specific choice of coupling $\yX$, chosen to fit the observed DM abundance, and vacuum masses $m_P, m_X$ and $m_f$. 
We note in passing that, in both panels, the reheating temperature is so low that reheating proceeds in regime \ref{it:reg1}, i.e. it can be described by means of perturbative techniques. 
\begin{figure}
    \centering
    \includegraphics[width=0.7\textwidth]{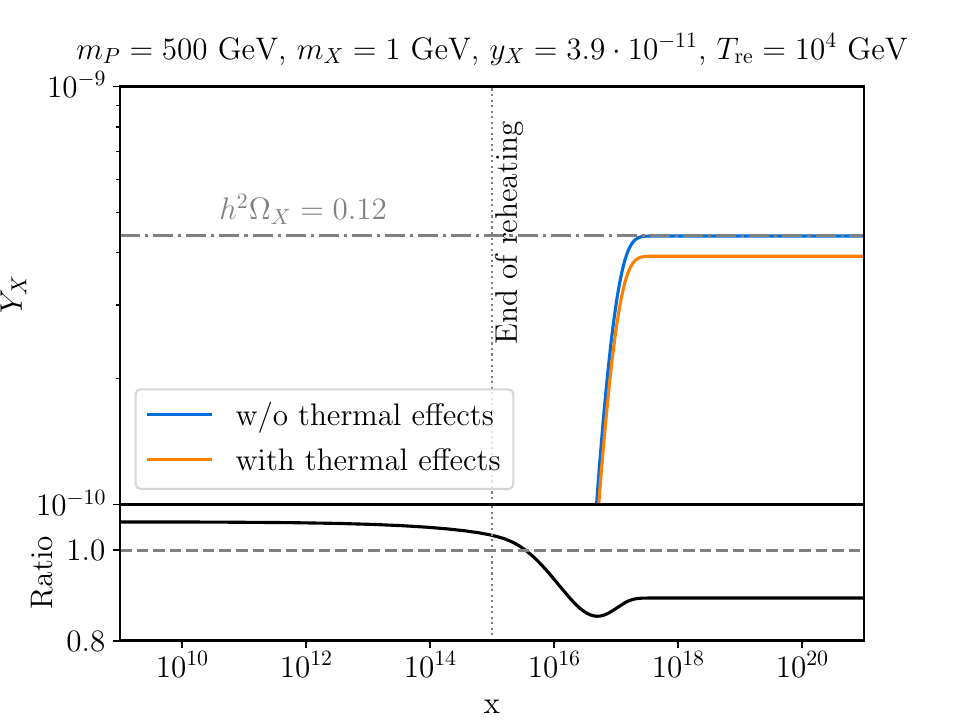}
    \includegraphics[width=0.7\textwidth]{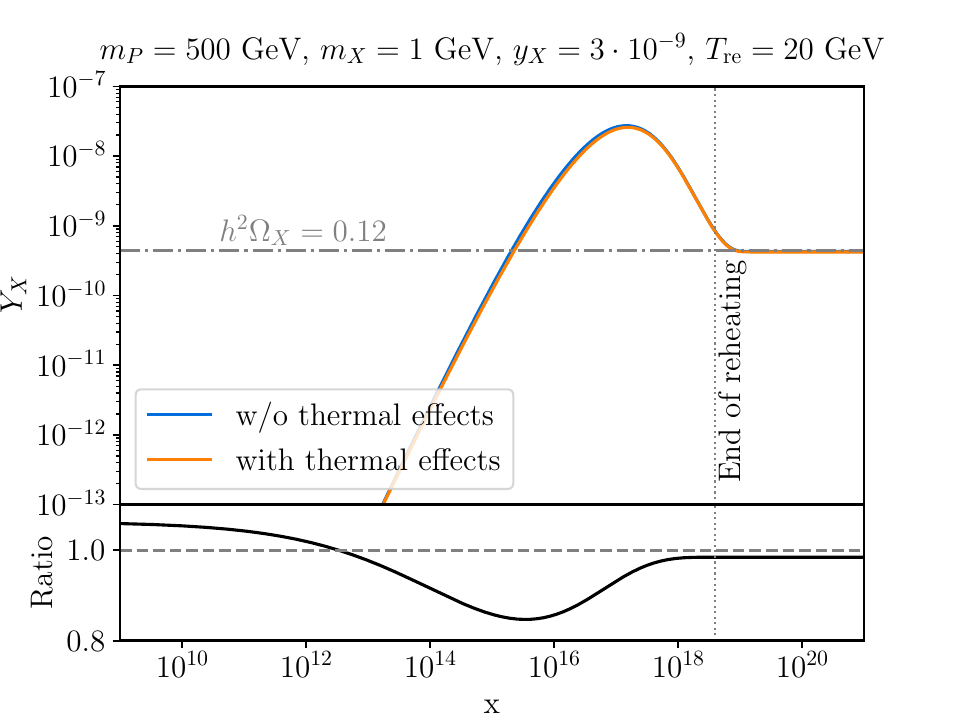}
    \caption{Time evolution of the DM yield for two different choices of reheating temperature $\Treh = 10^4$ GeV (upper panel) and $\Treh = 20$ GeV (lower panel). The blue curve represents the evolution obtained using the momentum averaging \eqref{MomAvColl} of the fully momentum-dependent rates \eqref{eq:DMprodrate_renormalisable_fullthermal}, whereas the orange curve displays the result obtained using the Maxwell-Boltzmann momentum-independent rates \eqref{eq:CDM_MBapprox}. The lower part of each panel highlights the time evolution of the ratio between these two results. Parameters chosen to perform this plot are specified in the title of the plot. In addition, we set $m_{f_R} = 0$. In the lower panel, the dashed horizontal line represents the observed present-day DM yield whereas the dotted vertical line denotes the end of the reheating era.}
    \label{fig:Thermal effects_DMprodrate}
\end{figure}
As one can see, the effect is of order 10-20$\%$ at most, in line with previous estimates obtained for radiation domination \cite{Becker:2023vwd} and consistent with the observation that, for non IR-enhanced processes (i.e. processes for which the characteristic momentum $\braket{p} \ll \pi T$), the ratio of Fermi-Dirac/Bose-Einstein distributions to Maxwell-Boltzmann distributions is bounded from above and below 
\begin{align}
\label{eq:MBapprox_quantifyEffect}
  0.95 \lesssim \frac{f_{\rm F/B}(E)}{f_{\rm MB}(E)} \leq 1.05 ~~\mbox{ for } E\gtrsim \pi T.
\end{align}
We note in passing that, in the upper panel of Fig.~\ref{fig:Thermal effects_DMprodrate}, the production of DM peaks for
\begin{align}
    T \in \left[m_P/5, m_P/2\right] =\left[100,250\right] \mbox{ GeV },
\end{align}
well after the end of the reheating era at $\Treh = 10^4$ GeV and the beginning of radiation domination. If one were to choose a lower reheating temperature, e.g.~$\Treh  = 20$ GeV, as illustrated in the bottom panel of Fig.~\ref{fig:Thermal effects_DMprodrate}, such that DM production truly peaks during the reheating epoch, the impact of these thermal effects would only be diminished. Indeed, the late decay of the inflaton condensate enhances the DM production cross-section for $T \lesssim m_P$ such that it peaks at lower temperatures than for typical freeze-in scenarios. At these low temperatures, the Maxwell-Boltzmann approximation is more justified.

This observation is also illustrated for a wider range of scalar particle's mass in Fig.~\ref{fig:mPvsdecaylength}.
In this figure, we display the theoretical prediction for the expected decay length in the $P$ rest frame $c\tau$ of $X$ assuming it reproduces the observed DM abundance, focusing on the collider-accessible mass range. 
We compare the results obtained using the complete production rate \eqref{eq:DMprodrate_renormalisable_fullthermal} (continuous lines) with the results obtained using the Maxwell-Boltzmann approximation \eqref{eq:CDM_MBapprox} (dotted lines) for three choices of reheating temperature $T_{\rm re} \in \{20,100, 10^4\}$ GeV.
It is clear that the difference between the dotted and continuous lines decreases as one decreases the reheating temperature.

\setcounter{footnote}{18}
\begin{figure}[!ht]
    \centering
    \includegraphics[width=0.7\textwidth,trim={.25cm 0 0 0}]{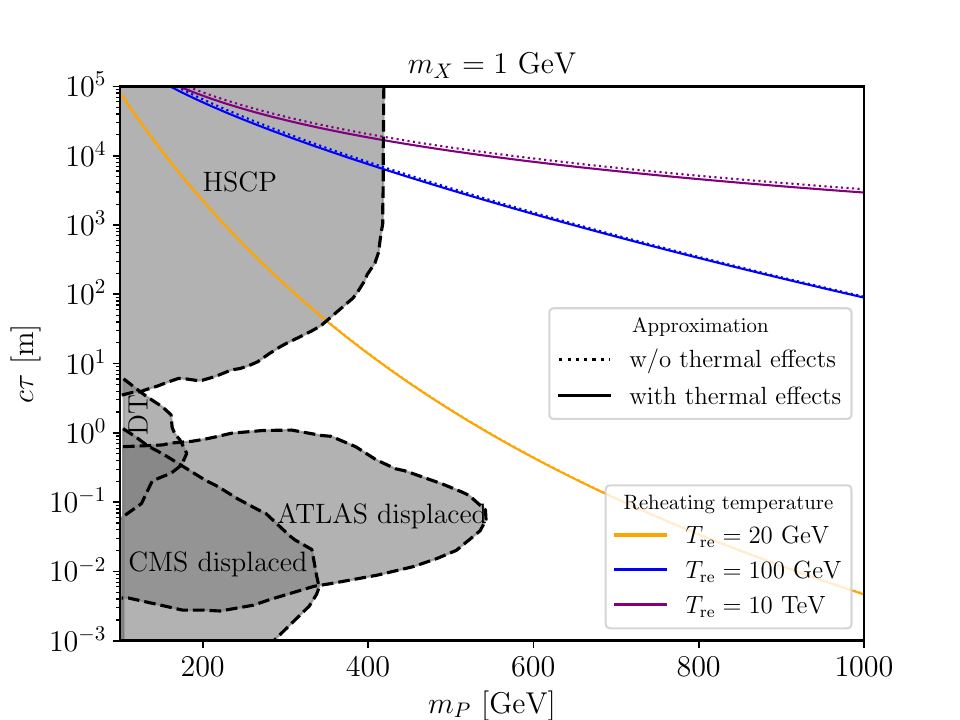}\\
    \includegraphics[width=0.7\textwidth,trim={.25cm 0 0 0}]{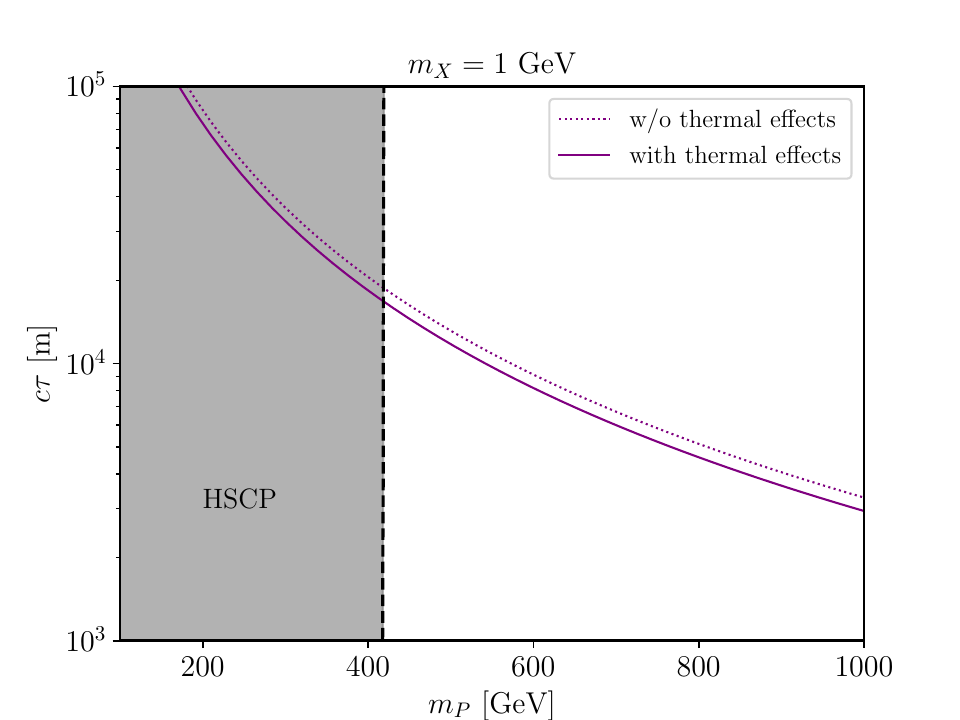}
    \caption{
    \emph{Upper panel}: The contours compare the values of the rest frame decay length $c\tau$ and mass $m_P$  accounting for the observed DM relic abundance without and with thermal corrections, based on \eqref{eq:CDM_MBapprox} and \eqref{MomAvColl} with \eqref{eq:DMprodrate_renormalisable_fullthermal}.
    We compare different choices of the reheating temperature, chosen as in \cite{Becker:2023tvd}, and the expected decay length with and without accounting for quantum statistical effects in the DM production rate. 
    \emph{Lower panel}: We zoom in on the results for $\Treh = 10$ TeV to highlight more clearly the impact of said thermal effects. 
    In gray, we highlight the parameter space region excluded for a muophilic DM candidate $X$ in the muonphilic Majorana DM model \cite{Calibbi:2021fld}, including searches for heavy stable charged particles (HSCP) at ATLAS \cite{ATLAS:2019gqq}, searches for disappearing tracks (DT) at ATLAS \cite{ATLAS:2017oal} and CMS \cite{CMS:2018rea,CMS:2020atg} as well as searches for displaced leptons at Atlas \cite{ATLAS:2020wjh} and CMS \cite{CMS:2014xnn,CMS:2016isf}.\protect\footnotemark~
    }
    \label{fig:mPvsdecaylength}
\end{figure}
\footnotetext{We note the existence of more recent studies updating these bounds, see e.g. \cite{ATLAS:2022rme,ATLAS:2024vnc}. However, updating the constraints shown in this plot would require a dedicated recasting of these analyses, following the procedure outlined in \cite{Calibbi:2021fld}. Such an analysis lies beyond the scope of the present work. Moreover, including these updates would not qualitatively affect the conclusions drawn from this figure.}

The impact of thermal corrections in Fig.~\ref{fig:mPvsdecaylength} is not only minuscule compared to experimental error bars, but also small compared to other effects that were neglected in the early universe computation.
For instance, the temperature at which $\GG = H$ is commonly used as proxy for the reheating temperature 
\eqref{eq:defreheatingtemp}
to obtain the approximate expression 
\eqref{eq:approxTr}.
This choice, however, typically overestimates the true reheating temperature and, for low reheating temperatures, the DM abundance as well. 
As an illustration, we estimated this effect for a benchmark with $m_P = 500$ GeV and $\Treh = 100$ GeV.
The reheating temperature $\Treh$ obtained from Eq.~\eqref{eq:approxTr} is about $\sim 30\%$ larger than the value found by solving the kinetic equations \eqref{eq:reducedform_withDM};
using the correct value in the approximation of instantaneous reheating made in appendix \ref{TrefromCMB} leads to an overestimate of the DM abundance by about a factor two, 
and an underestimate of the expected decay length for $P$ at colliders by a similar factor.
Dropping the instantaneous reheating assumption leads to a further correction that in principle should also be accounted for. However, all these approximations lead to errors that are smaller than the current observational uncertainties in Tab.~\ref{tab:sensitivities}. 

\setcounter{footnote}{19}
\begin{figure}
    \centering
    \includegraphics[width=0.7\textwidth,trim={.25cm 0 0 0}]{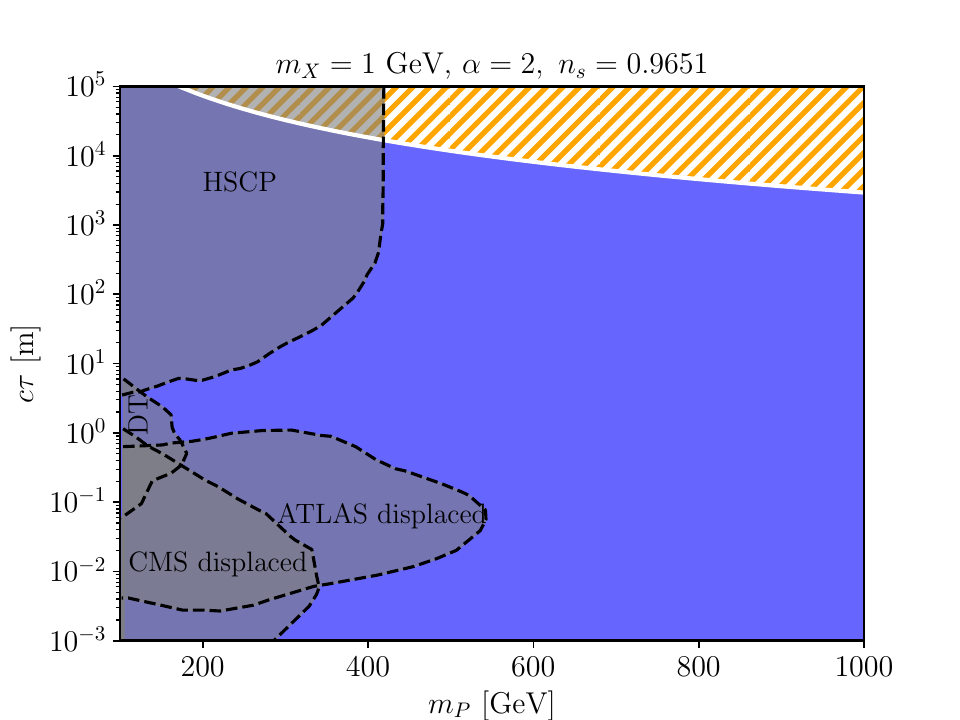}
    \caption{Forecast for constraints set on the model parameter space displayed in Fig.~\ref{fig:mPvsdecaylength} by the next-generation CMB experiment LiteBIRD in scenario \ref{it:A}, see Tab.~\ref{tab:inflation_benchmark}. 
    In dark blue shaded, we display the region that would be disfavoured at $3\sigma$ or more. 
    Since $m_P \ll T_{\rm re}$, the DM production is essentially insensitive to the reheating epoch, and the region consistent with CMB observation at the $2\sigma$ level is compressed into the immediate proximity of the region with instant reheating ($\Nreh=0$) indicated by the white line, making it indistinguishable from DM production in the radiation dominated epoch.
    The orange hatched area represents the parameter space region where the observed DM abundance can never be achieved independently of the value of the reheating temperature.\protect\footnotemark~
    }
\label{fig:scenA}
\end{figure}
\footnotetext{The reason is simply that, within the chosen setup for reheating, a low reheating temperature can only dilute the produced DM abundance but never enhance it.}

\begin{figure}
    \centering
\includegraphics[width=0.7\textwidth,trim={.25cm 0 0 0}]{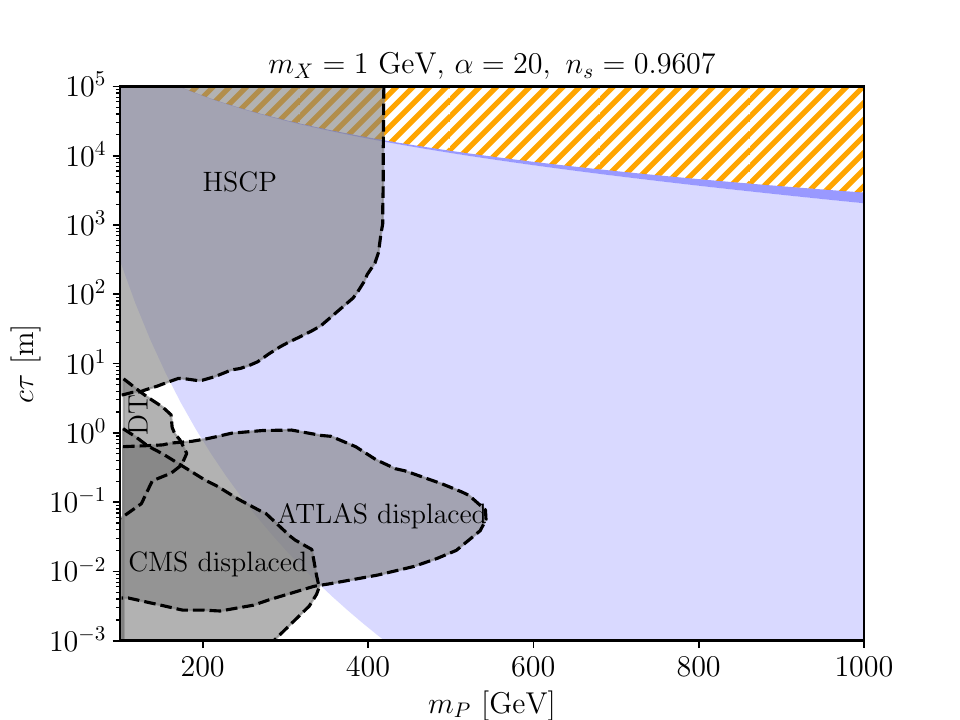}
    \caption{Constraints set on the model parameter space displayed in Fig.~\ref{fig:mPvsdecaylength} by the next-generation CMB experiment LiteBIRD in scenario \ref{it:B}, see Tab.~\ref{tab:inflation_benchmark}.
The white, light blue and darker blue areas correspond to the regions consistent with such observation at the $1\sigma$, $2\sigma$ and $3\sigma$ level. 
Similarly to the previous figure, the orange hatched area represents the parameter space region where the observed DM abundance can never be achieved independently of the value of the reheating temperature.
    }
\label{fig:scenB}
\end{figure}

\begin{figure}
    \centering
    \includegraphics[width=0.7\textwidth,trim={.25cm 0 0 0}]{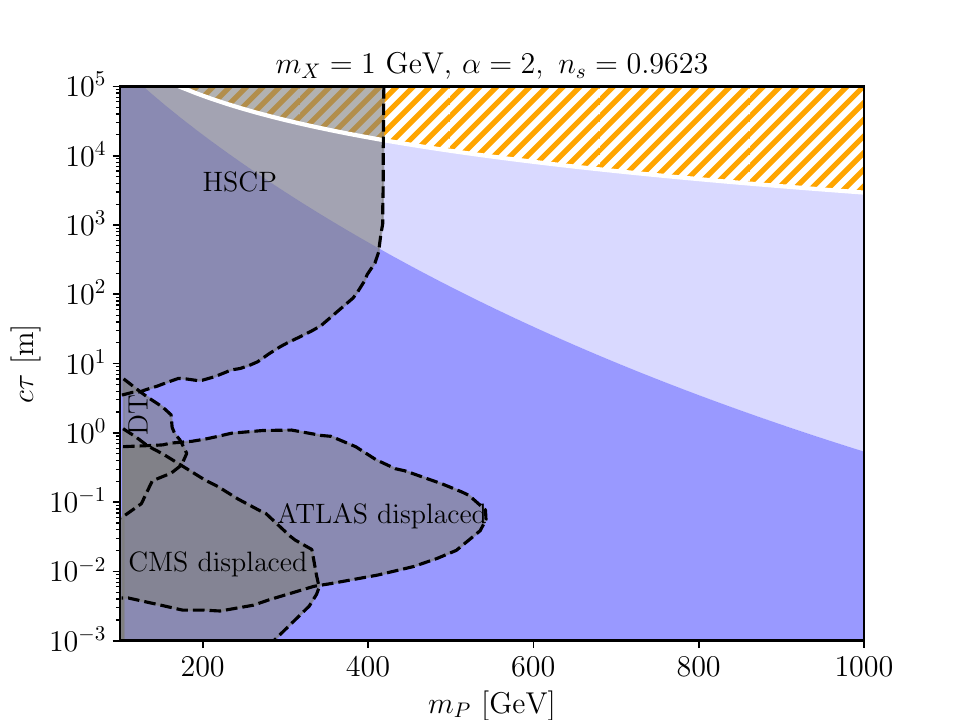}
    \caption{Constraints set on the model parameter space displayed in Fig.~\ref{fig:mPvsdecaylength} by the next-generation CMB experiment LiteBIRD in scenario \ref{it:C}, see Tab.~\ref{tab:inflation_benchmark}. 
The white, light blue and darker blue areas correspond to the regions consistent with such observation at the $1\sigma$, $2\sigma$ and $3\sigma$ level. As in Fig.~\ref{fig:scenA}, the $1\sigma$ region is compressed into the immediate vicinity
of the region with instant reheating ($\Nreh=0$) indicated by the white line, making it indistinguishable from DM production in the radiation dominated epoch. However, in scenario \ref{it:C} the $2\sigma$ region is distinguishable from instant reheating.
Similarly to the previous figures, the orange hatched area represents the parameter space region where the observed DM abundance can never be achieved independently of the value of the reheating temperature.
    }
\label{fig:scenC}
\end{figure}

\subsubsection{Collider predictions from CMB observations}

Finally, we close this subsection by highlighting in Figs.~\ref{fig:scenA}-\ref{fig:scenC} the complementarity between the constraints set by future CMB experiments, such as LiteBIRD, on the model parameter space and those from laboratory experiments.
We carry out this analysis for the three benchmark scenarios defined in table \ref{tab:sensitivities}, and the posteriors for $\Treh$ are obtained using the method introduced in \cite{Drewes:2022nhu}, see \eqref{Eq:Likelihood} in the appendix.

Fig.~\ref{fig:scenA} shows that CMB observations with LiteBIRD 
could reduce the viable parameter space for the inflationary scenario~\ref{it:A} to a line in the mass-lifetime plane of the mediator $P$, part of which is accessible to HSCP searches \cite{ATLAS:2019gqq} in the muonphilic Majorana DM model \cite{Calibbi:2021fld}.
However, it should be added that, in this 
scenario, the DM production is essentially insensitive to the reheating epoch, as the viable parameter space lies in the $m_P \ll T_{\rm re}$ region, where DM is mostly produced during radiation domination.  
As a result, the constraints expected from LiteBIRD would only be very mildly stronger than those from Planck and BICEP/Keck, which currently already imply $\Treh > 10^4$ GeV even if $\alpha$ is treated as a free parameter \cite{Liu:2025sut}. 
In fact, neither the current constraint nor those expected from LiteBIRD correspond to a true measurement of the reheating temperature in the sense that an upper and lower bound on $\Treh$ are both inferred from observational data; the error bar reported in Tab.~\ref{tab:sensitivities} originates from the combination of a lower observational bound and an upper bound from the consistency condition $\Nreh \geq 0$.
As a result, the posterior is highly non-Gaussian and asymmetric, making the interpretation of the error bar on $\Treh$ tricky.

The situation is very different in scenario \ref{it:B}, as illustrated in Fig.~\ref{fig:scenB}. In this case the posterior on $\Treh$ is in good approximation Gaussian and a measurement $
{\rm log}_{10}(T_{\text{re}}/\text{GeV}) =
-0.8^{+1.5}_{-1.0}$ can be obtained, 
 placing $y_\phi$ deep inside region \ref{it:reg1}.
In this scenario, which implies that inflation happens in a hidden or secluded sector that couples only very feebly to the sector containing the SM fields, DM production mostly happens during reheating. 
In this scenario, observations with LiteBIRD could corner the available model parameter space at the $1\sigma$ level into a region that is accessible to long-lived particle searches at the HL-LHC, as the sensitivity to $c\tau$ is expected to improve linearly with the integrated luminosity \cite{Drewes:2025ocf}.\footnote{The simple analytic sensitivity estimates in \cite{Drewes:2025ocf} are most accurate for lepton colliders, but can still be used as order of magnitude estimates for the HL-LHC, see Fig.~8 in \cite{Drewes:2019vjy}.} 
However, the $2\sigma$ region would still include a vast are that is inaccessible to colliders, a situation that could possibly be improved by adding constraints on $n_s$ from the EUCLID satellite or 21cm tomography \cite{Drewes:2022nhu}.

Finally, Fig.~\ref{fig:scenC} shows that in scenario \ref{it:C} 
the parameter space favoured by LiteBIRD observations covers regions where collider-accessible DM is produced during reheating as well as regions where it is produced during radiation domination. 
Within the $1\sigma$ posterior regime for $\Treh$, the production of collider-accessible DM necessarily proceeds during the radiation domination era (since $m_P \lesssim 1$ TeV $< \Treh$ at $1\sigma$). At the $2\sigma$-level, the lower bound on the reheating temperature relaxes, i.e., $\Treh \gtrsim 40$ GeV, and collider-accessible DM can be produced during reheating. 
We remark that part of this parameter space can be probed by HSCP searches at the LHC. 

\subsection{DM production from a Fermi-like interaction}
\label{sec:DMprodFermilike_noinflatonthermal}

As a second example, we consider the impact of quantum statistical effects on DM produced from non-renormalisable interactions. 
For simplicity, we consider a Lagrangian of the form 
\begin{align}
\label{eq:FermiLagrangian}
    \mathcal{L}_{X,int} = 2\sqrt{2}\mathcal{G}(\Psi_3 \gamma_\mu P_R X)(\Psi_1 \gamma^{\mu} P_R \Psi_2 ) \ ,
\end{align}
where $\Psi_2$ and $\Psi_{3}$ are two massless fermions, $X$ is the DM (of mass $m_X$) and $\Psi_1$ is another fermion of mass $m_{\Psi_1}$.
Furthermore, we assume $m_{\Psi_1} = m_X$.\footnote{While this choice might seem tuned, we only did so for simplicity in order to avoid having to consider decays altogether. } 
One may think of the interaction \eqref{eq:FermiLagrangian} as an effective field theory description of a gauge theory in terms of a generalised Fermi constant $\mathcal{G}$.
For a neutral $Z'$ boson with mass $m_{Z'}$ and coupling $g$, one would e.g.~find $\mathcal{G} = g^2/(4\sqrt{2} m_{Z'}^2)$.\footnote{In generic UV-completion of such theory such as Left-Right symmetric theories \cite{Pati:1974yy,Mohapatra:1974gc,Senjanovic:1975rk} or $U(1)_{B-L}$ theories \cite{Mohapatra:1980qe,Wetterich:1981bx,Buchmuller:1991ce,Buchmuller:1992qc}, multiple channels exist through which the DM candidate $X$ can interact. 
However, since our goal is merely to highlight the impact of thermal effects, 
and since the rates related to these additional channels will exhibit a similar scaling with $T$, there is no need to go into the details of potential UV completions of \eqref{eq:FermiLagrangian} for the purpose of the present work. }

In this scenario, DM is primarily produced through scatterings. The full thermal interaction rate in a Fermi-like theory can be expressed as \cite{Asaka:2006rw,Drewes:2025bbb} 
\begin{align}
\nonumber
    &\GX(q) = \frac{8\mathcal{G}^2 \fF^{-1}(-q_0)}{q_0} \int \frac{\d^3 \mathbf{p}_1}{(2\pi)^3 2E_1}\int \frac{\d^3 \mathbf{p}_2}{(2\pi)^3 2E_2}\int \frac{\d^3 \mathbf{p}_3}{(2\pi)^3 2E_3}\times\\
    \nonumber
    \times\Big[&(2\pi)^4\delta^4(p_1+p_2+p_3-q) \left(1-\fF(E_1)\right)\left(1-\fF(E_2)\right)\left(1-\fF(E_3)\right)\mathcal{T}(m_1,-m_2,m_3)\\
    \nonumber
    +&(2\pi)^4\delta^4(p_1+p_2-p_3-q) \left(1-\fF(E_1)\right)\left(1-\fF(E_2)\right)\fF(E_3)\mathcal{T}(m_1,-m_2,-m_3)\\
    \nonumber
    +&(2\pi)^4\delta^4(p_1+p_3-p_2-q) \left(1-\fF(E_1)\right)\fF(E_2)\left(1-\fF(E_3)\right)\mathcal{T}(m_1,m_2,m_3)\\
    \nonumber
    +&(2\pi)^4\delta^4(p_2+p_3-p_1-q) \fF(E_1)\left(1-\fF(E_2)\right)\left(1-\fF(E_3)\right)\mathcal{T}(-m_1,-m_2,m_3)\\
    \nonumber
    +&(2\pi)^4\delta^4(p_3-p_1-p_2-q) \fF(E_1)\fF(E_2)\left(1-\fF(E_3)\right)\mathcal{T}(-m_1,m_2,m_3)\\
    \nonumber
    +&(2\pi)^4\delta^4(p_2-p_1-p_3-q) \fF(E_1)\left(1-\fF(E_2)\right)\fF(E_3)\mathcal{T}(-m_1,-m_2,-m_3)\\
    +&(2\pi)^4\delta^4(p_1-p_2-p_3-q) \left(1-\fF(E_1)\right)\fF(E_2)\fF(E_3)\mathcal{T}(m_1,m_2,-m_3)\Big] \ ,
\label{eq:fullFermirates}
\end{align}
with
\begin{align}
\nonumber
\mathcal{T}(m_1,m_2,m_3) &\equiv \operatorname{Tr}\left[\left(\slashed{q}+m_X\right) \gamma^\mu P_R\left(\slashed{p}_3+m_3\right) \gamma^\nu P_R \right] \operatorname{Tr}\left[\left(\slashed{p}_2+m_2\right) \gamma_\mu P_R\left(\slashed{p}_1+m_1\right) \gamma_\nu P_R\right] \\
&= 8\left[\Big( (p_2 \cdot p_3) (p_1 \cdot q)+(p_1 \cdot p_3) (p_2 \cdot q)\Big)\right]\ ,
\end{align}
where one has to identify $m_1 = m_X$, $m_2 = m_3 = 0$ in the present case.
The Maxwell-Boltzmann approximated rate is then given by
\begin{align}
\nonumber
    &\GX(q) = \frac{8\mathcal{G}^2 \fF^{-1}(-q_0)}{q_0} \int \frac{\d^3 \mathbf{p}_1}{(2\pi)^3 2E_1}\int \frac{\d^3 \mathbf{p}_2}{(2\pi)^3 2E_2}\int \frac{\d^3 \mathbf{p}_3}{(2\pi)^3 2E_3}\times\\
    \nonumber
    \times\Big[&(2\pi)^4\delta^4(p_1+p_2+p_3-q) \mathcal{T}(m_1,-m_2,m_3)\\
    \nonumber
    +&(2\pi)^4\delta^4(p_1+p_2-p_3-q) e^{-E_3/T}\mathcal{T}(m_1,-m_2,-m_3)\\
    \nonumber
    +&(2\pi)^4\delta^4(p_1+p_3-p_2-q) e^{-E_2/T}\mathcal{T}(m_1,m_2,m_3)\\
    \nonumber
    +&(2\pi)^4\delta^4(p_2+p_3-p_1-q) e^{-E_1/T}\mathcal{T}(-m_1,-m_2,m_3)\\
    \nonumber
    +&(2\pi)^4\delta^4(p_3-p_1-p_2-q) e^{-(E_1+E_2)/T}\mathcal{T}(-m_1,m_2,m_3)\\
    \nonumber
    +&(2\pi)^4\delta^4(p_2-p_1-p_3-q) e^{-(E_1+E_3)/T}\mathcal{T}(-m_1,-m_2,-m_3)\\
    +&(2\pi)^4\delta^4(p_1-p_2-p_3-q) e^{-(E_2+E_3)/T}\mathcal{T}(m_1,m_2,-m_3)\Big] \ .
\label{eq:MBapprox_Fermirates}
\end{align}

In Fig.~\ref{fig:Ferminteraction_DMthermaleffect}, we compare the final DM abundance in the scenarios where the collision integral \eqref{CollInTermsOfGammaX} is computed with the rates \eqref{eq:fullFermirates}
to the scenario where one uses the Maxwell-Boltzmann approximation \eqref{eq:MBapprox_Fermirates}. 
Given how miniscule the difference is, it is not necessary to include screening effects. 
The model parameters are chosen in a way that the DM production peaks during the reheating era and the EFT remains valid throughout. As observed, even if the production peaks during reheating, the impact of thermal corrections is of the same order of magnitude as in the previous example, i.e. $\mathcal{O}(10\%)$, which is to be expected from Eq.~\eqref{eq:MBapprox_quantifyEffect} as the process is not IR enhanced. 
\begin{figure}
    \centering
    \includegraphics[width=0.75\textwidth]{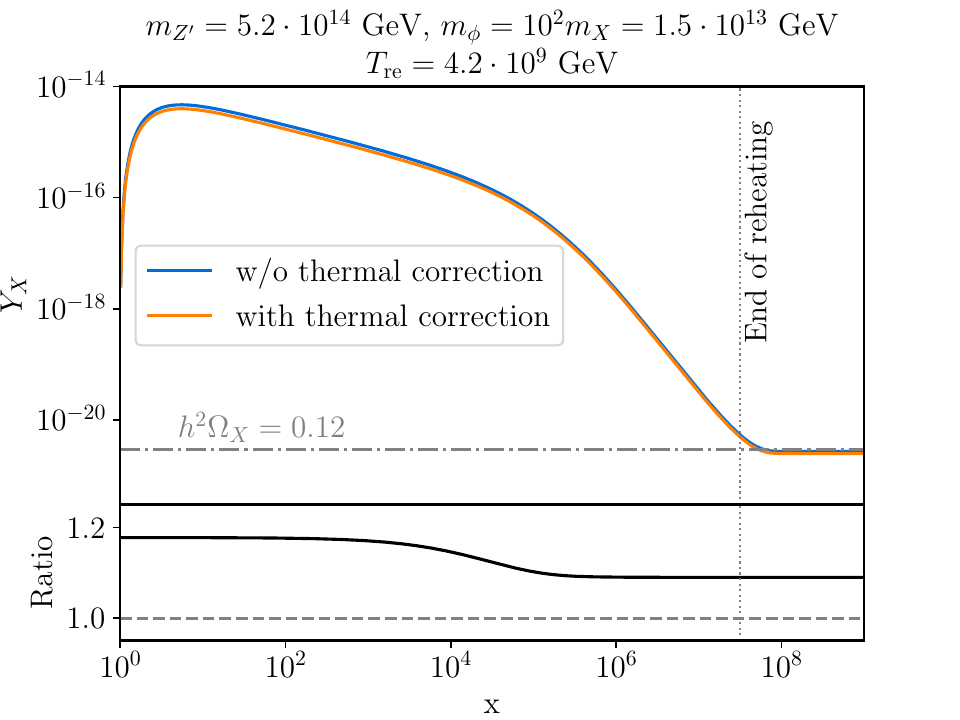}
    \caption{Impact of quantum statistical effects on the DM yield for a Fermi interaction. 
    The orange line has been obtained by using the full
    rates \eqref{eq:fullFermirates} 
    whereas the blue line has been derived using its Maxwell-Boltzmann approximation \eqref{eq:MBapprox_Fermirates}. In the bottom panel, we display the ratio of the DM yield obtained without including quantum statistical effects to the DM yield obtained while including quantum statistical effects.
    The dashed horizontal line represents the observed present-day DM yield whereas the dotted vertical line denotes the end of the reheating era.}
    \label{fig:Ferminteraction_DMthermaleffect}
\end{figure}

\subsection{General picture}

In the two examples at hand, thermal corrections are clearly small. This raises the question how general this conclusion is, and it is worthwhile understanding the physical reasons for this smallness.

A key assumption in both examples is that the DM particle is the only species which is out of equilibrium. In the example in Sec.~\ref{sec:1to2DMprod}, the freeze-in DM production is terminated by the freeze-out of the parent particle $P$, more precisely the Boltzmann suppression of its abundance. This is rather common for renormalisable interactions, where one typically finds $\GX\propto T$ in the regime $T \gg m_P$ on dimensional grounds, 
so that $\GX$ redshifts slower than the Hubble rate \eqref{eq:redshifting}. 
Hence, the freeze-in typically takes place at $T<m_P$
-- for $m_P /T\sim 2-5$ in the model considered here -- where quantum statistical effects are generally sub-leading \cite{Drewes:2015eoa}. 
The impact of screening effects on the DM production rate (and abundance) in this regime tends to be even smaller, as the thermal correction to particle dispersion relations $\sim g T$ exceeds the vacuum mass $m_P$ only when $T \gtrsim m_P/g$.  
At larger $T$, the thermal corrections to $\GX$ can be huge, but their impact on the DM relic density is diminished by the effects already discussed at the end of Sec.~\ref{sec:ThermaleffectsInflaton}: 
The temperature redshifts more slowly than during radiation domination,
and the DM density produced at those early times is diluted by the particles produced in inflaton decays at later times. The steeper temperature dependence $\GX \sim \mathcal{G}^2 T^5$ in our second example in Sec.~\ref{sec:DMprodFermilike_noinflatonthermal} is insufficient to overcome these obstacles, as one could already have predicted from \eqref{Gleichung39}. 
The bound \eqref{eq:MBapprox_quantifyEffect} also severely limits the impact of quantum statistical effects for non-IR enhanced thermal DM production, even in the ultra-relativistic regime.

Hence, we can conclude from this section as well as the reasoning presented in Sec.~\ref{sec:ThermaleffectsInflaton} that thermal corrections to $\GX$ and $\GG$ do not substantially modify the DM relic density in a wide class of scenarios.
There are several ways how this conclusion may be avoided: 
\begin{itemize}
    \item[$\star$] If the DM is predominantly produced from light particles coupling via sufficiently high-dimensional operators the freeze-in is UV dominated without further tunings, cf. Eq.~\eqref{Dfine1} and Sec.~\ref{Sec:dim9}, thereby enhancing the sensitivity of the present-day DM relic abundance to thermal corrections to $\GG$.
    \item[$\star$] If DM production involves particles with comparably large effective masses,  Boltzmann suppression of a coannihilating particle or phase space suppressions can enforce a UV-dominated freeze-in, cf.~Secs.~\ref{Sec:SuperHeavy} and \ref{Sec:Threshold}, respectively. As discussed in the previous point, this will enhance the sensitivity of the present-day DM relic abundance to thermal effects in $\GG$.
    \item[$\star$]  
    An effective coupling or mixing angle that depends on $T$ and/or $\phi$ that can lead to resonant conversion, as in the case of sterile neutrino DM \cite{Dodelson:1993je,Shi:1998km} or dark photon DM \cite{Xu:2025wlq}.
    \item[$\star$] In case the parent particles or mediators involved in DM production are themselves not in equilibrium, $\GX$ can be governed by their evolution rather than the radiation bath's temperature $T$, implying that $\GX$ can be dominated by thermal effects when the DM production peaks.
    \item[$\star$] If the DM is not produced from particle interactions, but instead non-thermally from a condensate (e.g., but not necessarily, the inflaton), DM production primarily depends on the evolution of that condensate (rather than $T$), including the possibility of non-perturbative DM production (e.g. in parametric or tachyonic resonances).
\end{itemize}
Nevertheless, for a conventional thermal history and thermally produced DM, the general rule is that the impact of thermal corrections to both $\GG$ and $\GX$ is small in regimes \ref{it:reg1} and \ref{it:reg2}. We present explicit exceptions from this rule in the following section, focusing on the first two possibilities.

\section{Counterexamples to the standard lore}
\label{sec:counterexamples}

In the previous sections \ref{sec:ThermaleffectsInflaton} and \ref{sec:ThermaleffectsDMprod}, we argued on general grounds that thermal corrections to the inflaton decay rate $\GG$ and the rate of DM production $\GX$ do not alter the relic abundance of thermally produced DM in the regime where they can be computed by means of perturbative finite-temperature field theory. 
We demonstrated this with two explicit examples in section \ref{sec:ThermaleffectsDMprod}.
A key point is that the majority of the DM relic abundance is usually produced towards the end of reheating or during radiation domination.
In the following, we present three concrete counterexamples based on \emph{UV-dominated freeze-in} to illustrate how this general conclusion can be avoided.
Note that all of these have been explicitly constructed as exceptions from the rule, and they do not invalidate its validity in general. 

For the sake of definiteness, we choose the potential 
\begin{equation}
\label{eq:potentialalphaattractor}
    \Vphi(\phi) = \M^4 \tanh^2\left[\frac{\phi}{\sqrt{6\alpha}\mpl}\right] \;,
\end{equation}
corresponding to a class of $\alpha$-attractor T-models.
We however emphasise again that our conclusions would essentially remain unchanged had we considered other popular plateau models, as argued on general grounds in appendix \ref{mphigenericPlateauPotential} and confirmed explicitly for the RGI and MHI models in appendix \ref{app:CMBconstraints_inflationaryparam}.
Fig.~\ref{fig:nsvsr_alphaattractor} shows that
constraints from the CMB essentially fix the inflaton mass $m_\phi$ to $1.5 \cdot 10^{13}$ GeV, consistent with \eqref{mPhiPlateau}.
We take the initial energy density of the inflaton to be $\Phi = 2\cdot 10^{10}$, close to its maximal value \eqref{PhiInitialMax} allowed by the upper bound on the tensor-to-scalar ratio in order to enhance the impact of thermal effects.

\begin{figure}
    \centering
    \includegraphics[width=0.49\textwidth]{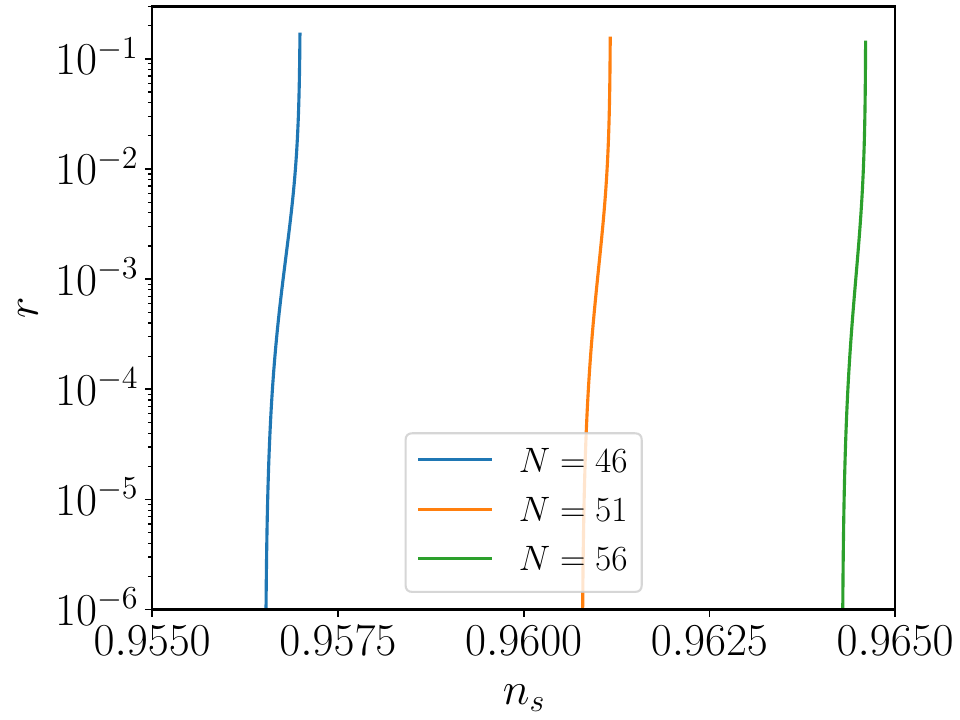}
    \includegraphics[width=0.49\textwidth]{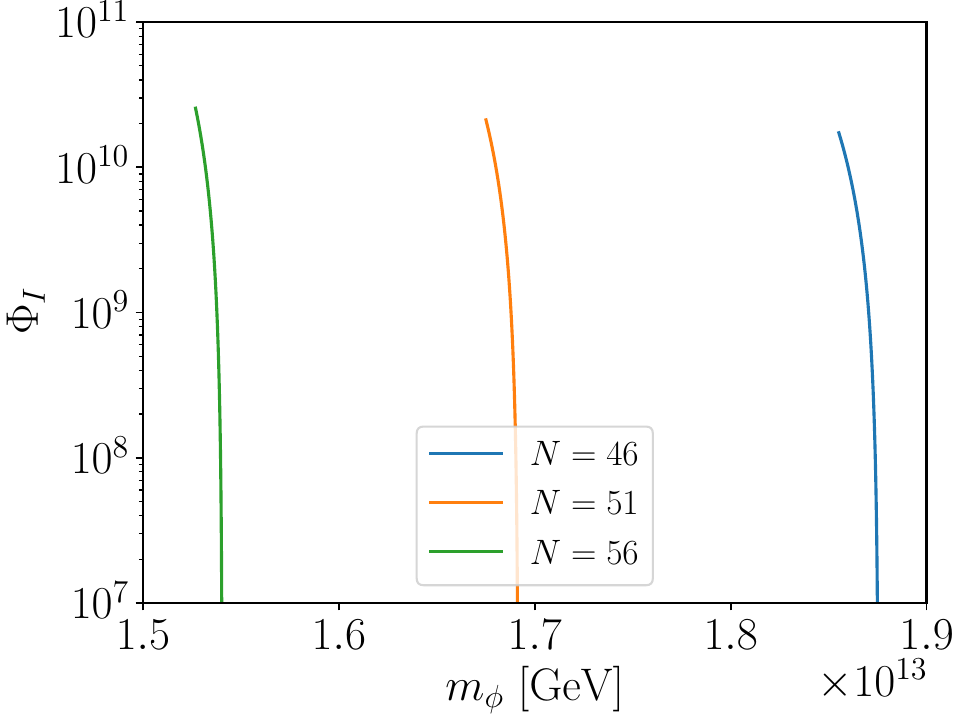}
    \caption{(\textit{Left}) Tensor-to-scalar ratio $r$ as a function of the spectral index $n_s$ for the $\alpha$-attractor T-model and fixed values of the number of e-folds $\Nk$, consistent with the Planck and BICEP/Keck data. The blue, orange and green lines correspond to $\Nk=46$, $\Nk=51$, and $\Nk = 56$, respectively. (\textit{Right}) Inflaton energy density $\Phi_I$ at the beginning of reheating as a function of the inflaton mass $m_\phi$ assuming the same choices for the number of e-folds and inflationary scenario. The colour scheme is identical to the left panel.}
    \label{fig:nsvsr_alphaattractor}
\end{figure}

As argued in \cite{Liu:2025sut} on the basis of the findings in 
\cite{Giudice:1999fb,Berges:2010zv,Fan:2021otj,Xu:2025wjq}, 
one way to extend regime \ref{it:reg2} to values of the coupling exceeding \eqref{YippieYaYaySchweinebacke} under relatively mild assumptions is to focus on scenarios where reheating is driven by a Yukawa coupling\footnote{One may argue that the Yukawa coupling in $\alpha$-T models should be defined in the original frame. In appendix \ref{YukawainalphaT}, we show that this practically makes no difference in the regime where \eqref{GeneralScalingSelf} is fulfilled.}  
\begin{align}\label{YukawaReheating}
    \mathcal{L}_\phi \supset y_\phi \phi \bar{\Psi}\Psi \ .
\end{align}
In this case, Pauli blocking tends to suppress the build-up of occupation numbers 
that are large enough to trigger feedback effects that impact on $\Nreh$ and thus $\Treh$ (e.g.~a parametric resonance), justifying to estimate $\GG$ by \eqref{GammaPerturbative} with $\{\g,\c\}=\{y_\phi,8\pi\}$, provided that the condition \eqref{GeneralScalingSelf} on the $\vv_i$ is fulfilled to prevent early fragmentation \cite{Garcia:2023dyf,Bhusal:2025oqg}.\footnote{Strictly speaking the Pauli blocking is of course also a feedback effect on $\GG$, assuming that the produced fermions do not rapidly decay due to their large effective mass $\sim y_\phi \phi$ during maximal $\phi$-elongation.} 
This relaxes the upper bound \eqref{YippieYaYaySchweinebacke} on the value of $\g=y_\phi$, though it is presently not known up to which value of $y_\phi$ the effective parametrisation \eqref{GammaPerturbative} can be used \cite{Garcia:2021iag}, as the discussion following \eqref{EffectiveYukawa} below shows.

For simplicity we assume a U(1) gauge interaction with elementary charge $\gPsi$ for the fermions $\Psi$.
The fermionic dispersion relation in principle is complicated and momentum dependent if the fermion has gauge interactions \cite{Weldon:1982bn};
we approximate it by 
\begin{eqnarray}\label{FermionMassFull}
M_\Psi^2 \simeq (m_\Psi +y_\phi \phi)^2 + (\gPsi T/2)^2,     
\end{eqnarray}
which can be justified if at any given moment one of the terms in the sum dominates. 
For definiteness, we take $\gPsi = 0.5$ in the following. 
The effective inflaton mass at order $\lambdaphi$ reads $M_\phi^2 = m_\phi^2 + \lambdaphi T^2/24 +  \lambdaphi \phi^2/2$,
with $m_\phi$ and $\lambdaphi$ the
coefficients of the quadratic and quartic terms in the Taylor expansion of the potential \eqref{eq:potentialalphaattractor}.

We shall in the following always neglect the fermions' vacuum mass, i.e., $m_\Psi \ll m_\phi$. Since $\phi_{\text{end}}$ can be close to $\mpl$, cf.~\eqref{phiEndalphaT}, 
while $\Tmax$ typically only slightly exceeds $m_\phi$,
the contribution  generated by the coupling to the inflaton condensate $y_\phi \phi$ usually exceeds the thermal mass $\gPsi T/2$ at early times, while the latter's relative importance increases during the reheating epoch. 
More precisely, in regime \ref{it:reg2} one can estimate from $\rho_R = \rho_\phi \sim m_\phi^2\phi^2$ and \eqref{eq:approxTr} that
the field value at the end of reheating is typically $\phi \sim y_\phi^2 \mpl $ 
while  $\Treh \sim y_\phi \sqrt{m_\phi \mpl}$, hence the thermal mass in \eqref{FermionMassFull} typically dominates for $y_\phi < 10^{-2}$. Then the thermally corrected  inflaton decay rate in \eqref{GammaConstraint} reads \cite{Drewes:2013iaa}\footnote{Here we only include quantum statistical and screening effects on the decay of single particle states. Similar effects on bound states \cite{Biondini:2025jvp,Binder:2026fwe} are of higher order, contributions from scatterings would dominate at $T\gg M_\phi$ \cite{Drewes:2013iaa,Drewes:2015eoa}.  }
\begin{align}
\label{eq:inflatondecayrate}
    \GG = \frac{y_\phi^2}{8\pi}M_\phi \left(1-2\left(\frac{\Mpsi}{M_\phi}\right)^2\right)^{1/2}\left(1-2\fF(M_\phi/2)\right).
\end{align}

In regime \ref{it:reg1}, the rate
\eqref{eq:inflatondecayrate} reduces to the form \eqref{GammaPerturbative}, and we can simply insert this expression into \eqref{eq:reducedform_withDM} to reconstruct the entire thermal history during reheating. 
Though the time-dependent mass $y_\phi \phi$ in \eqref{FermionMassFull} in principle dominates at early times, $\GG$ averages to the form \eqref{GammaPerturbative} when $\mathcal{R}\ll 1$ \cite{Garcia:2020wiy}, with
\begin{eqnarray}
    \mathcal{R} = 4y_\phi^2\frac{\phi^2}{m_\phi^2} \ ,
\end{eqnarray}
which corresponds to $\left(2M_\Psi/m_\phi\right)^2$ when the vacuum and thermal masses of $\Psi$ are subdominant. 
For the present example, we choose $y_\phi = 5\cdot 10^{-3}$
in order to fulfil the constraint \eqref{eq:lowerboundcoupling}.
In this case the condition $\mathcal{R}\ll 1$ is initially violated.
To accommodate for the kinematic suppression of $\GG$ in the moments of large $\phi$-elongation, one can introduce the effective Yukawa coupling \cite{Garcia:2020wiy}
\begin{eqnarray}\label{EffectiveYukawa}
    y_{\rm eff}^2 = 
       \begin{cases*}
       y_\phi^2 \quad {\rm if} \ \mathcal{R}\ll1 \\
       y_\phi^2 0.38 /\sqrt{\mathcal{R}} \quad {\rm if} \ \mathcal{R}\gg1
       \end{cases*} \ .
\end{eqnarray}
The two analytic approximations can be matched smoothly at $\mathcal{R}\approx 1/7$.
The expressions \eqref{EffectiveYukawa} has been derived in the framework of time-dependent perturbation theory. It does not include the effects of non-perturbative fermion production. Interestingly it turns out that, for $y_\phi = 10^{-4}$, neglecting the time-dependent mass $y_\phi \phi$ actually gives a better approximation for $\Tmax$  than including it when comparing to a non-perturbative computation, see figure 6 in \cite{Garcia:2021iag}. However, neither the perturbative computation from which \eqref{EffectiveYukawa} is obtained nor the non-perturbative result obtained in  \cite{Garcia:2021iag} include the fermions' gauge interactions, which can affect the efficiency of fermionic reheating (e.g. by re-distributing the energy density transferred into $\Psi$ into other degrees of freedom, or if the effective mass generated from forward scatterings terminates non-perturbative fermion production). This clearly indicates that the quantitative understanding of fermionic reheating in realistic models is incomplete at this stage. 
While this does not invalidate the conclusion drawn in \cite{Liu:2025sut} that regime \ref{it:reg2} can be extended to inflaton couplings exceeding the upper bound \eqref{GeneralScaling} for fermionic reheating, it implies that the use of perturbative methods to compute $\GG$ during the early stage of reheating is questionable whenever \eqref{GeneralScaling} is violated. 
This casts doubt on any studies computing the relic densities -- may it be for DM or gravitational waves (GW) -- by means of perturbation theory.
We nevertheless stick to the interaction \eqref{YukawaReheating} because the range of validity of perturbative methods 
for bosonic reheating is well-known to be even smaller, and the stronger feedback effects caused by induced transitions impose even stronger restrictions on the range of inflaton couplings $\g$ for which the full set of parameters needed to reconstruct the thermal history during reheating can be constrained from observations.

In the following, we compare the results of solving \eqref{eq:reducedform_withDM} with \eqref{eq:inflatondecayrate} and \eqref{EffectiveYukawa}. 
While none of these two approximations are strictly correct, this suffices to illustrate that the thermal history in regime \ref{it:reg2} can impact the DM relic density in some cases. Moreover, the difference between the results obtained under both approximations can be interpreted as a rough proxy for the modelling uncertainty in fermionic reheating. 

All parameters associated with this choice of inflation scenarios are summarised in Tab.~\ref{tab:inflation_benchmark}.\footnote{We note in passing that we chose $\alpha$ sufficiently large, i.e. $\alpha \gtrsim 0.25$, to avoid non-perturbative production of inflaton particles (fragmentation) due to higher-order terms in the potential \cite{Drewes:2019rxn}.}
For practical purposes, we can approximate $M_\phi = m_\phi$.
In addition, we assume that the DM is produced from particles in the early universe bath that are not directly coupled to the inflaton, such that one can safely neglect the impact of inflaton-generated effective masses on the DM production process itself.

\begin{table}[!t]
\begin{center}
\begin{tabular}{|c|c|c|c|c|c|c|c|}
\hline
Parameter & $m_\phi$ [GeV] & $\Phi$ & $y_\phi$ & $\alpha$ & $M$ [GeV] & $n_s$ & $r$\\
\hline
Value & $1.5\cdot 10^{13}$ & $2 \cdot 10^{10}$ & $5\cdot 10^{-3}$ & $2$ & $9.46\cdot 10^{15}$ & $0.965$ & $6.96 \cdot 10^{-3}$\\

\hline
\end{tabular}
\end{center}
\caption{Parameters characterising our benchmark for the inflation scenario used in the rest of this section. Note that all these parameters are not independent of each other and the first three parameters are sufficient to recover $\alpha$, $\M$, $n_s$ and $r$. }
\label{tab:inflation_benchmark}
\end{table}

\subsection{UV-dominated freeze-in from a dimension-9 operator
}\label{Sec:dim9}

In this first counter-example, we assume DM to be primarily produced from a dimension-9 operator where the masses of all fields except the DM $X$ can be neglected.
For instance, the Lagrangian could include the following interaction
\begin{align}\label{dimnine}
    \mathcal{L} \supset  \frac{\YX}{\Lambda^{5}}\left(\bar{X}\Psi_1\right) \left(\bar{\Psi}_2\Psi_3\right) \left(\bar{\Psi}_4\Psi_5\right)\ ,
\end{align}
where $\YX$ represents the effective coupling of this operator and the $\Psi_i$ are all massless fermions.
In this case, dimensional analysis implies
that the (momentum-averaged) DM production rate takes the form\footnote{
It is well known that graviton-mediated scatterings should produce DM already at dimension $8$. However, because gravitational interactions are much more severely suppressed by the Planck scale, it is therefore reasonable to neglect these for our choice of benchmark. This can be seen by comparing $\left(\Tmax/\Lambda\right)^{2(d-4)=10} \approx 1.7\cdot 10^{-7}$ with $\left(\Tmax/\mpl\right)^{2(d-4)=8} \approx 8.2\cdot 10^{-40}$, where $\Tmax$ was chosen for the scenario including thermal effects in the inflaton decay rate.
}
\begin{align}
\label{eq:dim9prodrate}
    \GX = c \YX^2 \frac{T^{2(d-4)+1}}{\Lambda^{2(d-4)}} = c \YX^2 \frac{T^{11}}{\Lambda^{10}}\ ,
\end{align}
with $c$ a numerical
factor. For concreteness, we choose
\begin{align}
    \YX = 3.8\cdot 10^{-2} ~~~ \mbox{ and } ~~~ c = 10^{-3}\ .
\end{align}
We also set $m_X = m_\phi/10^4$ as well as $\Lambda=10m_\phi$ for our EFT to remain valid during the entirety of the reheating epoch.

\begin{figure}
    \centering
    \includegraphics[width=0.49\textwidth,trim={.25cm 0 0 0}]{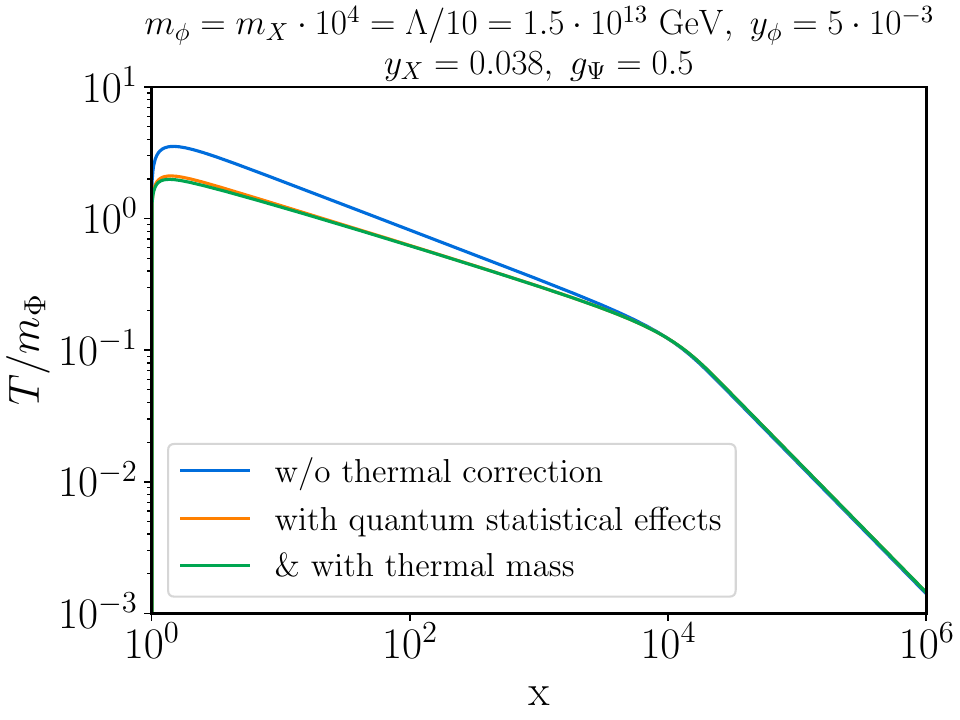}
    \includegraphics[width=0.49\textwidth,trim={.25cm 0 0 0}]{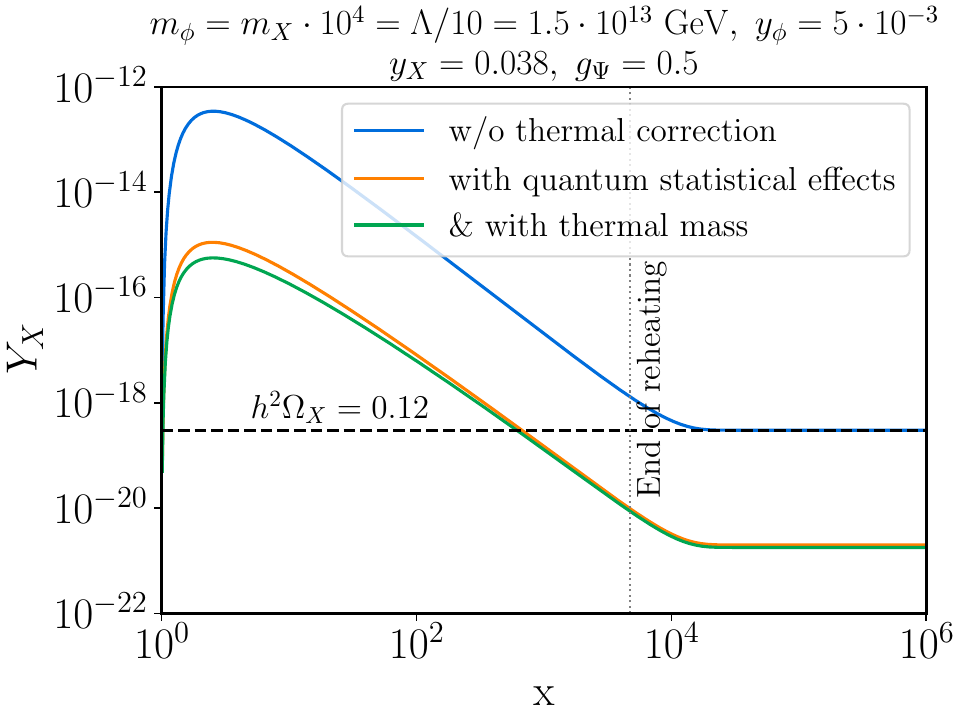}
    \caption{Time evolution 
    of the radiation bath's temperature $T$
    and DM yield $Y_X$
    assuming DM production from a dimension-9 operator, see Eq.~\eqref{eq:dim9prodrate}.
The green curve is based on solving the system of equations \eqref{eq:reducedform_withDM} with $\GG$ given by  given by \eqref{eq:inflatondecayrate} and $M_\Psi= \gPsi T/2$; for the orange line the thermal mass is neglected ($M_\Psi=0$), the blue line also neglects the quantum-statistical correction ($\fF=0$). In the right panel, the dashed horizontal line represents the observed present-day DM yield whereas the dotted vertical line denotes the end of the reheating era.}
    \label{fig:CounterExampleDim9}
\end{figure}

The resulting DM abundance as a function of the temperature is displayed in Figs.~\ref{fig:CounterExampleDim9} and \ref{fig:CounterExampleDim9_comparison}. 
Fig.~\ref{fig:CounterExampleDim9} is based on the approximation \eqref{eq:inflatondecayrate} for $\GG$, neglecting the impact of the effective mass $y_\phi \phi$ that fermions obtain due to their coupling to $\phi$ in \eqref{FermionMassFull}.   
As a result of our relatively large choice of coupling $y_\phi$, \eqref{eq:Tmaxrelation} would predict $\Tmax > m_\phi$ if thermal effects are neglected, but the inclusion of 
Pauli blocking reduces the maximal temperature reached by the early universe plasma by a factor of $\mathcal{O}(2)$. 
This is sufficient to modify the expected DM abundance by orders of magnitude, as is highlighted by the right panel in Fig.~\ref{fig:CounterExampleDim9}, because the DM production from \eqref{eq:dim9prodrate} is UV-dominated.
At the same time, the impact of the thermal fermion mass correction $\gPsi T/2$ is negligible. 
However, Fig.~\ref{fig:CounterExampleDim9_comparison} shows that the contribution $y_\phi \phi$ to the effective mass \eqref{FermionMassFull} can have an even larger impact than quantum statistics. The large discrepancy between the results obtained with the fundamental Yukawa coupling $y_\phi$ and the effective coupling $y_{\rm eff}$ in \eqref{EffectiveYukawa} indicates a significant uncertainty in the modelling of fermionic reheating.

Two conclusions can be drawn from this. Firstly, thermal corrections to $\GG$ -- in this case Pauli blocking -- can be relevant and affect the DM relic density by orders of magnitude for UV-dominated freeze-in. Secondly, the current uncertainty in the modelling of fermionic reheating introduces an even larger uncertainty in the relic abundance.

\begin{figure}
    \centering
    \includegraphics[width=0.49\textwidth,trim={.25cm 0 0 0}]{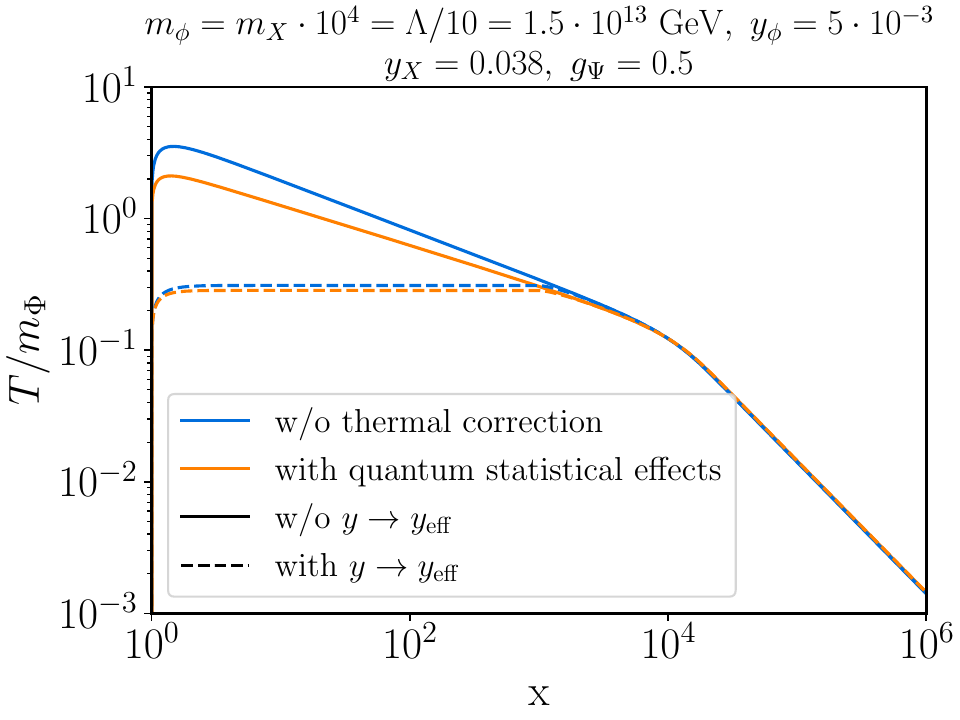}
    \includegraphics[width=0.49\textwidth,trim={.25cm 0 0 0}]{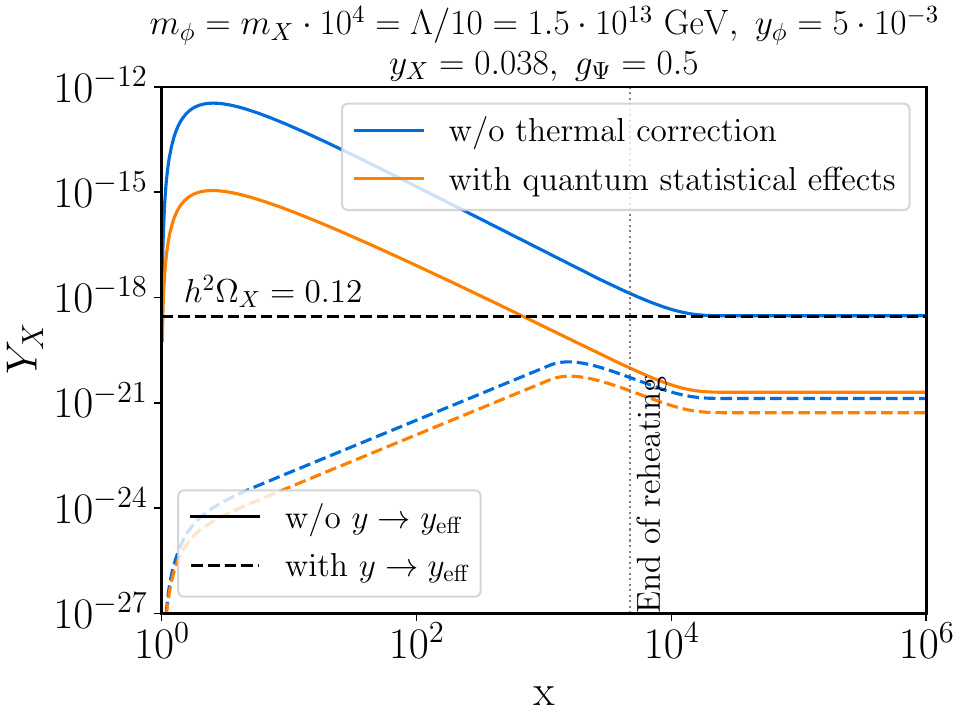}
    \caption{Time evolution of the radiation bath's temperature and DM yield assuming a production from a dimension-9 operator, see Eq.~\eqref{eq:dim9prodrate}.
    The solid lines are identical to Fig.~\ref{fig:CounterExampleDim9}, the dashed lines use the effective Yukawa coupling \eqref{EffectiveYukawa}.
    }
    \label{fig:CounterExampleDim9_comparison}
\end{figure}

\subsection{UV-dominated freeze-in 
due to Boltzmann suppression
}\label{Sec:SuperHeavy}

For our second counter-example,
we consider DM production from a Fermi-like interaction as in the 
Lagrangian \eqref{eq:FermiLagrangian}. 
The possibility to realise UV-dominated freeze-in from Boltzmann suppression has previously been 
observed in renormalisable models in \cite{Giudice:2000ex,Cosme:2023xpa}.
The key difference to Sec.~\ref{sec:DMprodFermilike_noinflatonthermal} is that  
we consider the case of a heavy mediator $X$, i.e. $m_X \gtrsim m_\phi$,
while we continue to 
assume that $m_{\Psi_1} = m_X$ in order to prevent the vacuum decay of $X$ or $\Psi_1$. 
The heavy masses $m_{\Psi_1}$ and $m_X$ exponentially suppress $\GX$ for $T\ll m_\phi$, implying that DM production is UV-dominated.
To be explicit, we choose 
\begin{align}
    m_{Z^\prime} \simeq 1.5 \cdot 10^{17} \mbox{ GeV} \gg T\ ,\\
    m_X = m_{\Psi_1} = 10 m_\phi\ .
\end{align}
Crucially, this scenario relies on the existence of another heavy particle $\Psi_1$, distinct from $X$ and already in thermal equilibrium in order to create the necessary Boltzmann suppression of the DM production rates during the reheating epoch. 
In that regard, a scenario where $\Psi_1$ is replaced by the out-of-equilibrium $X$ could limit the impact of thermal effects.

The resulting DM abundance as a function of the temperature is displayed in Figs.~\ref{fig:CounterExampleFermi} and \ref{fig:CounterExampleFermi_comparison}. 
As in the previous example in Sec.~\ref{Sec:dim9}, Fig.~\ref{fig:CounterExampleFermi} shows that thermal effects in the form of Pauli blocking lead to a reduction of $\Tmax$ by a factor of order $2$
when estimating $\GG$ by \eqref{eq:inflatondecayrate}.
This leads to a drastic reduction of the final DM abundance by about two orders of magnitude, which in this case is caused by the exponential dependence of $\GX$ on $T$. However, Fig.~\ref{fig:CounterExampleFermi_comparison} shows that this effect is outdwarfed by the uncertainty in the modelling of fermionic reheating; using the effective coupling \eqref{EffectiveYukawa} leads to a considerably stronger suppression of $\Tmax$ than Pauli blocking.   Qualitatively this resembles the behaviour observed for the previous example in Sec.~\ref{Sec:dim9}, but quantitatively the effect is even stronger due to the exponential dependence of $\GX$ on $T$, which has to be compared to the power-law \eqref{eq:dim9prodrate}.
\begin{figure}
    \centering
    \includegraphics[width=0.49\textwidth,trim={.3cm 0 0 0}]{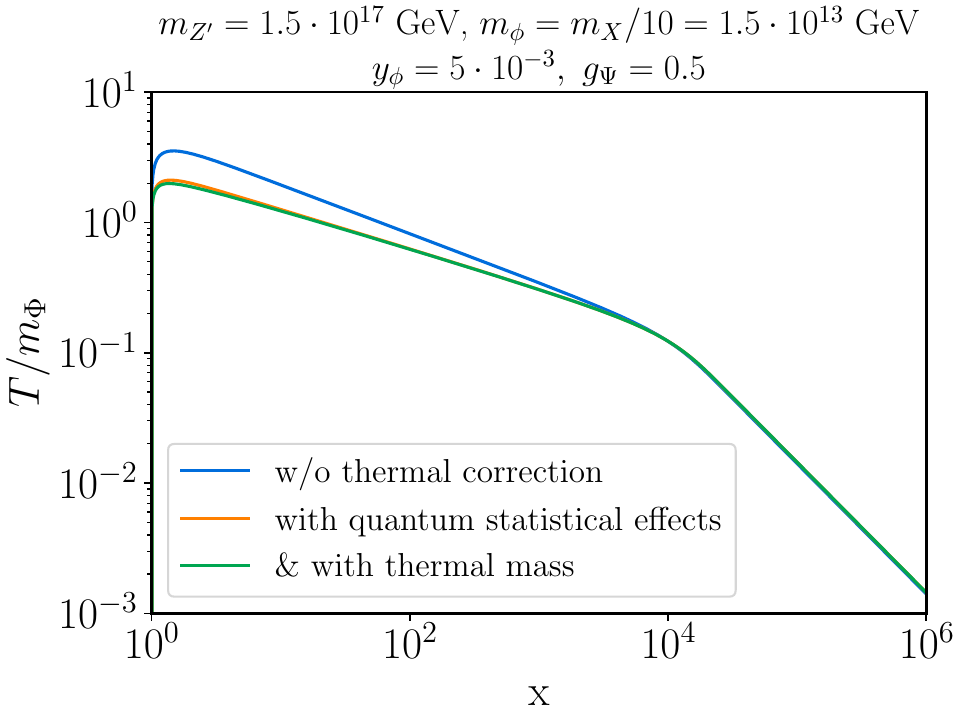}
    \includegraphics[width=0.49\textwidth,trim={.3cm 0 0 0}]{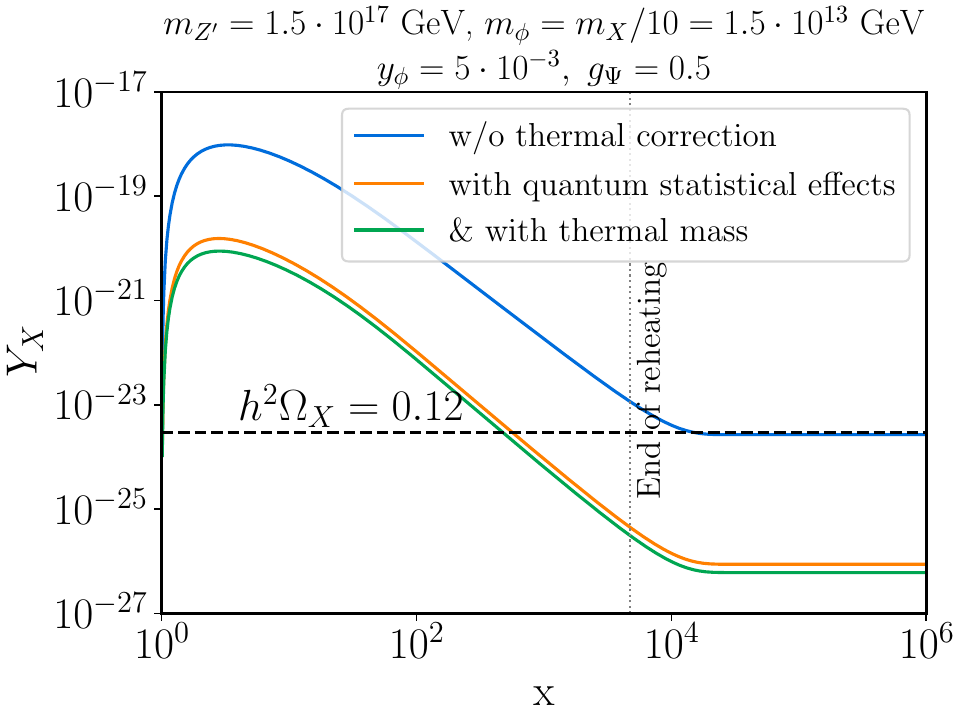}
    \caption{Time evolution (as function of the scale factor $x = am_\phi$) of the radiation bath's temperature $T$ and DM yield $Y_X$ assuming a production from a Fermi-like interaction, see Eq.~\eqref{eq:fullFermirates}. 
    In the right panel, the dashed horizontal line represents the observed present-day DM yield whereas the dotted vertical line denotes the end of the reheating era.}
    \label{fig:CounterExampleFermi}
\end{figure}

\begin{figure}
    \centering
    \includegraphics[width=0.49\textwidth,trim={.3cm 0 0 0}]{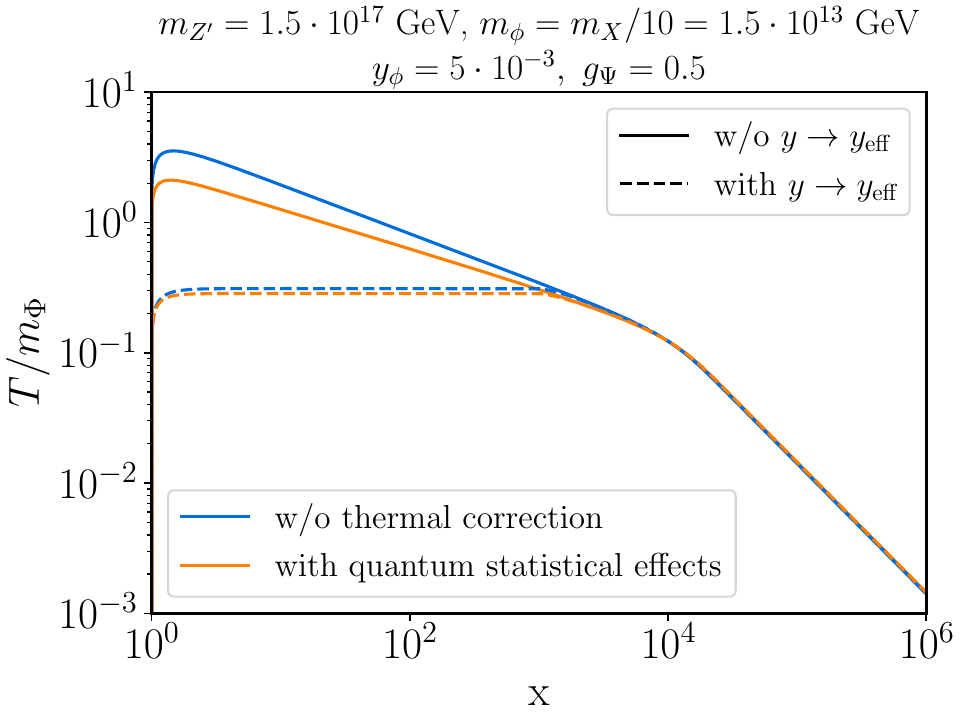}
    \includegraphics[width=0.49\textwidth,trim={.3cm 0 0 0}]{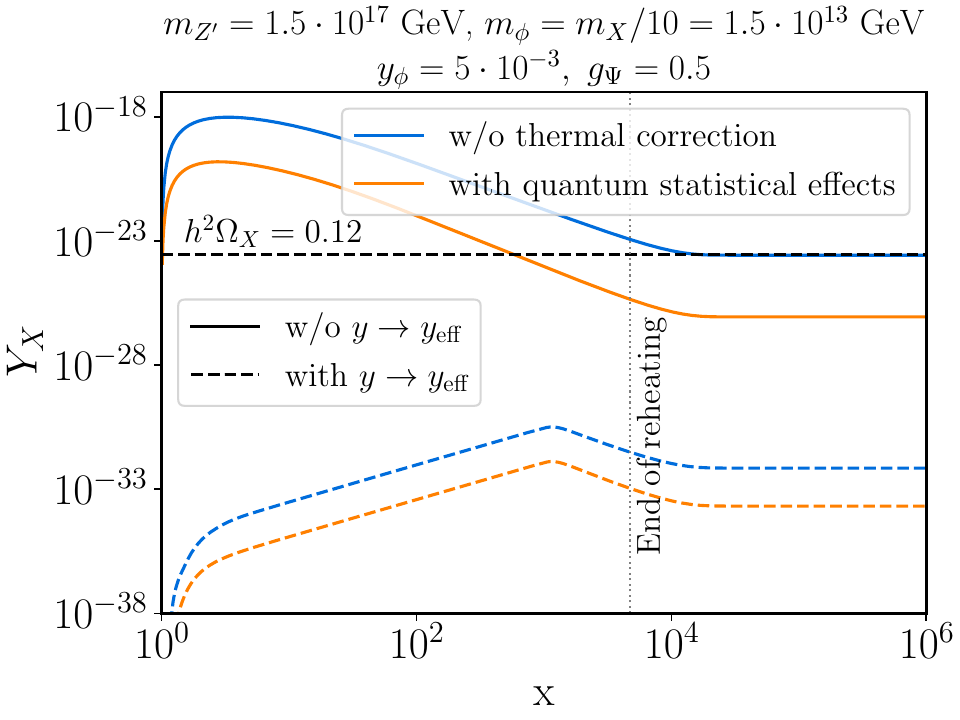}
    \caption{Time evolution (as function of the scale factor $x = am_\phi$) of the radiation bath's temperature $T$ and DM yield $Y_X$ assuming a production from a Fermi-like interaction, see Eq.~\eqref{eq:fullFermirates}.
    The solid lines are identical to Fig.~\ref{fig:CounterExampleDim9}, the dashed lines use the effective Yukawa coupling \eqref{EffectiveYukawa}.}
    \label{fig:CounterExampleFermi_comparison}
\end{figure}

Finally, let us briefly comment on 
the consistency of the scenario
presented here. As highlighted earlier, the order of magnitude impact of thermal effects on the DM abundance rely on the coupling of $X$ with a heavy particle $\Psi_1$ already in thermal equilibrium. Given that $\Psi_1$ is as heavy as $X$, it necessarily needs to possess additional interactions which rapidly brought it to equilibrium. These additional interactions, e.g. $(\bar{\Psi}_1 \ell) \phi $ where $\ell$ is another fermion and $\phi$ is another scalar field, could in principle open new production channels for the DM or even spoil its stability since decays of the form $X \rightarrow \Psi_3 + \Psi_2 + \ell + \phi$ could be allowed. Although it might not be impossible to build a scenario where these new channels are suppressed, this further strengthens our conclusion that it is in general difficult to build a 
realistic model in 
which thermal effects lead to a sizeable change to the DM abundance in the regime where they can be computed perturbatively.

\subsection{UV-dominated freeze-in from threshold effects
}\label{Sec:Threshold}

In our final counterexample, the UV-dominated DM production is achieved by kinematic effects that are sensitive to screening effects in the plasma.  
We consider the model described by Eq.~\eqref{eq:Lagrangian1to2decay}. In order to come into thermal equilibrium, the parent particle necessarily must have additional interactions, the impact of which on $\GX$ and the relic abundance we neglected in Sec.~\ref{sec:1to2DMprod}.
These interactions modify the in-medium dispersion relations of $P$-particles, which in the simplest case can be parametrised by a momentum-independent thermal mass $M_P$.
The impact of this on $\GX$ has been studied in Sec.~3 of \cite{Drewes:2015eoa}.
When the DM production is dominated by $1\to 2$ decays of $P$, this increases the DM production rate by decreasing the lifetime of $P$-particles, which is parametrically proportional to $1/M_P$. If they are gauge interactions, then at least one of the decay products $X$ and $f_R$ must be charged under the same gauge group and also receives a thermal mass from the same interaction, which we take to be $f_R$ here. Since the thermal mass generated for $P$ and $f_R$ is identical or similar in magnitude, the phase space for the decay 
$P \to X \ f_R$ becomes increasingly suppressed at high $T$, implying that scatterings become increasingly relevant at $T> m_P$. 
However, as a scalar, $P$ can have other self-interactions (e.g.~a quartic interaction $(P^\dagger P)^2$) or additional Yukawa-interactions with other fields. As a result, $M_P$ may exceed the thermal mass of the decay products, assuring that the $1\to 2$ decay remains the dominant contribution to $\GX$ even at $T>m_P$.
In this setup it is conceivable that the decay is kinematically forbidden in vacuum ($m_P < m_X + m_{f_R}$), but becomes allowed at high temperature when all masses are replaced by their thermal counterparts ($M_P > M_X + M_{f_R}$).
This can lead to UV-dominated freeze-in.

For the purpose of illustration, we choose a simple parametrisation $M_P^2 = m_P^2 + \lambda_P T^2/24 $ and
\begin{align}
    m_P = m_\phi/5\ \ , \
    m_{f_R} = m_\phi/8\ \ , \
    m_X = m_\phi/10\  \ , \ \yX = 6\cdot 10^{-7}\ \ , \ \lambda_P = 0.25\ .
\end{align}
In general, the DM production rate takes the form displayed in Eq.~\eqref{eq:DMprodrate_renormalisable_fullthermal}, which is what we will use in the rest of this section.
The resulting DM abundance as a function of the temperature is displayed in Fig.~\ref{fig:CounterExample1to2decay}.
The left panel of shows that the DM production is shut down during reheating when the decay $P\to X \ f_R$ becomes kinematically forbidden.
Hence, DM production freezes in earlier and undergo a larger period of dilution when one systematically includes Pauli blocking in the inflaton decay rate, ultimately leading to more than two orders of magnitude of difference in the overall yield.
In this case it is the combination of the Pauli blocking and the thermal mass, i.e., a screening effect, that makes the relic abundance sensitive to thermal corrections. 
However, as in the previous two examples, the uncertainty in the modelling of fermionic reheating is larger than the impact of thermal corrections.
In Fig.~\ref{fig:CounterExample1to2decay}, $\GG$ was computed based on \eqref{eq:inflatondecayrate}; when using \eqref{EffectiveYukawa} no DM is produced at all in $P$-decays because $T$ is never large enough for $M_P$ to exceed the DM mass.

\begin{figure}
    \centering
    \includegraphics[width=0.49\textwidth]{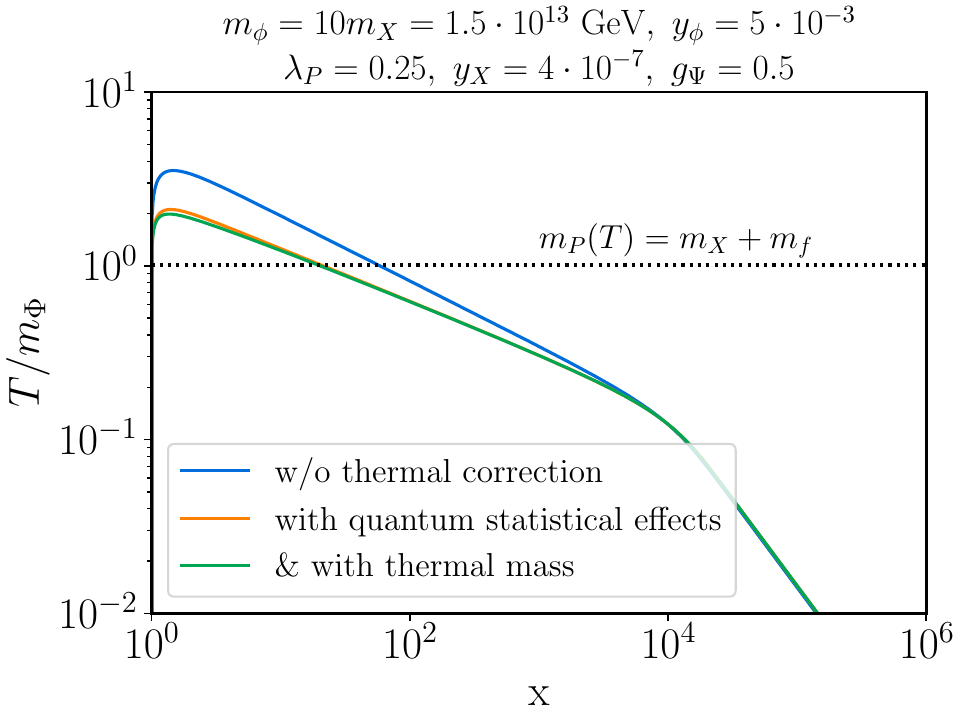}
    \includegraphics[width=0.49\textwidth]{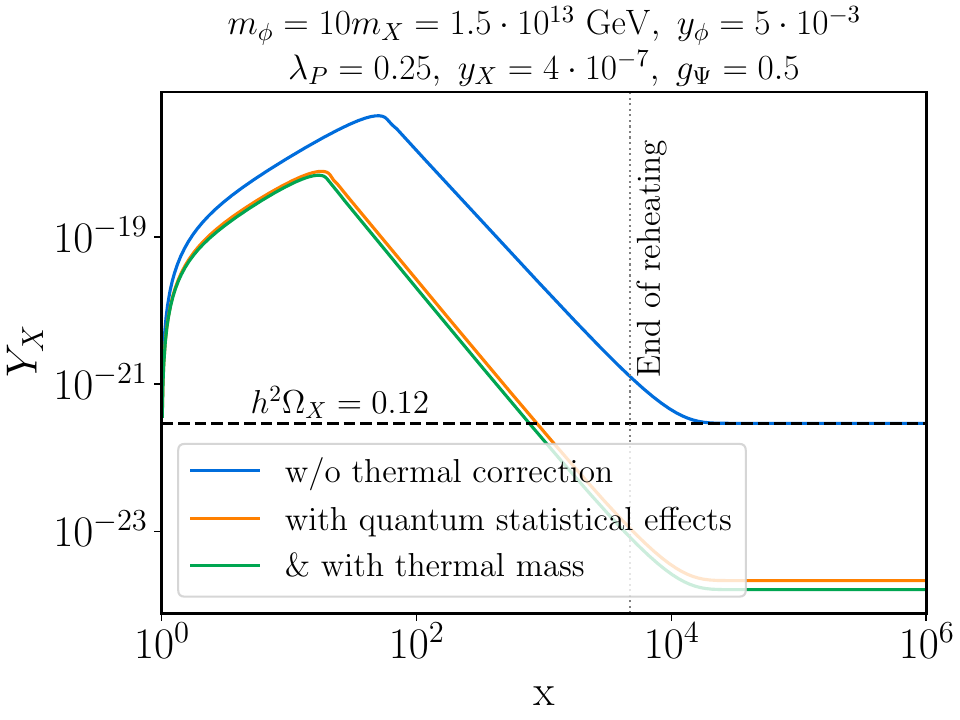}
    \caption{Time evolution (as function of the scale factor $x = am_\phi$) of the radiation bath's temperature $T$ and DM yield $Y_X$. The dotted line in the left panel denotes the threshold temperature below which the DM production is kinematically forbidden. In the right panel, the dashed horizontal line represents the observed present-day DM yield whereas the dotted vertical line denotes the end of the reheating era.}
    \label{fig:CounterExample1to2decay}
\end{figure}

Finally, we note in passing that the mechanism presented here completely neglects $P$- and $f_R$-mediated scatterings that could deplete or increase the DM abundance. 
While scatterings are typically subdominant to decays in the scenario described in Sec.~\ref{sec:1to2DMprod}, this does not necessarily mean that they cannot contribute sizeably to the final DM yield once decays are kinematically forbidden.
Moreover, interactions that are strong enough to give a large thermal mass to $P$ can be expected to also mediate scatterings that soon dominate over the decay unless the thermal correction in $M_P$ primarily comes from self-interactions \cite{Drewes:2015eoa}. 
 We however leave a more quantitative analysis of their impact for a future work. 
This difficulty further strengthens our claim that it is in general difficult to build a complete UV model of DM production during perturbative reheating for which thermal effects lead to a sizeable change to the DM abundance.

\section{Discussion and conclusion}
\label{sec:conclusion}

We studied the impact of finite-temperature corrections to thermal DM production during cosmic reheating. We focussed on quantum statistics and screening effects in the primordial plasma that modify either the inflaton decay rate $\GG$ or the rate  of DM production $\GX$. An important motivation for this is provided by the perspective to constrain the thermal history during reheating with upcoming CMB observations, which opens up the possibility to make predictions for collider experiments based on the requirement to explain the observed DM relic density. 

Our main finding in Secs.~\ref{sec:ThermaleffectsInflaton} and~\ref{sec:ThermaleffectsDMprod}  is that, as a general rule, finite-temperature corrections to $\GG$ and $\GX$ do not modify the DM relic density significantly in the regime where they can be computed by means of 
finite-temperature QFT. This is illustrated in Figs.~\ref{fig:Thermal effects_DMprodrate} and \ref{fig:Ferminteraction_DMthermaleffect}. 
In particular, they are typically sub-dominant when the reheating process is sufficiently simple so that all parameters needed to reconstruct the thermal history during reheating can be constrained from observations. 

In Figs.~\ref{fig:mPvsdecaylength}, \ref{fig:scenA}, \ref{fig:scenB} and \ref{fig:scenC} we illustrate how current and future CMB observations can provide complementary information to laboratory experiments on the DM particles and the mediators involved in their production in specific models. For instance, Fig.~\ref{fig:scenB} shows that CMB observations could  disfavour large fractions of the parameter space that is inaccessible to collider experiments. 

In Sec.~\ref{sec:counterexamples} we provide a few counterexamples to the aforementioned general rule in which thermal effects can change the DM relic density by orders of magnitude, see Figs.~\ref{fig:CounterExampleDim9}-\ref{fig:CounterExample1to2decay}.  
However, they are not counterexamples in the strict sense because our general rule referred to the perturbative regime, 
and we find that determining the thermal history during reheating by means of perturbative methods is questionable even for fermionic reheating and even in the regime where their use can be justified for the computations of the expansion history and the reheating temperature $\Treh$. 
The resulting modelling uncertainty outdwarfs 
the impact of thermal corrections on the relic density. This casts doubt on the validity of DM or GW relic density computations by means of perturbation theory and provides a strong motivation to improve the quantitative understanding of reheating in realistic models, including fermionic reheating.

Finally, in the course of this work we found a general estimate \eqref{mPhiPlateau} for the inflaton mass in plateau models that has, to the best of our knowledge, not been reported previously. We explicitly confirm it for three models in appendix \ref{app:PlateauModels}.

\section*{Acknowledgements}

The authors thank
Marcos Garcia, Xun-jie Xu
and Yong Xu for helpful discussions regarding preheating. 
They would also like to thank Lei Ming for providing the code to derive the expected constraints set by LiteBIRD on the reheating temperature.
The work of Y.G. has been supported by the French Community of Belgium through the FRIA grant No. 1.E.063.22F and by the World
Premier International Research Center Initiative (WPI), MEXT, Japan (Kavli IPMU). The work of S.Z. was supported by the European Research Council Gravites Horizon Grant AO number: 850 173-6. M.A.S.M. acknowledges the Conseil de l'action internationale (CAI) de l'UCLouvain for its support through the Coopération au développement grant. He also acknowledges support from the Erasmus+ programme of the European Union through Agence francophone pour l’éducation et la formation tout au long de la vie (AEF-Europe) at UCLouvain, and is grateful to the Technical University of Munich Physics Department for its hospitality during the completion of this work. M.A.S.M. further acknowledges the Abdus Salam International Centre for Theoretical Physics (ICTP) for its continuous support throughout his PhD and its generous hospitality. 

\textbf{Disclaimer:} Funded by the European Union. Views and opinions expressed are however those of
the authors only and do not necessarily reflect those of the European Union or European Research
Council. Neither the European Union nor the granting authority can be held responsible for them.

\appendix

\section{Relation between $\Treh$ and CMB observables}\label{TrefromCMB}

The relationship between $\Treh$ and CMB observables are well known and have been reviewed many times; here we give them for completeness following \cite{Ueno:2016dim}.
$\Nreh$ can be related to 
the number of $e$-folds $\Nk$ between 
the end of inflation and
the horizon-crossing of perturbations with wave number $k$  via
\begin{equation}
	\label{Nre}
	\begin{split}
		N_{\rm re} &= \frac{4}{3\wrehbar-1}\bigg[\Nk+\ln\left(\frac{k}{a_0 T_0}\right)+\frac{1}{4}\ln\left(\frac{40}{\pi^2g_\star}\right)
        +\frac{1}{3}\ln\left(\frac{11g_{\star}}{43}\right)-\frac{1}{2}\ln\left(\frac{\pi^2\mpl^2r A_s}{2\sqrt{\Vend}}\right)\bigg].
	\end{split}
\end{equation}
The above equation implicitly assumes that the effective numbers of degrees of freedom contributing to the energy and entropy density, respectively, are equal and constant over time, i.e., $g_\rho(T) = g_s(T) = g_\star$. 
$T_0=2.725~{\rm K}$ is the temperature of the CMB at the present time and $a_0$ the current scale-factor. 
$\Nk$ can be obtained from
\begin{equation}\label{Nk}
	\Nk=\ln\left(\frac{a_{\rm end}}{a_k}\right)=\int_{\phi_k}^{\phi_{\rm end}}\frac{H \d\phi}{\dot{\phi}}
	\approx\frac{1}{\mpl^2}\int_{\phi_{\rm end}}^{\phi_k}\d\phi\frac{\V }{\partial_\phi \V },
\end{equation}
where $\phi_k, H_k$, etc.~denote the values of  
$\phi, H$, etc.~at the horizon-crossing of the scale $k$. One can find $\phi_k$ in terms of 
$n_s$ and $r$ 
by solving
\begin{equation}
	n_s=1-6\epsilon_k+2\eta_k~, \quad r=16\epsilon_k.
	\label{nANDr}    
\end{equation}
In the slow-roll regime $\epsilon,\eta \ll 1$
\begin{equation}
	\label{H_k}
	H^2_k=\frac{\V(\phi_k)}{3\mpl^2}~
	=\pi^2 \mpl^2\frac{r A_s}{2}.
\end{equation}
By inserting \eqref{Nre} with \eqref{Nk} into \eqref{Tre}, $\Treh$ can be expressed in terms of the observables $\{A_s, n_s, r\}$, 
while $\phi_k$ is determined by solving \eqref{nANDr} for $\phi_k$. $\Vend$ and $\phi_{\rm end}$ are found by solving $\epsilon=1$ for $\phi$.
Using \eqref{nANDr} yields
\begin{equation}
	\epsilon_k=\frac{r}{16}~,\quad \eta_k=\frac{n_s-1+3r/8}{2}.
\end{equation}
Together with the definitions of $\epsilon$ and $\eta$ this gives
\begin{equation}\label{TakaTukaUltras}
\frac{\partial_\phi \V}{\V}\Bigl|_{\phi_k}=\sqrt{\frac{r}{8\mpl^2}} \ , \quad
\frac{\partial^2_\phi \V}{\V}\Bigl|_{\phi_k}=\frac{n_s-1+3r/8}{2\mpl^2}.
\end{equation}
Combining this with \eqref{H_k}, we obtain three equations which relate the potential and its derivatives to the observables $\{A_s,n_s,r\}$. In this way, the quantities $\wrehbar$ and $\Nreh$ in \eqref{Nre} in \eqref{GammaConstraint} can be expressed in terms of observables.

In practice  fixing $A_s$ to its best fit value provides a good approximation \cite{Drewes:2023bbs}. We can then quantify the knowledge gain about $\X=\log_{10}(\Treh/{\rm GeV})$  from data $\mathcal{D} = \{n_s,r\}$ in terms of a posterior distribution $P(\X|\mathcal{D})=P(\mathcal{D}|\X)P(\X)/P(\mathcal{D})$, where 
\begin{align}
    P(\mathcal{D})=\int d\X P(\mathcal{D}|\X)P(\X)\ ,
\end{align}
and the likelihood is approximated by 
\begin{eqnarray}\label{Eq:Likelihood}
	P(\mathcal{D}|\X) = C_2\mathcal{N}(n_s,r|\nsbar,\sigmans;\rbar,\sigmar)\theta(r). 
\end{eqnarray}
Demanding $\int P(\mathcal{D}|\X)d\mathcal{D} = 1$ fixes the constant $C_2$. 
The function $\mathcal{N}(n_s,r|\nsbar,\sigmans;\rbar,\sigmar)$ can be well approximated  by a two-dimensional Gaussian \cite{Drewes:2023bbs} that is parametrised by experimental sensitivities $\sigmans$ and $\sigmar$ for given fiducial values $r=\rbar$ and $n_s=\nsbar$. 
Expressing $\Treh$ and $\Nreh$ in terms of $\X$, we use a flat prior $P(\X) = \theta(\Nreh)\theta(\Treh - T_{\rm BBN})$, with $T_{\rm BBN} = 5.96$ MeV to assure successful primordial nucleosynthesis \cite{Barbieri:2025moq}.\footnote{We note that the precise lower bound on $T_{\rm re}$ depends on the details of how the inflaton couples to the SM. The above constraint has been obtained assuming that it does not couple to the hadronic sector but relaxing this assumption is only expected to slightly soften the lower bound on the reheating temperature, see e.g.~\cite{Kawasaki:1999na,Kawasaki:2000en,Hasegawa:2019jsa,deSalas:2015glj} for more detailed discussions.}

\section{Plateau models of inflation}\label{app:PlateauModels}
The general arguments presented in the main text only rely on the estimates \eqref{Mgeneral} and \eqref{mPhiPlateau}, which apply to a wide class of plateau models.
The generality of \eqref{Mgeneral}  directly follows from \eqref{H_k}, for \eqref{mPhiPlateau} we demonstrate it in the following Sec.~\ref{mphigenericPlateauPotential}.
Connecting CMB observables to $\Treh$ and $\g$, however, requires specifying a model of inflation. In Sec.~\ref{app:CMBconstraints_inflationaryparam} we provide the required relations for three concrete models, for which we verify \eqref{Mgeneral} and \eqref{mPhiPlateau} explicitly.  

\subsection{Inflaton mass in two-parameters plateau models}\label{mphigenericPlateauPotential}

We assume the inflationary potential is described by 
\begin{equation} \label{genericPlateauPotentialExponential}
    \Vphi(\phi) = M^4 f \left(\frac{\phi}{m_{\text{CMB}}}\right)\;,
\end{equation}
where $M$ and $m_{\text{CMB}}$ are two mass scales. Since the number of e-folds can be expressed as
 \begin{equation}
    \Nk = \frac{1}{\mpl} \int_{\phi_{\rm end}}^{\phi_k} \mathrm{d} \phi \frac{1} {\sqrt{2 \epsilon}}  \;,
\end{equation}
we approximate $\phi_k \sim m_{\text{CMB}}$ to estimate
\begin{equation} \label{epsilonApp}
    \epsilon_k \sim \frac{m_{\text{CMB}}^2}{\Nk^2 \mpl^2} \;.
\end{equation}
As evident from the example below, this scaling is correct up to logarithmic corrections. 
Imposing $\Vphi/\epsilon_k =24 \pi^2 A_s\mpl^4$, see Eqs.~\eqref{nANDr} and \eqref{H_k}, now fixes 
\begin{equation} \label{normalizationApp}
\frac{M^2}{m_{\text{CMB}}} \sim  \frac{\sqrt{24 \pi^2 A_s}\mpl}{\Nk}  \;.
\end{equation}
As a result, expanding the potential around the origin gives 
\begin{equation} \label{mPhiPlateauAppendix}
      m_\phi \sim \sqrt{\partial_{\phi\phi} \Vphi} \sim \frac{M^2}{m_{\text{CMB}}} \sim  \frac{\sqrt{24 \pi^2 A_s} \mpl}{\Nk} \;,
\end{equation}
which matches  Eq.~\eqref{mPhiPlateau} shown in the main part. It is important to point out that the dimensionless prefactor in the above equation can be zero. In this case, the inflaton is massless during reheating and Eq.~\eqref{mPhiPlateau} does not apply. This is e.g.~realised in $\alpha$-attractor models (cf.~Eq.~\eqref{eq:potentialalphaattractor}) if the exponent of the hyperbolic tangent is larger than $2$. 

Finally, we illustrate how Eq.~\eqref{mPhiPlateauAppendix} arises in a class of plateau models described by
\begin{equation} \label{genericPlateauPotential}
    \Vphi(\phi) = \M^4 \left(1-c\, \exp\left(-\frac{\phi}{\sqrt{2}m_{\text{CMB}}}\right)\right)^2 \;,
\end{equation}
where $c>0$ is a dimensionless coefficient and the quadratic exponent ensures that the inflaton has a mass during reheating. Approximating
\begin{equation}
    \frac{\partial_\phi \Vphi}{\Vphi} \approx \frac{\sqrt{2} c}{m_{\text{CMB}}}\exp\left(-\frac{\phi}{\sqrt{2} m_{\text{CMB}}}\right) \;,
\end{equation}
we compute the field value as a function of the number of e-folds $\Nk$ from Eq.~\eqref{Nk}
\begin{equation}
    \exp\left(-\frac{\phi_k}{\sqrt{2} m_{\text{CMB}}}\right) \approx \frac{m_{\text{CMB}}^2}{c \Nk \mpl^2} \;.
\end{equation}
Hence, the first slow-roll index $\epsilon_k \equiv \frac{\mpl^2}{2}\left(\frac{\partial_\phi \Vphi}{\Vphi}(\phi = \phi_k)\right)^2$ can be computed explicitly
\begin{equation}
    \epsilon_k \approx \frac{c^2 \mpl^2}{m_{\text{CMB}}^2}  \exp\left(-\sqrt{2}\frac{\phi_k}{m_{\text{CMB}}}\right) \approx  \frac{m_{\text{CMB}}^2}{\Nk^2 \mpl^2} \;,
\end{equation}
and Eqs.~\eqref{nANDr} and \eqref{H_k} yield
\begin{equation}
\frac{M^2}{m_{\text{CMB}}} =  \frac{\sqrt{24 \pi^2 A_s}\mpl}{\Nk}  \;.
\end{equation}
Again we arrive at Eq.~\eqref{mPhiPlateauAppendix} and Eq.~\eqref{mPhiPlateau} shown in the main part. As a final cross-check, the $\alpha$-attractor potential \eqref{eq:potentialalphaattractor} corresponds to
\begin{equation}\label{UniversalMphi}
    c=2 \;, \qquad m_{\text{CMB}}=\frac{\sqrt{3}}{2} \sqrt{\alpha} \mpl \;,
\end{equation}
and so Eq.~\eqref{mPhiPlateauAppendix} matches Eqs.~\eqref{mPhiAlphaAttractor}, \eqref{mphiRGI} and \eqref{mphiMHI} derived in appendix~\ref{app:CMBconstraints_inflationaryparam}
and is consistent with what was mentioned in \cite{Cado:2025orb}.

\subsection{Useful relations for specific plateau models}
\label{app:CMBconstraints_inflationaryparam}

In this appendix, we summarise the constraints set by CMB observations on various motivated two-parameter inflationary scenarios. In particular, we show that these all predict an inflaton mass of $m_\phi \gtrsim \mathcal{O}(10^{13})$ GeV.

\subsubsection{$\alpha$-Attractor T-Model Inflation}

Originally developed in \cite{Kallosh:2013yoa}, $\alpha$-attractor models represent a broad class of chaotic inflation scenarios \cite{Linde:1983gd}.
In this work, we focus on so-called T-models whose potential takes the form \eqref{eq:potentialalphaattractor}, i.e.,
\begin{equation}
    \Vphi(\phi) = \M^4 \tanh^2\left[\frac{\phi}{\sqrt{6\alpha}\mpl}\right] \;,
\end{equation}
where $\M$ is a mass scale and $\alpha>0$ a real number.

In order to extract the constraints set by CMB observations, let us start with a brief inflationary analysis. The slow-roll indices are
	\begin{align}
		\epsilon &\equiv \frac{\mpl^2}{2}\left(\frac{\partial_\phi \Vphi}{\Vphi}\right)^2 = \frac{4 }{3\alpha \sinh^2\left(\frac{\sqrt{2} \phi }{\sqrt{3\alpha }\mpl}\right)} \;,\\
		\eta &\equiv  \mpl^2\frac{\partial_{\phi\phi} \Vphi}{\Vphi} = -\frac{4 \left(\cosh \left(\frac{\sqrt{2} \phi }{\sqrt{3\alpha }\mpl}\right)-2\right) }{3 \alpha \sinh^2\left(\frac{\sqrt{2} \phi }{\sqrt{3\alpha }\mpl}\right)} \ .
	\end{align}
Solving the condition $\epsilon=1$, we find the approximate field value at the end of inflation
	\begin{equation} \label{phiEndalphaT}
		\phi_{\text{end}} = \sqrt{\frac{3 \alpha}{2}} \mpl\, \text{arcsinh}\left(\frac{2}{\sqrt{3\alpha
		}}\right) \;.
	\end{equation}
    We note that $\phi_{\text{end}}\approx \sqrt{2} \mpl$ for $\alpha \gtrsim 1$, while $\phi_{\text{end}}$ can be smaller for small $\alpha$.
    Evaluating the number of e-folds \eqref{Nk}
    \begin{align}
    \Nk  = 	\frac{3}{4} \alpha  \left(\cosh \left(\frac{\sqrt{2}\phi_k}{\sqrt{3\alpha }\mpl}\right)-\sqrt{\frac{4}{3 \alpha }+1}\right) \;,
\end{align}
we solve
	\begin{equation}
		\phi_k = \sqrt{\frac{3 \alpha}{2}} \mpl \, \text{arccosh}\left(\sqrt{\frac{4}{3 \alpha }+1} + \frac{4 \Nk}{3 \alpha }\right) \ .
	\end{equation}
   This relation automatically implies
    \begin{align}
		\epsilon_k &= \frac{3 \alpha }{3 \alpha +4 \Nk^2+2  \alpha \Nk \sqrt{\frac{12}{\alpha }+9}
		} \;,\\
		\eta_k & = - \frac{4 \Nk- \alpha\left(6-\sqrt{\frac{12}{\alpha }+9} \right)}{3 \alpha +4
			\Nk^2+2 \alpha  \Nk \sqrt{\frac{12}{\alpha }+9} } \;.
	\end{align}

It is now possible to evaluate CMB observables as a function of $\alpha$ and $\Nk$
\begin{align}
		r &= 16 \epsilon_k =
    \frac{48 \alpha }{3 \alpha +4 \Nk^2+2  \alpha \Nk \sqrt{\frac{12}{\alpha }+9}
		} \;,\\
		n_s & = 1 -\frac{8}{4 \Nk - 3 \alpha +  \alpha\sqrt{\frac{12}{\alpha }+9}}\;.
\end{align}
We note in passing that in a wide range of parameter space, we can approximate
    \begin{equation}
        r \sim \frac{12 \alpha}{\Nk^2} \;, \qquad 1-n_s \sim \frac{2}{\Nk} \;,
    \end{equation}
    but we will use the exact formulas for all computations. When we match with CMB observations \cite{Planck:2018vyg},\footnote{Recent results from the Atacama Cosmology Telescope \cite{AtacamaCosmologyTelescope:2025blo} and the South Pole Telescope \cite{SPT-3G:2025bzu} have hinted at a preference for larger values of $n_s$ compared to those inferred from Planck. However, it has been recently shown \cite{McDonough:2025lzo} that this shift in $n_s$ is primarily driven by a tension between the CMB dataset and the DESI BAO measurements, which is why we  chose not to consider these results for the moment.} the $2\sigma$-constraint $n_s \gtrsim 0.957$ implies
    \begin{equation}
        \Nk \gtrsim 46 \;,
    \end{equation}
    while the bound \cite{BICEP:2021xfz} $r\lesssim 0.03$ yields
    \begin{equation}
        \alpha \lesssim 3\cdot 10^{-3} \Nk^2 \;.
    \end{equation}
    All these arguments did not rely on the observed amplitude of CMB perturbations. Evaluating
	\begin{equation} \label{VOverEps}
		\frac{\Vphi}{\epsilon_k} = \frac{M^4 \left(4\Nk^2 + 3 \alpha +2 \alpha \Nk \sqrt{\frac{12}{\alpha }+9} \right) \tanh ^2\left(\frac{1}{2} \text{arcsech}\left(\frac{3 \alpha
			}{4\Nk+\sqrt{\frac{12}{\alpha }+9} \alpha}\right)\right)}{3 \alpha } \;.
	\end{equation}
    we see that matching the observed value \cite{Planck:2018vyg}, $\Vphi/\epsilon_k \approx 5\cdot 10^{-7} \mpl^4$, fixes $M$ as function of $\Nk$ and $\alpha$. Combined with $r\lesssim 0.03$, this yields a universal upper bound
    \begin{equation}
        \Nk \lesssim 56 \;,
    \end{equation}
    corresponding to instantaneous reheating. We can use Eq.~\eqref{H_k} to evaluate $M$ explicitly: 
\begin{equation}
    \M =
     \mpl
     \left(
     \frac{3\pi^2}{2}
     A_s r
     \right)^{1/4}
     \tanh^{-1}\left(
     \frac{\phi_k}{\sqrt{6\alpha}\mpl}
     \right) \ ,
    \label{M3}
\end{equation}
which is consistent with \eqref{Mgeneral}.

    Finally, we require the vacuum mass of $\phi$. Expanding the potential \eqref{eq:potentialalphaattractor} gives
    \begin{equation} \label{mPhiAlphaAttractor}
        m_\phi = \frac{M^2}{\sqrt{3\alpha} \mpl} \;,
    \end{equation}
    which again can be viewed as function of $\alpha$ and $\Nk$. Importantly, it is evident from Eq.~\eqref{VOverEps} that $M^2/\sqrt{\alpha}$ is approximately fixed by CMB normalisation. 
    Therefore, there is little room to vary $ m_\phi$, up to changing $\Nk$. 
    In order to illustrate this observation, we show plots of $r$ as function of $n_s$ and of $\Phi_I\equiv \Vend/m_\phi^4$ as function of $m_\phi$ in Fig.~\ref{fig:nsvsr_alphaattractor}. 
    In addition, we checked numerically that the approximations made in this derivation hold to a good accuracy. This is illustrated in Fig.~\ref{fig:PhiIvsmphi_alphaattractor_noapprox}, where we show the complete parameter space in the $\Phi_I-m_\phi$ plane consistent with the Planck and BICEP/Keck data at the $2\sigma$ level. The colour indicates the number of e-folds associated with each parameter space point.

\begin{figure}
    \centering
    \includegraphics[width=0.7\textwidth]{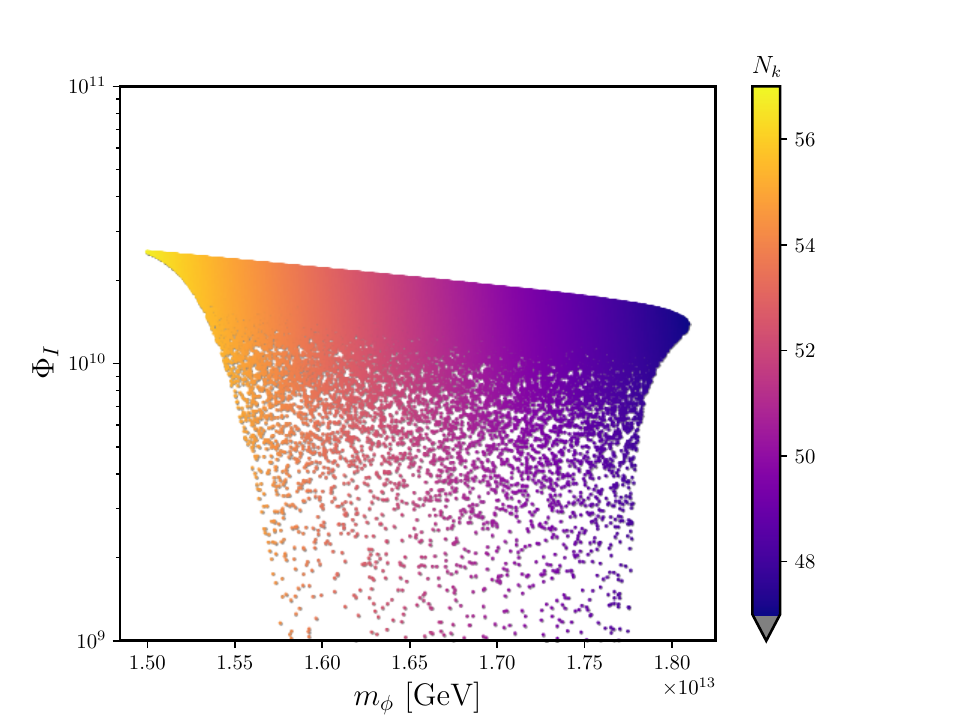}
    \caption{Region of the $\alpha$-attractor T-model parameter space in the $\Phi_I-m_\phi$ plane consistent with the Planck and BICEP/Keck data at the $2\sigma$ level. The colour indicates the number of e-folds $\Nk$ of inflation for a given parameter space points.}
    \label{fig:PhiIvsmphi_alphaattractor_noapprox}
\end{figure}

\subsubsection{Radion Gauge Inflation}

Originally developed in \cite{Fairbairn:2003yx} as a variation over the idea of gauge inflation \cite{Arkani-Hamed:2003wrq,Kaplan:2003aj,Arkani-Hamed:2003xts}, its potential takes the form
\begin{equation}
    \Vphi(\phi) = M^4 \left[ \frac{\left(\phi/\mpl\right)^2}{\alpha + \left(\phi/\mpl\right)^2} \right] \;.
\end{equation}
Proceeding analogously as above, we compute 
\begin{align}
		\epsilon_k &=  \frac{2 \alpha ^2 \mpl^2}{\phi ^2 \left(\alpha + \left(\phi/\mpl\right)^2\right)^2} \;,\\
		\eta_k & = \frac{2 \alpha \mpl^2 \left(\alpha -3 \left(\phi/\mpl\right)^2\right)}{\phi ^2 \left(\alpha + \left(\phi/\mpl\right)^2\right)^2}  \;,
	\end{align}
	and hence $\epsilon=1$ yields
	\begin{equation} \label{phiEndRGI}
		\phi_{\text{end}} = \mpl \frac{\left(\alpha ^2 \tilde{\alpha}\right)^{1/3}-\alpha
   }{\sqrt{3} \left(\alpha ^2 \tilde{\alpha}\right)^{1/6}} \;,
	\end{equation}
    where we defined
    \begin{equation}
        \tilde{\alpha} \equiv \alpha +3 \left(\sqrt{6 \alpha +81}+9\right) \;.
    \end{equation}
As for the $\alpha$-attractor model, we get $\phi_{\text{end}}\approx \sqrt{2} \mpl$ for $\alpha \gtrsim 1$, while smaller $\alpha$ leads to smaller $\phi_{\text{end}}$.

Computing the number of e-folds,
\begin{equation}
    \Nk = \frac{\phi_k^4 - \phi_{\text{end}}^4}{8 \alpha }+\frac{\phi_k^2 - \phi_{\text{end}}^2}{4} \;,
\end{equation}
we solve
\begin{equation}
    \phi_k =2 \mpl \sqrt{\tilde{N}}\;, 
\end{equation}
with
\begin{equation}
   \tilde{N} \equiv \frac{\alpha}{12}  \left(\sqrt{\left(\frac{\alpha }{ \tilde{\alpha}}\right)^{2/3}+2 \left(\frac{\alpha }{ \tilde{\alpha}}\right)^{1/3}+2\left(\frac{\tilde{\alpha}}{\alpha }\right)^{1/3}+\left(\frac{\tilde{\alpha}}{\alpha }\right)^{2/3}+\frac{72 \Nk}{\alpha }+3}-3\right) \;.
\end{equation}
For large values of $\alpha$, we get $\tilde{N} \approx \Nk$ and so $\phi_k \approx 2 \sqrt{\Nk}\mpl$. Evaluating
\begin{align}
    \epsilon_k & = \frac{\alpha ^2}{2 \tilde{N} (\alpha +4 \tilde{N})^2} \;,\\
    \eta_k & =  \frac{\alpha  (\alpha -12 \tilde{N})}{2 \tilde{N} (\alpha +4 \tilde{N})^2} \;,
\end{align}
we arrive at the observables
 \begin{align}
		r &= \frac{8 \alpha ^2}{\tilde{N} (\alpha +4 \tilde{N})^2} \;,\\
		n_s & = 1-\frac{2 \alpha  (\alpha +6 \tilde{N})}{\tilde{N} (\alpha +4 \tilde{N})^2}\;.
	\end{align}
    For large $\alpha$, we can approximate
    \begin{equation}
        r \sim \frac{8}{\Nk^2} \;, \qquad 1-n_s \sim \frac{2}{\Nk} \;.
    \end{equation}
Since $n_s$ has the same scaling as for the $\alpha$-attractor, we also get 
\begin{equation}
    46 \lesssim \Nk \lesssim 56 \;.
\end{equation}
In contrast, small $\alpha$ gives
\begin{equation}
        r \sim \frac{\sqrt{2\alpha}}{\Nk^{3/2}} \;, \qquad 1-n_s \sim \frac{3}{2 \Nk} \;.
    \end{equation}
    Therefore, the observational constraint $0.957 \lesssim n_s \lesssim 0.973$ can be satisfied in the wider window 
\begin{equation}
    35 \lesssim \Nk \lesssim 56 \;.
\end{equation}
We see that that $n_s$ and $r$ do not bound $\alpha$, but its value influences the admissible range of $\Nk$.

Finally, matching the CMB amplitude with
\begin{equation}
		\frac{\Vphi}{\epsilon_k} = \frac{8 M^4 \tilde{N}^2 (\alpha +4 \tilde{N})}{\alpha ^2} \;,
\end{equation}
we can fix $M$ as a function of $\alpha$ and $\Nk$. The requirement $M\lesssim \mpl$ gives a mild upper bound $\alpha \lesssim 10^{10}$, while for small $\alpha$ we get
\begin{equation}
   	\frac{\Vphi}{\epsilon_k} \approx \frac{8 \sqrt{2}  M^4 \Nk^{3/2}}{\sqrt{\alpha }} \;.
\end{equation}
As before, we can use Eq.~\eqref{H_k} to evaluate $M$ explicitly,
\begin{equation}
    \M =
     \mpl
     \left(
     \frac{3\pi^2}{2} r A_s
     \left(1 + \alpha \frac{\mpl^2}{\phi_k^2}\right)
     \right)^{1/4} \ ,
    \label{M2}
\end{equation}
and expanding the potential around the origin, we obtain
 \begin{equation}\label{mphiRGI}
        m_\phi = \frac{\sqrt{2} M^2}{\sqrt{\alpha} \mpl} \;.
    \end{equation}
For $\alpha \gtrsim 4 N$, we see that CMB normalisation fixes $M^2/\sqrt{\alpha}$ and hence $ m_\phi$, as was also the case for the $\alpha$-attractor. However, smaller values of $\alpha$ allow for different scalings leading to larger values of $ m_\phi$. 
We display plots of $r$ as function of $n_s$ and of $\Phi_I$ as function of $m_\phi$ in Fig.~\ref{fig:nsvsr_RGI}. Moreover, similarly as for the $\alpha$-attractor T-model, we display in Fig.~\ref{fig:PhiIvsmphi_RGI_noapprox} the complete parameter space in the $\Phi_I-m_\phi$ plane consistent with the Planck and BICEP/Keck data at the $2\sigma$ level. The colour indicates the number of e-folds associated with each parameter space point. As one can see, the slight approximations made in this section hold well.

    \begin{figure}
    \centering
    \includegraphics[width=0.49\textwidth]{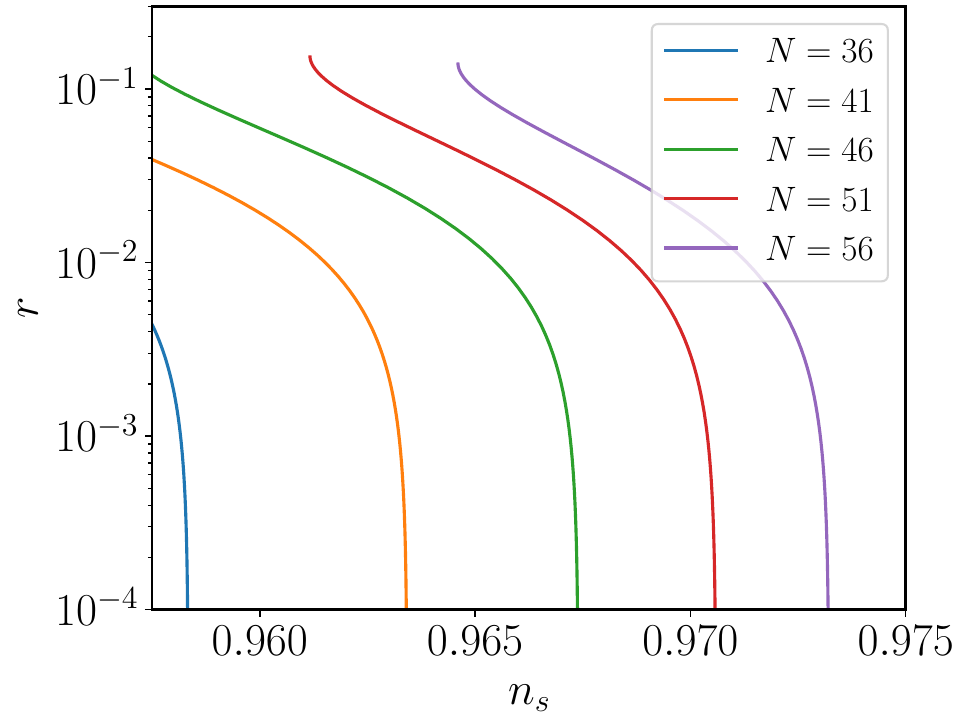}
    \includegraphics[width=0.49\textwidth]{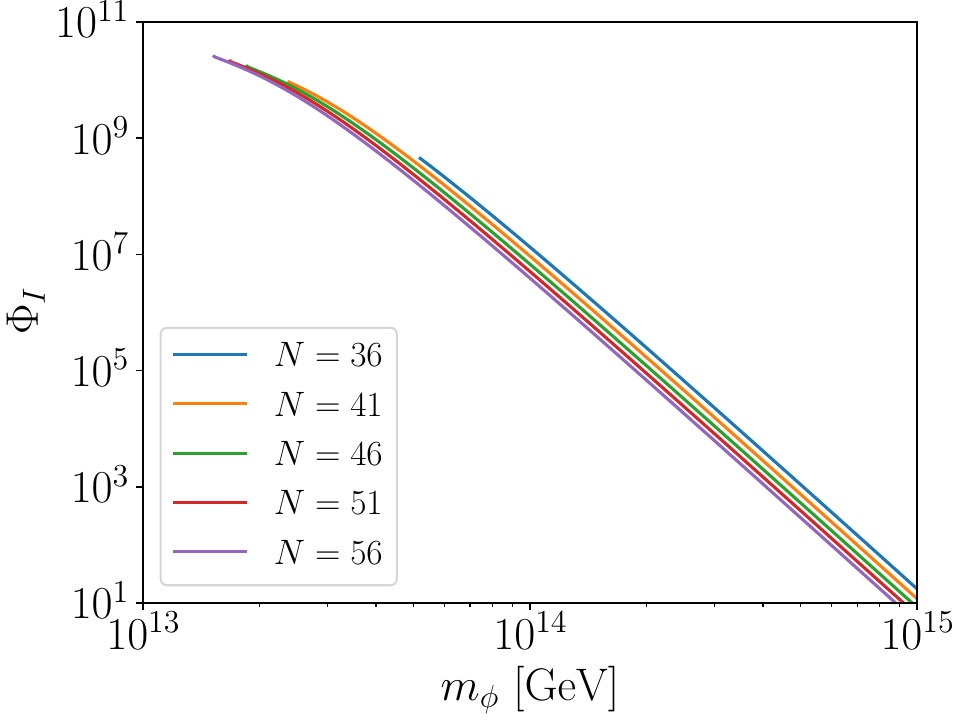}
    \caption{(\textit{Left}) Tensor-to-scalar ratio $r$ as a function of the spectral index $n_s$ for the radion gauge inflation model and fixed values of the number of e-folds $\Nk$, consistent with the Planck and BICEP/Keck data. The purple, red, blue, orange and green lines correspond to $\Nk=36$, $\Nk=41$, $\Nk=46$, $\Nk=51$, and $\Nk = 56$, respectively. (\textit{Right}) Inflaton energy density $\Phi_I$ at the beginning of reheating as a function of the inflaton mass $m_\phi$ assuming the same choices for the number of e-folds and inflationary scenario. The colour scheme is identical to the left panel.}
    \label{fig:nsvsr_RGI}
\end{figure}

\begin{figure}
    \centering
    \includegraphics[width=0.7\textwidth]{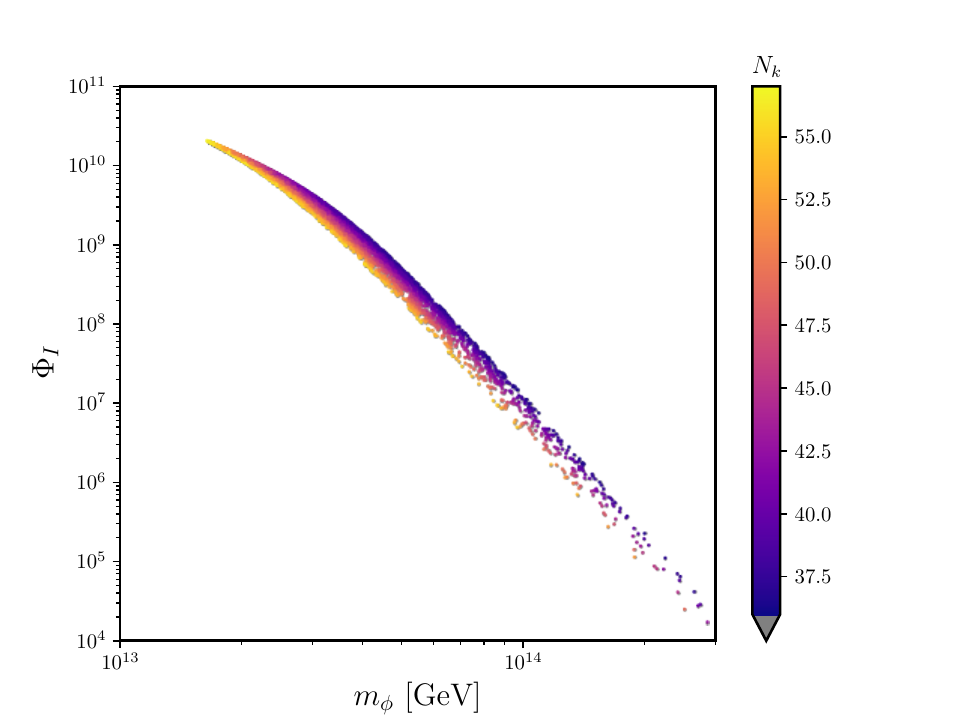}
    \caption{Region of the radion gauge inflation model parameter space in the $\Phi_I-m_\phi$ plane consistent with the Planck and BICEP/Keck data at the $2\sigma$ level. The colour indicates the number of e-folds $\Nk$ of inflation for a given parameter space points.}
    \label{fig:PhiIvsmphi_RGI_noapprox}
\end{figure}

\subsubsection{Mutated Hilltop Inflation}

Originally developed in \cite{Pal:2009sd,Pal:2010eb} as a refined version of hilltop inflation \cite{Boubekeur:2005zm,Kohri:2007gq}, its potential takes the form
\begin{equation}
    \Vphi(\phi) = M^4 \left[ 1-\frac{1}{\cosh\left(\frac{\phi}{\alpha \mpl}\right)} \right].
\end{equation}
Again performing the same analysis as above, we get
\begin{align}
		\epsilon_k &=  \frac{1}{2 \alpha^2 \cosh^2\left(\frac{\phi}{\alpha \mpl}\right)\tanh^2\left(\frac{\phi}{2 \alpha \mpl}\right)} \;, \label{epsilonPhiMHI}\\
		\eta_k & = -  \frac{\cosh\left(\frac{\phi}{\alpha \mpl}\right)-3 }{4 \alpha^2 \cosh^2\left(\frac{\phi}{\alpha \mpl}\right)\sinh^2\left(\frac{\phi}{2 \alpha \mpl}\right)} \;, \label{etaPhiMHI}
\end{align}
and the condition $\epsilon=1$ gives \cite{Drewes:2023bbs}
\begin{equation}
		\phi_{\text{end}} = \alpha \mpl   \text{arccosh}\left(\frac{2^{2/3}\left(3 + 2 \alpha ^2\right) +2^{1/3} b^{2/3}+2 \alpha  b^{1/3}}{6 \alpha  b^{1/3}}\right)\;,
	\end{equation}
    where
    \begin{equation}
        a = \sqrt{4 \alpha ^4+22 \alpha ^2-1} \;, \qquad b = 3\sqrt{6} a +36 \alpha +  4 \alpha ^3 \;.
    \end{equation}
    We see that $b$ is real only if $\alpha$ is larger than
    \begin{equation}
        \alpha_{\text{crit}} = \frac{\sqrt{5 \sqrt{5}-11}}{2} \approx 0.21 \;.
    \end{equation} 

For the subsequent analytic analysis, we shall assume that $\alpha > \alpha_{\text{crit}}$. Then we get for the number of e-folds
\begin{equation}
    \Nk = \alpha ^2 \left(2 \ln \left(\cosh \left(\frac{\phi_{\text{end}}}{2 \alpha\mpl }\right)
   \text{sech}\left(\frac{\phi_k}{2 \alpha \mpl}\right)\right)-\cosh \left(\frac{\phi_{\text{end}}}{\alpha \mpl}\right)+\cosh \left(\frac{\phi_k}{\alpha \mpl}\right)\right) \;.
\end{equation}
Unlike in the other inflationary scenarios studied before, we now need to use an approximation in order to solve for $\phi$. 
However, one can easily verify that $\cosh(x) \gg 2\ln \cosh(x/2)$ for any values of $x$ such that we can neglect all logarithmic terms.
We then obtain
\begin{align}
    \phi_k &\approx \alpha \mpl  \text{arccosh}\left(\frac{\Nk}{\alpha
   ^2} + \cosh \left(\frac{\phi_{\text{end}}}{\alpha \mpl}\right)\right)\nonumber\\
    & = \alpha \mpl  \text{arccosh}\left(\frac{\Nk}{\alpha ^2} + \frac{2^{2/3} \left(3 + 2 \alpha ^2\right) +2^{1/3} b^{2/3}+2 \alpha  b^{1/3}}{6 \alpha  b^{1/3}}\right)\;.
\end{align}
Plugging this into Eqs.~\eqref{epsilonPhiMHI} and \eqref{etaPhiMHI} determines the slow-roll indices as function of $\Nk$, and we can approximate for $\alpha<1$:
 \begin{equation}
        r \sim \frac{8 \alpha^2}{\Nk^2} \;, \qquad 1-n_s \sim \frac{2}{\Nk} \;.
    \end{equation}
   Again $n_s$ has the same scaling as in the scenarios studied before, and so we get 
\begin{equation}
    46 \lesssim \Nk \lesssim 56 \;.
\end{equation}
    
Still using the same approximation of small $\alpha$, we obtain for the amplitude of CMB perturbations
\begin{equation}
		\frac{\Vphi}{\epsilon_k} = \frac{2 M^4 \Nk^2}{\alpha ^2} \;,
\end{equation}
which determines $M$ as function of $\alpha$ and $\Nk$. Explicitly, Eq.~\eqref{H_k} gives
\begin{equation}
    \M=\mpl\sqrt[4]{\frac{3\pi^2r A_s}{2[1-1/{\rm cosh}(\phi_k/(\alpha \mpl))]}} \ .
    \label{M1}
\end{equation}
As last step, we expand the potential around the origin to obtain
 \begin{equation}
        m_\phi = \frac{M^2}{\alpha \mpl} \;,
    \end{equation}
    Thus, CMB normalisation determines 
    \begin{equation}\label{mphiMHI}
        m_\phi =  \frac{1.2\cdot 10^{15}\, \text{GeV}}{\Nk} \;,
    \end{equation}
    and so -- within the approximations we made -- $m_\phi$ must lie in the small interval between $2.1\cdot 10^{13}\, \text{GeV}$ and $2.6\cdot 10^{13}\, \text{GeV}$. 
    We conclude that we cannot go to small $m_\phi$ as long as $\alpha>\alpha_{\text{crit}}$. In this parameter space, we display plots of $r$ as function of $n_s$ and of $\Phi_I\equiv \Vend/m_\phi^4$ as function of $m_\phi$ in Fig.~\ref{fig:nsvsr_MHI}.
    Moreover, similarly as in the previous two models, we show in Fig.~\ref{fig:PhiIvsmphi_MHI_noapprox} the complete parameter space in the $\Phi_I-m_\phi$ plane consistent with the Planck and BICEP/Keck data at the $2\sigma$ level. 
    The colour indicates the number of e-folds associated with each parameter space point. 
    We note in passing that the parameter space region with large $\Phi_I$ and relatively small $m_\phi$, which appears to clearly deviate from the expectation based on our analytical approximations, cf.~Fig.~\ref{fig:nsvsr_MHI}, corresponds to points with large values of $\alpha \gtrsim 1$, for which the approximations made in this section do not necessarily hold.

        \begin{figure}
    \centering
    \includegraphics[width=0.49\textwidth]{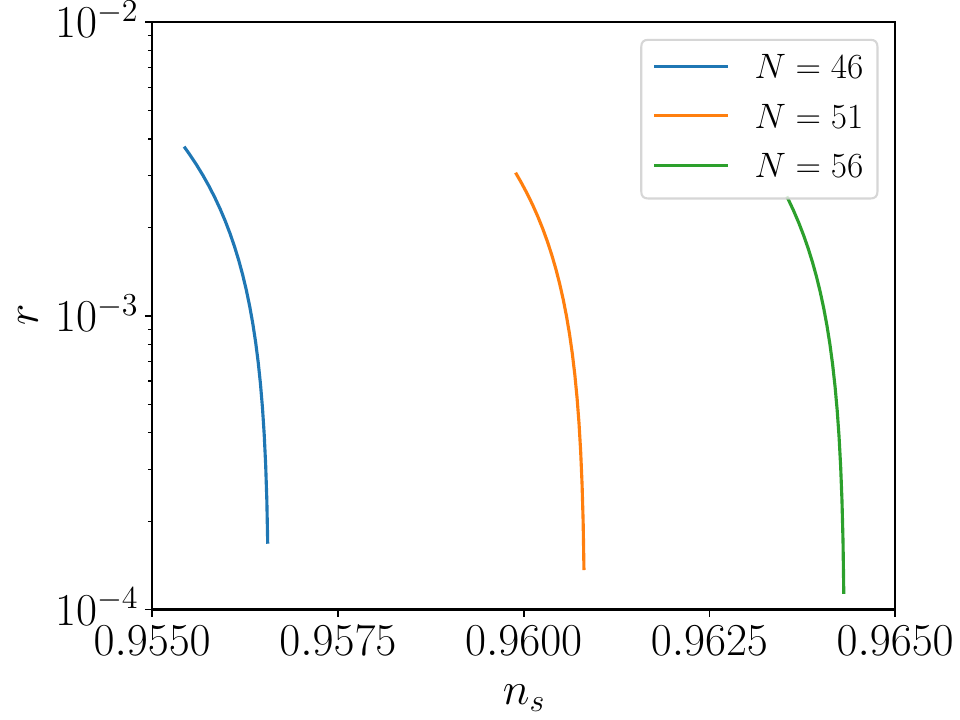}
    \includegraphics[width=0.49\textwidth]{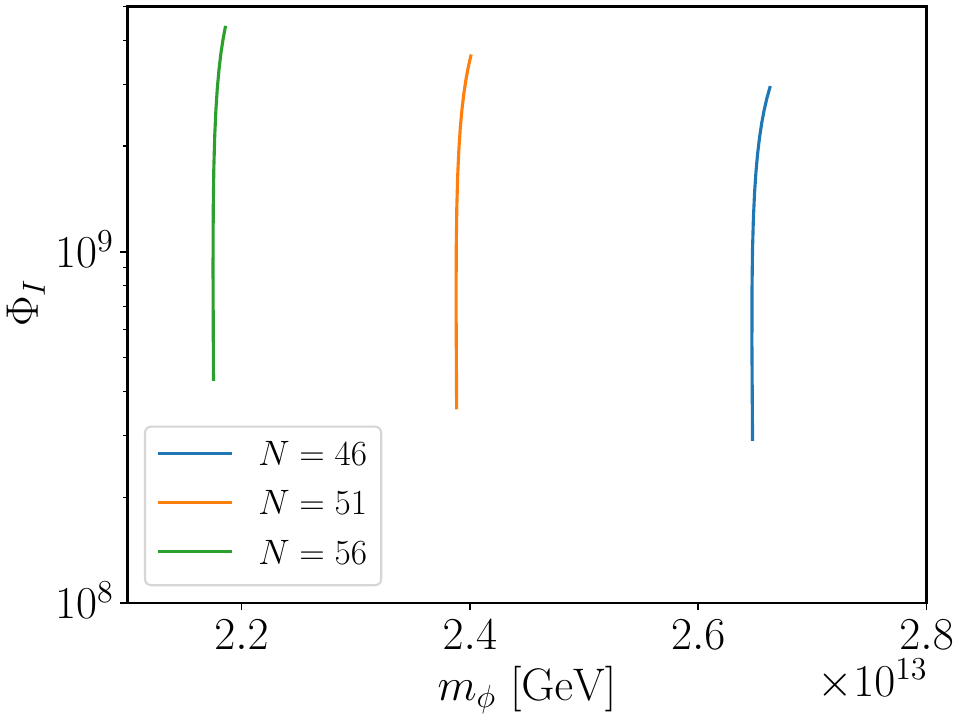}
    \caption{(\textit{Left}) Tensor-to-scalar ratio $r$ as a function of the spectral index $n_s$ for the mutated hilltop inflation model and fixed values of the number of e-folds $\Nk$, consistent with the Planck and BICEP/Keck data. The blue, orange and green lines correspond to $\Nk=46$, $\Nk=51$, and $\Nk = 56$, respectively. (\textit{Right}) Inflaton energy density $\Phi_I$ at the beginning of reheating as a function of the inflaton mass $m_\phi$ assuming the same choices for the number of e-folds and inflationary scenario. The colour scheme is identical to the left panel.}
    \label{fig:nsvsr_MHI}
\end{figure}

\begin{figure}
    \centering
    \includegraphics[width=0.7\textwidth]{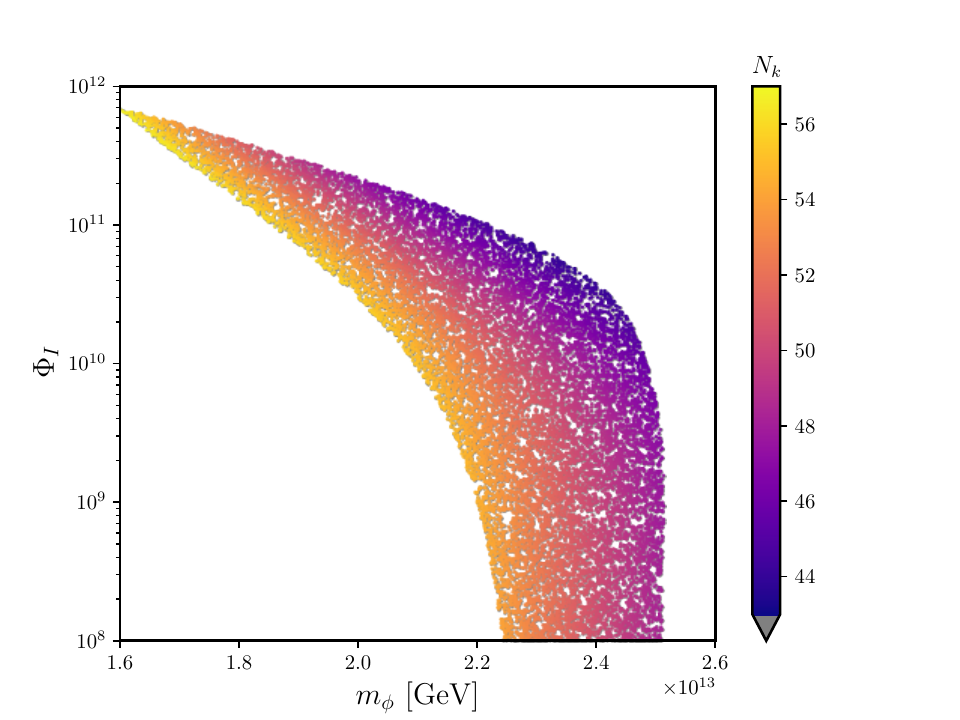}
    \caption{Region of the mutated hilltop inflation model parameter space in the $\Phi_I-m_\phi$ plane consistent with the Planck and BICEP/Keck data at the $2\sigma$ level. The colour indicates the number of e-folds $\Nk$ of inflation for a given parameter space points.}
    \label{fig:PhiIvsmphi_MHI_noapprox}
\end{figure}

\section{Fundamental coupling of $\alpha$-attractors}\label{YukawainalphaT}
In the original proposal of the $\alpha$-attractor T-model, the fundamental action is \cite{Kallosh:2013hoa}
\begin{equation} \label{alphaAttractorFundamentalAction}
S = \int \d^4x\,\sqrt{-g}\,\Bigg[
\frac{1}{12}\big(\chi^2 - \varphi^2\big) R
+ \frac{1}{2} \partial_\mu\chi\partial^\mu\chi
- \frac{1}{2} \partial_\mu\varphi\partial^\mu\varphi
-f\left(\frac{\varphi}{\chi}\right)\big(\chi^2 - \varphi^2\big)^2
\Bigg] \;.
\end{equation}
Here $\chi$ and $\varphi$ are two scalar fields, and the action is fixed by two requirements. The first one is Weyl invariance, and the second one is $SO(1,1)$-symmetry between $\chi$ and $\varphi$, which is only mildly broken by a non-constant $f(\varphi/\chi)$. After fixing the gauge via $\chi^2 - \varphi^2 = 6\mpl^2$, the action becomes
\begin{equation}
S = \int \d^4x\,\sqrt{-g}\,\Bigg[
\frac{1}{2} \mpl^2 R
-\frac{1}{2}
\frac{6\mpl^2 }{\varphi^2+6\mpl^2}
\partial_\mu\varphi\partial^\mu\varphi
-36\mpl^4
f\left(\frac{\varphi}{\sqrt{\varphi^2+ 6\mpl^2}}\right)
\Bigg] \;.
\end{equation}
Introducing the canonically normalised field $\phi$ via 
\begin{equation}
\phi =\sqrt{6} \mpl\, \mathrm{arcsinh}\left(\frac{\varphi}{\sqrt{6}\mpl}\right) \;,
\end{equation}
we arrive at
\begin{equation} \label{alphaAttractorEinstein}
S = \int \d^4x\,\sqrt{-g}\,\left[
\frac{1}{2}\mpl^2 R -\frac{1}{2}
\partial_\mu\phi \partial^\mu\phi
- 36 \mpl^4 f\Big(\tanh \left(\frac{\phi}{\sqrt{6}\mpl}\right)\Big)
\right] \;.
\end{equation}
For $f(x)=\lambda x^{2n}$, this gives the inflaton potential
\begin{equation}
    \Vphi = M^4 \tanh^{2n} \left(\frac{\phi}{\sqrt{6}\mpl}\right) \;, \qquad M\equiv \lambda^{1/4} \sqrt{6}\mpl \;.
\end{equation}
Famously, there was no $\alpha$ in \cite{Kallosh:2013hoa} -- this was only introduced later \cite{Kallosh:2013yoa}.

Now it is clear how to add a fermion. In the action \eqref{alphaAttractorFundamentalAction}, we add
\begin{equation}
    S \rightarrow S +  \int \d^4x\,\sqrt{-g}\,\Bigg[\left(i \bar{\Psi} \slashed{D} \Psi + \text{h.c.}\right) + y \sqrt{\chi^2-\varphi^2} g\left(\frac{\varphi}{\chi}\right) \bar{\Psi}\Psi \Bigg] \;.
\end{equation}
After gauging fixing and transforming to a canonical scalar field, the Einstein-frame action \eqref{alphaAttractorEinstein} is supplemented as
\begin{equation}
    S \rightarrow S + \int \d^4x\,\sqrt{-g}\,\Bigg[\left(i \bar{\Psi} \slashed{D} \Psi + \text{h.c.}\right) + \sqrt{6} y \mpl g\Big(\tanh \left(\frac{\phi}{\sqrt{6}\mpl}\right)\Big)\bar{\Psi}\Psi \Bigg] \;.
\end{equation}
To match a Yukawa coupling, we choose $g(x)=x$ and get
\begin{equation}
    \mathcal{L}_\phi \supset \sqrt{6} y \mpl \tanh \left(\frac{\phi}{\sqrt{6}\mpl}\right) \bar{\Psi}\Psi \;.
\end{equation}
By analogy to how the inflaton potential is generalised to arbitrary $\alpha$ \cite{Kallosh:2013yoa}, this should give\footnote{We note that in inflationary models based on a non-minimally coupling to gravity, the matter sector is generically strongly altered by the conformal transformation that is necessary to derive the Einstein-frame action \cite{Rigouzzo:2025hza,Rigouzzo:2025ycb}.}
\begin{equation}
    \mathcal{L}_\phi \supset \sqrt{6\alpha} y \mpl \tanh \left(\frac{\phi}{\sqrt{6\alpha}\mpl}\right) \bar{\Psi}\Psi \;.
\end{equation}
In the case of the $\alpha$-attractor model, we know that $\alpha$ must be sufficiently large, i.e. $\alpha \gtrsim 0.25$, to fulfil condition \eqref{GeneralScalingSelf}.
In  this regime, the field value during reheating remains roughly sub-Planckian $\phi \lesssim \phi_{\rm end} \simeq \sqrt{2}\mpl$ during the entirety of the reheating era such that the $\tanh$ can be expanded around its minimum to a good approximation.
Essentially, the strength of the higher-order corrections to the inflaton–fermion Yukawa coupling is controlled by the same parameter, i.e., $\phi_{\rm end}/\sqrt{6\alpha}\mpl$, that determines when reheating enters the non-perturbative regime.
As an illustration, for the benchmark chosen in Sec.~\ref{sec:counterexamples} verifying $\alpha = 2$, the relative difference between the full $\tanh$ and its first order expansion is of order $5\%$.

\bibliographystyle{JHEP}
\bibliography{main}

\end{document}